\PassOptionsToPackage{table}{xcolor}
\documentclass[aps,prx,twocolumn,superscriptaddress,10pt]{revtex4-2}

\usepackage{csquotes}
\usepackage{booktabs}
\usepackage{multirow}
\usepackage[inline]{enumitem}
\setlist[enumerate]{label=(\roman*)}
\usepackage{mathtools}
\usepackage[colorlinks]{hyperref}
\usepackage[capitalise]{cleveref}
\usepackage[caption=false]{subfig}
\usepackage{xcolor}
\usepackage{tikz}

\usepackage{qcd}
\newenvironment{subcaptions}[1]{%
        \begin{tikzpicture}[every node/.style={inner sep=0}]%
        \node[use as bounding box,anchor=north west] (image) at (0,0) {#1};%
        \begin{scope}[x={(image.north east)},y={(image.south west)}]%
    }{%
        \end{scope}\end{tikzpicture}%
}
\newcommand*{\oversubcaption}[2]{\node[anchor=north west, transparent] at (#1) {\subfloat[#2]{\hspace{2ex}}};}
\newcommand\sref[1]{(\protect\subref{#1})}

\begin{document}

\title{Hadrons in (1+1)D Hamiltonian hardcore lattice QCD}

\author{Marco Rigobello}
\email{marco.rigobello.2@phd.unipd.it}
\affiliation{Dipartimento di Fisica e Astronomia ``G. Galilei'', Universit\`a di Padova, I-35131 Padova, Italy.}
\affiliation{Padua Quantum Technologies Research Center, Universit\`a degli Studi di Padova.}
\affiliation{Istituto Nazionale di Fisica Nucleare (INFN), Sezione di Padova, I-35131 Padova, Italy.}

\author{Giuseppe Magnifico}
\affiliation{Dipartimento di Fisica e Astronomia ``G. Galilei'', Universit\`a di Padova, I-35131 Padova, Italy.}
\affiliation{Padua Quantum Technologies Research Center, Universit\`a degli Studi di Padova.}
\affiliation{Istituto Nazionale di Fisica Nucleare (INFN), Sezione di Padova, I-35131 Padova, Italy.}
\affiliation{Dipartimento di Fisica, Universit\`a di Bari, I-70126 Bari, Italy.}

\author{Pietro Silvi}
\affiliation{Dipartimento di Fisica e Astronomia ``G. Galilei'', Universit\`a di Padova, I-35131 Padova, Italy.}
\affiliation{Padua Quantum Technologies Research Center, Universit\`a degli Studi di Padova.}
\affiliation{Istituto Nazionale di Fisica Nucleare (INFN), Sezione di Padova, I-35131 Padova, Italy.}

\author{Simone Montangero}
\affiliation{Dipartimento di Fisica e Astronomia ``G. Galilei'', Universit\`a di Padova, I-35131 Padova, Italy.}
\affiliation{Padua Quantum Technologies Research Center, Universit\`a degli Studi di Padova.}
\affiliation{Istituto Nazionale di Fisica Nucleare (INFN), Sezione di Padova, I-35131 Padova, Italy.}

\begin{abstract}
    We study 2-flavor Hamiltonian lattice QCD in (1+1)D with hardcore gluons, at zero and finite density, by means of matrix product states. We introduce a formulation of the theory where gauge redundancy is absent and construct a gauge invariant tensor network ansatz. We show that the model is critical in an extended subregion of parameter space and identify at least two distinct phases, one of which embeds the continuum limit location. We reconstruct a subset of the particle spectrum in each phase, identifying edge and bulk gapless modes. We thereby show that the studied model provides a minimal SU(3) gauge theory whilst reproducing known phenomena of (3+1)D QCD. Most notably, it features charged pions.
\end{abstract}

\maketitle

\section{Introduction}

\begin{figure}[t]
    \includegraphics[width=0.88\columnwidth]{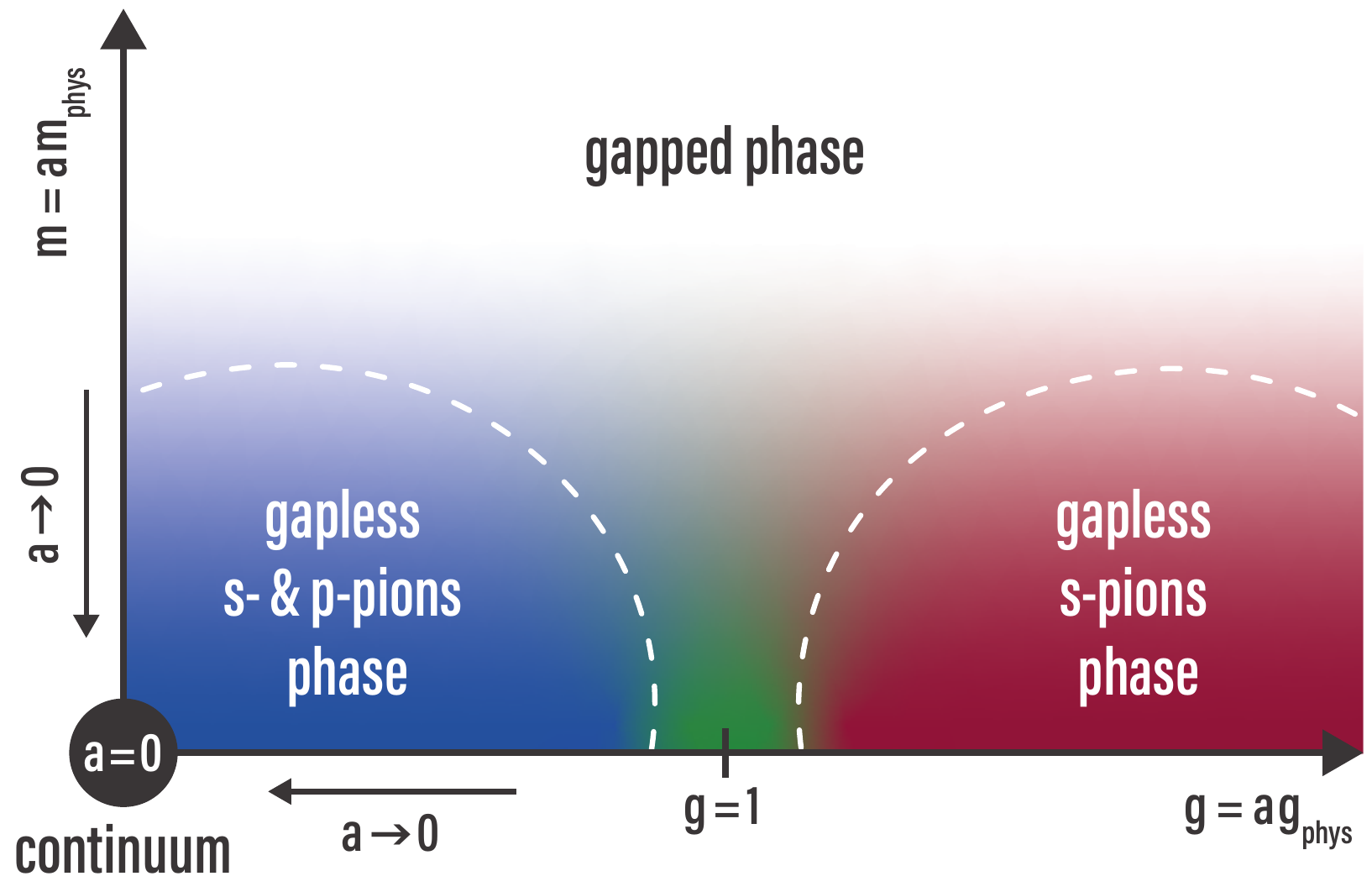}
    \caption{\label{fig:phases}%
        Sketch of the phase diagram of \themodel{} in the $(\cpl,m)$-plane, $m=m_u=m_d$ (quark masses).
        The arrows direct towards the continuum, where the lattice correlation length $\corrlen=\corrlen_{\text{phys}}/a$ of physical excitations diverges.
        The model is gapped above a threshold bare quark mass of order $m\sim10^{-2}$, and gapless below.
        Within the gapless region, the dashed ellipses highlight the weak ($\cpl\ll1$, blue) and strong ($\cpl\gg1$, red) coupling phases, which attracted most of this work's focus.
        At their interface lays the intermediate coupling regime ($\cpl\approx1$, green).
        Both the small and large $\cpl$ phases feature gapless charged pions $\pi^{\pm}$ ($s$- and $p$-wave for $\cpl\ll1$, only $s$-wave for $\cpl\gg1$).
        Circumstantial evidence suggests the neutral pions $\pi^0$ are also gapless; protons, neutrons and delta baryons were found to be gapped (at finite $g$).
    }
\end{figure}

Quantum Chromodynamics (QCD) \cite{Fritzsch1973AdvantagesColorOctet,Wilson1974ConfinementQuarks,Brambilla2014QcdStronglyCoupled} is the sector of the Standard Model of particle physics responsible for the description of the quark and gluon fields, and their strong interactions.
Asymptotic freedom ensures that, at short length scales, these fields manifest as almost free parton particles, thus allowing for a perturbative expansion in a small coupling parameter.
Conversely, at wavelengths of the order of the size of a proton, quarks and gluons confine, perturbative techniques become unviable, and a plethora of color-neutral hadron bound states and resonances emerge.
Hadron masses make up the majority of visible matter and can be determined directly via collider experiments \cite{Duerr2008AbInitioDetermination}.
Consequently, understanding QCD at these scales is essential to our knowledge of the physical universe and to validate high-energy theories against experimental evidence, \eg{} from the Large Hadron Collider (LHC).
Lattice Monte Carlo (MC) numerical methods have long been employed in attacking this formidable challenge,
producing outstanding predictions for hadron masses and decay rates,
and elucidating the mechanisms of color confinement and chiral symmetry breaking, as well as the thermal properties of QCD \cite{Lin2022HadronSpectroscopyStructure,Detar2004LatticeQuantumChromodynamics}.
Despite this vast array of successes, MC methods are plagued with the notorious sign problem in a variety of physically relevant scenarios, such as finite baryon number density and real-time dynamics \cite{Detar2004LatticeQuantumChromodynamics,Grabowska2013SignProblemsNoise,Nagata2022FiniteDensityLattice}.
Especially in these regimes, there is strong demand for alternative, non-perturbative strategies aiming for a complete characterization of QCD's collective phenomena, such as its phase diagram \cite{Kogut2003PhasesQuantumChromodynamicsa,Banuls2021TensorsCastTheir,Borsanyi2020QcdCrossoverFinite}.

In this work, we numerically characterize the phase diagram and spectral properties of Hamiltonian Lattice QCD with two matter flavors in one spatial dimension, under a hardcore gluon approximation.
Our results, summarized in \cref{fig:phases}, are obtained by means of Tensor Networks (TN) \cite{Verstraete2008MatrixProductStates,Orus2014PracticalIntroductionTensor,Montangero2018IntroductionTensorNetwork,Silvi2019TensorNetworksAnthology}, a framework that has shown vast potential for quantum many-body physics in the last decades.
Rather than sampling partition functions in Euclidean space,
numerical TN methods rely on the canonical formalism to variationally optimize many-body wave functions; they are thus immune to sign problems \cite{Emonts2023FindingGroundState}.
TN tame the exponential growth of the Hilbert space with the system size by efficiently compressing the wave function.
Multiple families of TN sate ans\"atze have been developed.
Their effectiveness (and limitations) are rooted in results from quantum information theory regarding the entanglement content of physically relevant many-body states, with each ansatz being tailored to a
specific pattern of correlations \cite{Eisert2010ColloquiumAreaLaws}.
Notable examples include: Matrix Product States (MPS) \cite{Accardi1981TopicsQuantumProbability,Fannes1989ExactAntiferromagneticGround,Oestlund1995ThermodynamicLimitDensity}, encoding one-dimensional (1D) area law entanglement; PEPS \cite{Verstraete2004RenormalizationAlgorithmsQuantum,Verstraete2006CriticalityAreaLaw}, generalizing the MPS construction to higher dimensional lattices; and finally TTN \cite{Fannes1992GroundStatesVbs,Felser2021EfficientTensorNetwork,Ferrari2022AdaptiveWeightedTree} and MERA \cite{Vidal2007EntanglementRenormalization}, capturing critical 1D correlations.
Originally conceived in the context of quantum spin chains \cite{Verstraete2008MatrixProductStates,Fannes1992FinitelyCorrelatedStates}, TN quickly spread to a diverse set of problems in condensed matter physics, statistical physics \cite{Levin2007TensorRenormalizationGroup}, and recently even  quantum chemistry and machine learning \cite{Stoudenmire2016SupervisedLearningTensor,Felser2021QuantumInspiredMachine}.
Of particular relevance here, is their application to Abelian and non-Abelian Lattice Gauge Theories (LGT).
With it, approximatively a decade ago, TN methods started leaking into high-energy physics domain \cite{Rico2014TensorNetworksLattice,Tagliacozzo2014TensorNetworksLattice,Silvi2014LatticeGaugeTensor,Pichler2016RealTimeDynamics, Ercolessi2018PhaseTransitionsZn, Magnifico2020RealTimeDynamics, Banuls2017EfficientBasisFormulation,Banuls2019TensorNetworksTheir,Banuls2020SimulatingLatticeGauge,Zohar2021WilsonLoopsArea,Felser2021QuantumInspiredMachine,Montangero2022LoopFreeTensor,Bloch2022GrassmannTensorNetwork,DiMeglio2023QuantumComputingHigh,Bauer2023QuantumSimulationHigh}.

Recent years have witnessed
a surge in proposals
\cite{Hauke2013QuantumSimulationLattice,Banerjee2013AtomicQuantumSimulation,Zohar2016QuantumSimulationFundamental,GonzalezCuadra2022HardwareEfficientQuantum,Kruckenhauser2022HighDimensionalSo4,Knaute2022RelativisticMesonSpectra,Nguyen2022DigitalQuantumSimulation,Davoudi2023GeneralQuantumAlgorithms,Bazavan2023SyntheticZ2Gauge,Turco2023TowardsQuantumSimulation}
and realizations
\cite{Martinez2016RealTimeDynamics,Bruzewicz2019TrappedIonQuantum,Wintersperger2020RealizationAnomalousFloquet,Bluvstein2021ControllingQuantumMany,Mueller2022QuantumComputationDynamical,Pomarico2023DynamicalQuantumPhase}
of quantum simulation and computation platforms.
On par with TN, one of the goals of this program is that of providing non-perturbative sign-problem-free routes for quantum many-body computations, a prominent example being lattice QCD
\cite{Wiese2014TowardsQuantumSimulating,Funcke2023ReviewQuantumComputing,Farrell2023PreparationsQuantumSimulationsa,Farrell2023PreparationsQuantumSimulations,Chawdhry2023QuantumSimulationColour,Bauer2023QuantumSimulationFundamental,Ciavarella2023QuantumSimulationLattice,Kadam2023LoopStringHadron,Wang2023QuantumSimulationHadronic,Qian2022SolvingHadronStructures,Yao2022QuantumSimulationLight,Barata2022MediumInducedJet,Barata2021QuantumStrategyCompute,Berges2021QcdThermalizationAb,NuQSCollaboration2020PartonPhysicsQuantum}.
Quantum-based approaches, especially analog quantum simulation, are also traditionally formulated in the canonical picture \cite{Farrelly2020DiscretizingQuantumField}.
Because of this, they share with TN some of their strengths as well as many challenges.
The implications for TN are twofold: on one hand, they become natural candidates to benchmark early-stage quantum devices; on the other hand, they provide a bridge between classical and quantum simulation, where some of the advances on one front transfer to the other.
Areas where such cross-fertilization played a role are the protection of gauge symmetries \cite{Halimeh2022GaugeProtectionNon,Raychowdhury2020SolvingGausssLaw,Mathew2022ProtectingLocalGlobal} and the truncation of the unbound gauge fields in LGTs with continuous groups \cite{Jakobs2023CanonicalMomentaDigitized,Bauer2023NewBasisHamiltonian,Alexandru2023FuzzyGaugeTheory}.
Schemes put forward in the context of quantum simulation have found thriving applications in TN algorithms; these encompass
\begin{enumerate*}
    \item quantum link models (QLMs) \cite{Horn1981FiniteMatrixModels,Chandrasekharan1997QuantumLinkModels,Chandrasekharan1997QuantumLinkModels,Wiese2021QuantumLinkModels},
    \item discrete subgroup approximations \cite{Horn1979HamiltonianApproachZn,Notarnicola2015DiscreteAbelianGauge,Alexandru2022SpectrumDigitizedQcd},
    \item $q$-deformation of Lie algebras \cite{Bimonte1996Suq2LatticeGauge,Zache2023QuantumClassicalSpin,Hayata2023StringNetFormulation,Hayata2023BreakingNewGround},
          and
    \item projection onto low-dimensional irreducible representations (irreps) of the gauge group \cite{Zohar2015FormulationLatticeGauge}.
\end{enumerate*}

Here we introduce a simplified model of (1+1)D 2-flavor QCD and study it by means of MPS methods, most notably the Density Matrix Renormalization Group (DMRG) \cite{White1992DensityMatrixFormulation,Schollwoeck2011DensityMatrixRenormalization,Hauschild2018EfficientNumericalSimulations}.
To make QCD amenable to TN methods or quantum simulation, we combine the Hamiltonian formulation of LGT by Kogut and Susskind \cite{Kogut1975HamiltonianFormulationWilsons,Banks1976StrongCouplingCalculations,Susskind1977LatticeFermions} with a gauge field truncation in irrep space.
We consider only the strictest possible truncation, here labeled \emph{hardcore gluon} approximation in analogy to atomic physics, and do not attempt a finite truncation extrapolation.
We lay out a recipe supplying the building blocks for a gauge invariant TN state ans\"atze or quantum simulation protocol.
In the light of future quantum simulation implementations, it is crucial to single out and characterize models of minimal complexity which share as many features as possible with the theory of interest\emdash{}here QCD.
Our TN analysis shows that the maximally-truncated model studied here provides a minimal realization of a \gaugegroup{} gauge theory reproducing part of the particle spectrum of (1+3)D QCD\emdash{}namely, charged pions.

The manuscript is organized as follows: in \cref{sec:model} we define the Hamiltonian of the (untruncated) model, reviewing the key ingredients of Kogut-Susskind LGT.
In \cref{sec:methods} we truncate the gauge field, construct the gauge invariant TN ansatz, and sketch the TN toolbox employed.
\Cref{sec:results} is devoted to the analysis of numerical results:
we show that the model admits a continuum limit (\cref{sec:results-criticalty}) and compare its particle spectrum to that of ordinary QCD (\cref{sec:excitations}).
\Cref{sec:discussion} compares our results with those from alternative QCD-like models studied in literature and gives an outlook.

\section{Model}\label{sec:model}
We study a truncated version of 2-flavor quantum chromodynamics
\cite{Fritzsch1973AdvantagesColorOctet}
in 1+1 spacetime dimensions (\qcd{}).
\qcd{} is a Yang-Mills theory \cite{Yang1954ConservationIsotopicSpin} with gauge group \gaugegroup{}-color, coupled with Dirac fermion matter.
In the Kogut-Susskind formalism, its lattice Hamiltonian reads
\cite{Kogut1975HamiltonianFormulationWilsons,Banks1976StrongCouplingCalculations,Susskind1977LatticeFermions}
\begin{multline}\label{eq:lattice_yang_mills_hamiltonian}
    H =
    \sum_{\xx,\ff,\cc[1;2]} \bigg[
        - \frac{i}{2} \matt \comp \matt{\xx+1}|\cc[2]|*
        + \hc
        \bigg]
    \\\hphantom{H =} {}
    + \sum_{\xx,\ff,\cc} \mass\stgsgn\matt*\matt
    + \sum_{\xx} \frac{\cpl^2}{2} \casimir
    \,.
\end{multline}
We chose Planck units, while parameters and operators have been made dimensionless by rescaling by appropriate powers of the lattice spacing $\spc$.
In terms of the (dimensionless) bare masses $\mass$ and coupling $\cpl$, the continuum limit reads $\mass,\cpl\to0$ \cite{Hamer1982MassiveSchwingerModel}.

The staggered fermion field $\matt*$ acts on \emph{matter quark} degrees of freedom (\dofs{}) living on lattice \emph{sites} $\xx$; it carries a flavor index
\begin{math}
    \ff\in\{u,d\}
\end{math},
plus a color index
\begin{math}
    \cc\in\{\str,\stg,\stb\}
\end{math}
in the fundamental representation of local \gaugegroup{};
it obeys canonical anticommutation relations.
The site Hilbert space is obtained acting repeatedly with $\matt*$ on a Fock vacuum;
it decomposes in a (finite) direct sum of irreducible representations (irreps) of \gaugegroup{}
and basis states are labeled
$\ket{j, \lambda, m}$,
where $j$ is an irrep, $\lambda$ is a degeneracy index,
and $m$ enumerates the states inside irrep $j$ \cite{Zohar2015FormulationLatticeGauge}.

\emph{Gauge gluon} \dofs{} sit on lattice \emph{links} where the chromoelectric energy density operator $\casimir$ and the parallel transporter $\comp$ act; the latter
transforms with the fundamental irrep and its dual at the left and right ends of the link respectively, making the hopping term in \cref{eq:lattice_yang_mills_hamiltonian} gauge invariant.
Operators acting on different links commute with each other and with matter fields.
The link Hilbert space is generated by $\comp$ and it is spanned by states $\ket{j,m,m'}$,
with $j$ an irrep
label and $m, m'$ indices of states in $j$ and its dual $\bar{j}$ respectively.
$\elecSym^2$ is diagonal in this basis:
\begin{equation}\label{eq:casimir}
    \elecSym^2\ket{j,m,m'}=C_2(j)\ket{j,m,m'}
    \ ,
\end{equation}
$C_2(j)$ being the quadratic Casimir of irrep $j$.
In principle gauge bosons can occupy states in every possible irrep $j$, resulting in an \emph{infinite dimensional} link space \cite{Zohar2015FormulationLatticeGauge}.
The physical Hilbert space is the \emph{gauge invariant subspace} \cite{Henneaux1992QuantizationGaugeSystems} of the many body Hilbert space\emdash{}namely, of the tensor product of all matter and gauge local \dof{}.
The gauge invariant sector is singled out by the \emph{Gauss law} constraint
\begin{math}
    \gauss^{\mu} \ket{\pphys} = 0\ \forall \xx, \mu \in \{1..8\}
\end{math},
where $\gauss^{\mu}$ are the generators of local \gaugegroup{} transformations at site $\xx$ \cite{Silvi2014LatticeGaugeTensor}.
When implementing the numerics, we reformulate the model using dressed sites:
the basis consists only of gauge invariant states, and Gauss law is replaced with a set of simpler Abelian link selection rules \cite{Silvi2014LatticeGaugeTensor}.

\qcd{} was already studied in the single flavor case in \cite{Silvi2019TensorNetworkSimulation}.
Nevertheless, generalizing to $N_{\ff}>1$ is particularly convenient for spectral investigations.
Indeed, in the absence of the electroweak force, each quark flavor number is separately conserved.
This singles out rest states of flavored particles as ground states in the appropriate symmetry sectors.

\section{Methods}\label{sec:methods}
We perform numerical Tensor Network (TN) simulations (\cref{sec:tn}) on a finite chain with open boundary conditions, at zero and finite density.
To this aim, the Hilbert space of the model has to be truncated to a finite dimensional one (\cref{sec:truncation}), which can be further compressed exploiting some of the available symmetries (\cref{sec:gauss}).

\subsection{Hardcore Gluons}\label{sec:truncation}

We truncate the infinite dimensional link Hilbert space to a $19$-dimensional one, keeping only the trivial irrep, the fundamental and its dual.
We label this truncation \emph{hardcore gluon} approximation, in analogy to lattice quantum physics of atoms, as the truncated space is spanned acting on the vacuum $\ket{0,0,0}$ with (at most) a single application of the parallel transporters $\field{\compSym}{\cc;\cc[2]}{}$ or $\field{\compSym}{\cc;\cc[2]}{}*$.
Such a projection preserves exact gauge invariance but spoils the unitarity of parallel transporters.
Because higher \gaugegroup{} irreps have a larger quadratic Casimir, by \cref{eq:casimir}, at strong coupling $g\gg1$ the truncation acts effectively as an energy cutoff.
In the weak coupling (or continuum) limit the truncation yields a different model, which we refer to as \emph{\themodel{}}.

Clearly, recovering the physics of the original continuum theory requires relaxing the link truncation while sending $\spc\to0$ \cite{Orland1990LatticeGaugeMagnets,Banuls2017EfficientBasisFormulation}.
A meaningful extrapolation to the untruncated theory should be possible with a finite number of irreps.
The computational Hilbert space dimensions resulting from the inclusion the next few irreps are reported in \cref{app:basis}.
In the present work, however, we focus solely on the maximally-truncated model and on the features it shares with (1+3)D QCD, beyond \gaugegroup{} gauge invariance and confinement.

\subsection{Gauss law}\label{sec:gauss}

In a gauge theory, physical quantities are strictly gauge invariant.
In numerical simulations it is desirable to exploit this fact to lower the dimensionality of the computational space by \emph{discarding unphysical states}.
At the same time, it is imperative to \emph{preserve the local structure} of the global computational Hilbert space, namely its realization as a tensor product space, upon which TN techniques rely.
It is in general not obvious how to enforce Gauss law in a local fashion because
gauge transformations at neighbouring sites involve non disjoint subsets of local degrees of freedom (they share a link).
Ingenious strategies to remove entirely either the matter or the gauge fields have been put forward but, in their current formulation, they do not work for the multi-flavor non-Abelian model studied here \cite{Banuls2017EfficientBasisFormulation,Zohar2019RemovingStaggeredFermionic}.

\begin{figure}
    \includegraphics[width=\columnwidth]{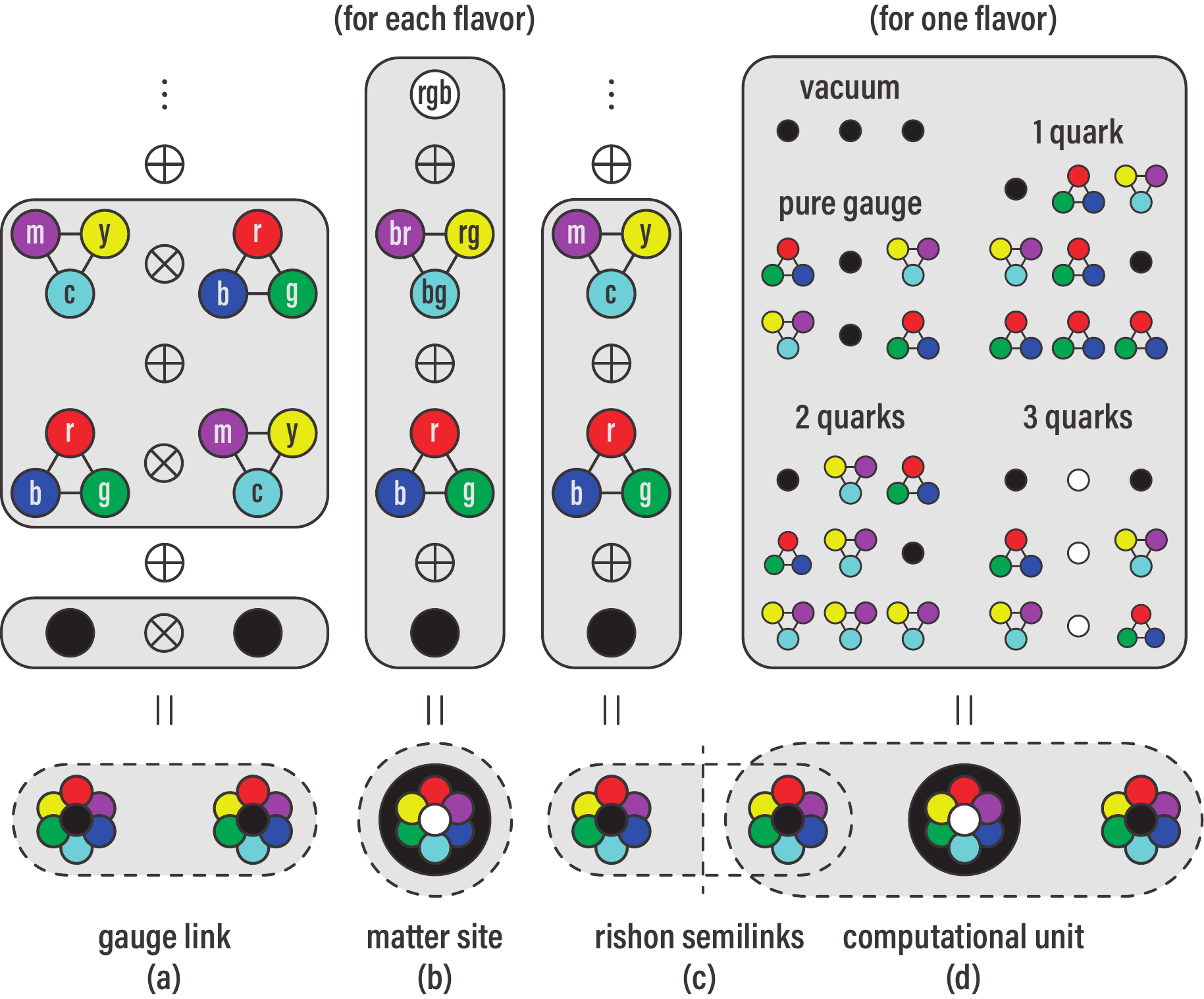}
    \caption[list=off]{\label{fig:building_blocks}%
        Construction of the building blocks for a \gaugegroup{} gauge singlet TN or quantum simulator.
        Cartoon representation of the simplified case of a single quark flavor:
        \begin{enumerate*}[label=(\alph*)]
            \item color irrep decomposition and truncation of the gauge boson link space (irreps are grouped by their quadratic Casimir eigenvalue);
            \item decomposition of the matter fermion site;
            \item splitting of the links in rishon semilink spaces; and finally
            \item composite rishon-matter-rishon computational unit with its $12$ color singlets.
        \end{enumerate*}
        The full recipe is detailed in \cref{app:basis}.
    }
\end{figure}

Our approach to enforce Gauss law consists of three steps, outlined in the following (details in \cref{app:basis}) and depicted pictorially in \cref{fig:building_blocks}.
In the first step we draw inspiration from \cite{Brower1999QcdAsQuantum} and decompose each link in a pair of \emph{rishons}: new \dofs{} each residing on one end of the link and accounting for the respective \gaugegroup{} transformations.
Such semilink Hilbert spaces are generated by operators $\risl*$ and $\risr*$ respectively;
adopting the usual labeling convention, states $\ket{j,m}$ provide a basis for each semilink.
Finally, $\comp$ is identified with $\risl*\field{\rishSym}{R,\xx+1,\cc[2]}{}$, restricted to the subspace $j_L^{\vphantom{*}}=j_R^*$.
The latter requirement ensures that rishons from each pair are in mutually conjugate irreps and thus specify a valid gauge link state.
This is achieved putting a local Abelian
symmetry constraint on the links.
In the third and last step, a composite or \emph{dressed} site is forged fusing together a matter site and its attached semilinks.
Gauge invariance mandates that these are collectively in a \emph{color singlet} state, thus Gauss law is recast as an internal constraint on the dressed site.
We solve it,
obtaining a 54-dimensional local computational basis.

The above procedure replaces Gauss law with a simpler \emph{Abelian} selection rule on each pair of neighboring sites.
As a byproduct, it yields a computational site which embeds both matter and gauge \dof{} but is still smaller than the original
$2^{N_{\ff} \cdot N_{\cc}}
    =64$-dimensional matter site alone.

\subsection{Symmetric tensor networks}\label{sec:tn}
Hamiltonian lattice gauge theories are many-body quantum systems.
In this work, we tackle the exponential growth of the many-body Hilbert space with the system size by means of TN methods \cite{Orus2014PracticalIntroductionTensor,Montangero2018IntroductionTensorNetwork,Silvi2019TensorNetworksAnthology}.
We use DMRG \cite{White1992DensityMatrixFormulation,Schollwoeck2011DensityMatrixRenormalization,Hauschild2018EfficientNumericalSimulations}, a deterministic energy minimization algorithm over the MPS variational class \cite{Rommer1997ClassAnsatzWave}, to efficiently construct ground states of the Hamiltonian in \cref{eq:lattice_yang_mills_hamiltonian} and probe their properties.

The DMRG implementation we employ protects internal Abelian symmetries \cite{Hauschild2018EfficientNumericalSimulations}.
We exploit this feature to fix:
\begin{enumerate*}
    \item $j_L^{\vphantom{*}}=j_R^*$ on each link;
    \item the up-quark number, that is, the number of up-quarks minus the number of up-antiquarks $N_u = \sum_{x,c} ( \matt[u]*\matt[u] - 1/2 )$;
    \item the down-quark number $N_d$ (as above); and
    \item speed-up the numerical simulation \cite{Silvi2019TensorNetworksAnthology}.
\end{enumerate*}
By controlling $(N_u,N_d)$ we can study the model's vacua as well as its flavored excitations.
Targeting a given charge sector amounts to starting the variational optimization from an MPS with the desired quantum numbers.
Note that, thanks to the locality of DMRG updates, when running $\ell$-site DMRG it is sufficient to impose $\ell+1$ independent link constraints, rather than an extensive number of them \cite{Tschirsich2019PhaseDiagramConformal,Silvi2019TensorNetworkSimulation}.
Finally, flavorless particle states (such as a $\pi^0$ meson)
can also be found by looking for intra-sector excitations \cite{Banuls2013MassSpectrumSchwinger}; yet, that has a significantly higher computational cost.

In this work we will often target critical phases, which violate the MPS area law entanglement bound.
Accordingly, the MPS bond dimension $\chi$, which controls the accuracy and the computational cost of the TN approximation, has to be increased polynomially with the system size \cite{Pollmann2009TheoryFiniteEntanglement,Pirvu2012MatrixProductStates,Stojevic2015ConformalDataFinite}.
Bond dimensions as high as $\chi=8192 = 2^{13}$ were used in this study.
Information on the convergence of our simulations is reported in \cref{app:convergence}.

\section{Results}\label{sec:results}
We compute and inspect \themodel{} vacua and excitations across the parameter space
$(\mass,\cpl)$, gaining insights on the phase diagram of the model.
First, we verify that the model admits a continuum limit
(\cref{sec:results-criticalty}).
Next, we focus on some candidate stable particles of the model and show that charged pions are physical in the continuum limit (\cref{sec:excitations}).

\subsection{Criticality}\label{sec:results-criticalty}

The continuum limit of a lattice model is approached when the lattice spacing $\spc$ becomes much smaller than the physical correlation lengths $\corrlen_{\text{phys}}$ of the propagating degrees of freedom \cite{Hernandez20111LatticeField}:
in order for $\corrlen_{\text{phys}}=a\corrlen$ to be finite (or $\infty$) when $a\to0$, the lattice correlation length $\xi$ has to diverge.
We check that \themodel{} possesses the expected continuum limit by verifying that $\mass,\cpl\to0$ is a critical point.
The long distance universal properties of a critical phase are encoded by a conformal field theory (CFT).
It is a well known CFT result \cite{Holzhey1994GeometricRenormalizedEntropy,Calabrese2004EntanglementEntropyQuantum} that, for an infinite 1D critical system in its ground state, the entanglement entropy $S$ of a large subregion grows logarithmically with the length of the subregion.
Contrarily, 1D area law implies that $S$ is bounded by a constant in a gapped phase \cite{Srednicki1993EntropyArea}.
There are finite-size corrections to the critical behaviour: length is replaced by the chord length; moreover, a (possibly oscillating) term decaying as a power law away from the boundary has been observed in Luttinger liquids \cite{Laflorencie2006BoundaryEffectsCritical,Calabrese2010ParityEffectsScaling,Xavier2012FiniteSizeCorrections}.
Ultimately, for a bipartition obtained cutting an open chain of length $L$ at $x$, in the $\xx, L - \xx \gg 1$ limit,
we have
\begin{equation}\label{eq:ll_entanglement_entropy}
    S(x) \simeq \frac{c}{6} \log \ell + c' + c'' F(\ell/L) \cos(2x\kappa) \abs{\ell}^{-p}
    \ .
\end{equation}
Here $\ell=(L/\pi)\sin(\pi x / L)$ is the chord distance of the cut from the boundary and $\kappa$ is the Fermi momentum; the central charge $c$, the critical exponent $p$ and the scaling function $F$ are universal (\idest{}, they are properties of the CFT alone), while $c'$ and $c''$ are model-dependent (thus non-universal) constants.

\begin{figure}
    \begin{subcaptions}{\includegraphics{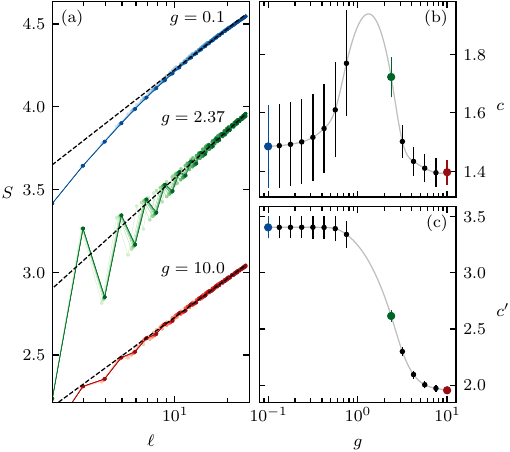}}
        \oversubcaption{0.0, 0.0}{\label{fig:criticality_massless_S}}
        \oversubcaption{0.5, 0.0}{\label{fig:criticality_massless_c}}
        \oversubcaption{0.5, 0.5}{\label{fig:criticality_massless_c''}}
    \end{subcaptions}
    \caption{\label{fig:criticality_massless}%
        Entanglement entropy $S$ of a bipartition as a function of the chord distance of the cut $\ell$ \sref{fig:criticality_massless_S} for: weak, intermediate, and strong coupling $\cpl$, massless quarks, and many system sizes $L\in[24, 80]$ (darker tones correspond to longer chains).
        Data points from all $L$ are linearly interpolated to the first two terms in \cref{eq:ll_entanglement_entropy}, prioritizing large $x$ points via weights $\ell^w$.
        For each $g\in[0.1,10]$ and $w\in\{0,1,\ldots,4\}$, a fit is preformed and assigned a weight $(1-R^2)^{-1}$, $R^2$ being its coefficient of determination.
        Estimates of the central charge $c$~\sref{fig:criticality_massless_c} and $c'$~\sref{fig:criticality_massless_c''} as a function of $g$ are obtained averaging over the relevant fits (the gray spline is just for visual aid).}
\end{figure}
In \cref{fig:criticality_massless} we fit $S(x)$ to the first two terms in \cref{eq:ll_entanglement_entropy} on the whole $\mass=0$ line subregion of parameter space.
We find that the model is always critical in the massless regime and identify two distinct
phases with an interface at $g\sim1$.
Although we did not fit the $c''$ term, we observe that the weak coupling phase\emdash{}where the continuum physics is expected to lie\emdash{}is compatible with $\kappa=0$, while clearly $\kappa=\pi/2$ for $g\gtrsim1$.
Oscillations are particularly pronounced at the phases' interface.

\begin{figure}
    \begin{subcaptions}{\includegraphics{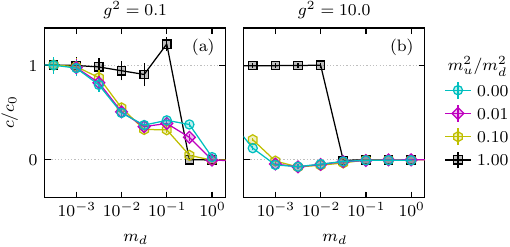}}
        \oversubcaption{0.0, 0.0}{\label{fig:criticality_massive_weak}}
        \oversubcaption{0.5, 0.0}{\label{fig:criticality_massive_strong}}
    \end{subcaptions}
    \caption{\label{fig:criticality_massive}%
        Estimated central charge $c$ versus the heavy quark mass $m_d$, in units of its $\mass=0$ value $c_0$; $c\ll1$ signals a gapped phase.
        The plots involve $L=60$ simulations at weak \sref{fig:criticality_massive_weak} and strong \sref{fig:criticality_massive_strong} coupling $g$, and a variety of $m_u/m_d$ each.
        The computation of $c$ follows the procedure detailed in \cref{fig:criticality_massless}.
    }
\end{figure}
In \cref{fig:criticality_massive} we study how robust criticality is to nonzero bare quark masses.
Decreasing $\mass$, the massless phenomenology is eventually recovered at weak coupling, while at strong coupling this happens only for $m_u=m_d$ (in the scanned mass range), when the up and down quarks form a global $\SUN(2)$-isospin doublet.
That degenerate quark masses favour criticality can be partially understood in the framework of chiral perturbation theory \cite{Ecker1995ChiralPerturbationTheory},
where explicit isospin-breaking is known to induce a correction to the $\pi^0$ pion mass.
The splitting between the $\pi^{\pm}$ and $\pi^0$ masses implies that they cannot be simultaneously gapless.
We expect an analogous phenomenology to arise in the presence electromagnetic interactions, in which case it is the $\pi^{\pm}$ mass that gets a correction \cite{Das1967ElectromagneticMassDifference}.
The onset of criticality is abrupt if $m_u=m_d$, while the growth of $c$ is otherwise gradual and mostly controlled by the mass $m_d$ of the heavy quark.
Since $c$ roughly \enquote{counts} the number of gapless degrees of freedom \cite{Zamolodchikov1986IrreversibilityFluxRenormalization},
the above discussion implies that
\begin{enumerate*}
    \item there are multiple massless particles at $\mass=0$ and
    \item their gaps close at different mass scales when $m_d \neq m_u$.
\end{enumerate*}
\Cref{sec:excitations} is devoted precisely to the classification of such gapless particles; pinning the eventual degeneracies of the vacuum sector is a prerequisite and is carried out in \cref{app:vacuum}, where we show that the model has a unique vacuum and find indirect evidence of the presence of massless flavor-neutral excitations.

\subsection{Edge and bulk excitations}\label{sec:excitations}

We now turn to characterizing some of the particle excitations of \themodel{}, focusing on modes which survive in the continuum\emdash{}\idest{} those whose mass gap $M = 1/\corrlen$ measured in lattice units closes when approaching the continuum limit \cite{Hernandez20111LatticeField}.
The argument is the dual to that for the correlation length:
if a gap does not close, $M_{\text{phys}} = M / a$ diverges when $a\to0$ and the particle is effectively pushed out of the spectrum.
Gauge theories are known to confine in (1+1)D \cite{Creutz1980MonteCarloStudy,Abdalla2001NonPerturbativeMethods,Borla2020ConfinedPhasesOne}, therefore only color-neutral excitations are viable particle candidates.
Among these, we investigate the fate of the charged pion $\pion=u\bar{d}$, proton $\proton=uud$, and Delta baryon $\deltabaryon=uuu$ gaps,
working at $\mass = 0$.
The extension to $\pion*=d\bar{u}$, $\proton*=udd$, and $\deltabaryon*=ddd$ follows by flavor parity symmetry.
More exotic hadrons, such as tetra- and pentaquarks \cite{Jaffe1977MultiQuarkHadrons,Bicudo2022TetraquarksPentaquarksLattice,Atas2023SimulatingOneDimensional}, could also be studied with the same techniques, provided enough flavors are included in the Hamiltonian.

Gapless modes are identified by either \cite{Itou2023CalculatingCompositeParticle}
\begin{enumerate*}
    \item \label{item:gampless_modes_gaps}\relax %
          computing directly particles' rest states (ground states in the appropriate symmetry sectors) and their gaps; or
    \item \label{item:gampless_modes_correlators}\relax %
          studying the vacuum two-point function of fields with the desired quantum numbers.
\end{enumerate*}
An advantage of the former approach is that, in TN calculations, energies are much less sensitive than correlators to the MPS bond dimension.
On the other hand, via correlation functions, information about many different particle types can be efficiently extracted from a single vacuum MPS.
Additionally, correlation functions always probe the bulk physics while eventual edge modes have to be detected and discarded by hand when working at finite quark number.

\subsubsection{Inter-sector excitations}\label{sec:excitations:sectors}

\begin{figure}
    \includegraphics{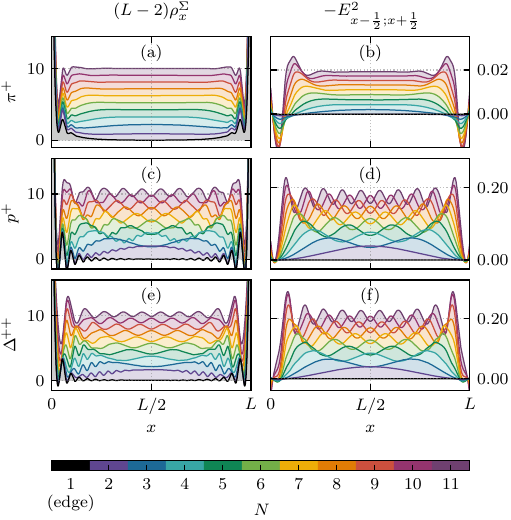}
    \caption{\label{fig:excitation_profiles}%
        Local observables evaluated on states of up to $N=11$ excitations of specie
        \begin{math}
            \specie\in\pion\!,\;\proton\!,\;\deltabaryon
        \end{math}
        (rows).
        Specifically: particle number density $\chargedens$ rescaled by the bulk size $L-2$ (left),
        and gauge field energy density in units of $\cpl^2$, $\casimir$ (right).
        The $N=0$ (vacuum) contribution is subtracted in $\casimir$ plots.
        Points are spline interpolated to make the plots more readable.
        Fixed parameters: $L=48$, $\cpl=0.1$, $\mass=0$.
    }
\end{figure}

We obtain finite density states with $N$ particles of type
\begin{math}
    \specie\in\{\pion\!,\,\proton\!,\,\deltabaryon\}
\end{math}
by constraining DMRG to the flavor symmetry sector $\charge[]=N\charge$, where
\begin{math}
    \charge = (\charge{u},\charge{d})
\end{math}
is the flavor charge of \specie{}, \eg{}
\begin{math}
    \smash{\charge[\pion]} = (+1, -1)
\end{math}.
For each specie, we start from the vacuum ($N=0$) and increase $N$ until the band is completely filled.
All the results of this subsection are for the weak coupling phase, $\cpl=0.1$.

\Cref{fig:excitation_profiles} shows the particle number densities
\begin{equation}
    \textstyle
    \chargedens =
    \left.\sum_{\ff} \charge\fdens\right.
    \left(\sum_{\ff} \abs{\charge}\right)^{-1}
    \ ,
\end{equation}
with \cite{note_particle_number}
\begin{equation}\label{eq:fdens}
    \fdens = \sum_{\cc}(\matt*\matt-1/2)
    \ ,
\end{equation}
and link energy densities $\casimir$ up to $N=11$, taking $N=1$ as a reference state.
For all species \specie{}, the $N=1$ densities decay away from the boundaries, a strong signature of a low-energy edge excitation\emdash{}an edge zero mode.
Conversely, $N>1$ states are manifestly bulk excitations of $N-1$ hardcore particles (see first column).
For \proton{} and \deltabaryon{}, the $N$th profiles are approximatively reproduced stacking the first $N-1$ free particle-in-a-box probability density functions, suggesting that they interact weakly.
Friedel oscillations are also present in density profiles of fermionic modes (\proton{}, \deltabaryon{}) but absent in bosonic ones (\pion{}) \cite{DallaTorre2016FriedelOscillationsAs}.

\begin{figure}
    \includegraphics{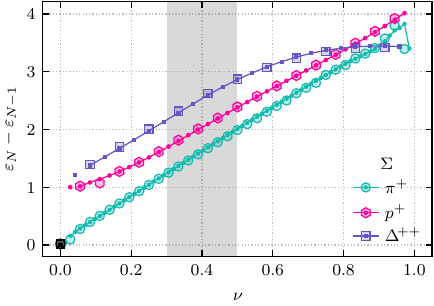}
    \caption{\label{fig:excitation_gaps}%
        Energy gaps between states at different \pion{}, \proton{} and \deltabaryon{} filling $\nu$.
        At $\cpl=0.1$, $\mass=0$.
        Bigger points are obtained at $L=24$, while lines and smaller markers come from $L=48$.
        Black points at the origin refer to the edge mode ($N=1$) gaps with respect to the vacuum.
        The shading emphasizes the alignment of all species' slopes at intermediate $\nu$.
    }
\end{figure}
\begin{figure}
    \includegraphics{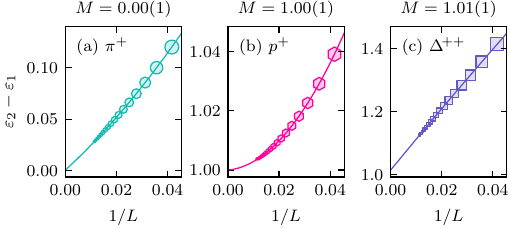}
    \caption{\label{fig:mass_gaps}%
        Finite size scaling of the gap $\varepsilon_2-\varepsilon_1$ of the first \pion{}, \proton{} and \deltabaryon{} bulk modes;
        $L\in[24, 88]$ (smaller markers for longer chains).
        The lines show the interpolation with a degree $2$ polynomial, the extrapolated mass is also reported.
        The error estimate comes from the comparison with the estimate from a linear interpolation.
        Fixed parameters: $\cpl=0.1$, $\mass=0$.
    }
\end{figure}

The energy gaps $\multigap$ between subsequent states are plotted in \cref{fig:excitation_gaps} as a function of the up quark filling fraction, adjusted by discarding the edge modes' contribution:
\begin{math}
    \nu = (\charge{u}[] - \charge{u}) / (3L/2)
\end{math}
\cite{note_particle_number}.
This ensures that, for each specie \specie{}, the first nonzero $\nu$ corresponds to the first bulk mode;
in the thermodynamic limit, the associated gap is the mass gap $M$ of \specie{}.
Subsequent gaps measure the energy cost of adding one particle to the system at finite density, and thus can be interpreted as a finite-size chemical potential.
On the other hand, in a weakly interacting picture, adding particles means progressively exciting higher wavenumber
modes.
Then, $\nu$ is the highest occupied wavenumber and\emdash{}neglecting the interaction energy\emdash{}each curve in \cref{fig:excitation_gaps} mimics the dispersion relation of the corresponding specie.
Corroborating a weakly interacting explanation are
\begin{enumerate*}
    \item the collapse of data from different system sizes $L=24,48$;
    \item the linearity of the \pion{} case, compatible with a vanishing mass gap (\cref{fig:mass_gaps}); and
    \item the common slope of curves from different species at intermediate momenta.
\end{enumerate*}
The latter suggests the emergence of a \enquote{speed of light}, hinting at the restoration of Lorentz invariance\emdash{}at least away from cutoff effects (infrared and ultraviolet) and eventual interactions with the edge mode at low $\nu$.

\begin{table*}
    \setlength\tabcolsep{2ex}
    \begin{tabular}{c>{$}c<{$}}
            \toprule
        excitation                      & \text{operator $\corrfield$}
        \\ \midrule\addlinespace[4pt]
        on-site $s$-wave \pion          & \sum_{\cc} \matt[u] \matt[d]*                                                                                                        \\[4pt]
        nearest-neighbor $s$-wave \pion & \sum_{\cc,\cc[2]} (\matt[u]\comp\matt{\xx+1}[d]|\cc[2]|* + \matt{\xx+1}[u]\comp*\matt[d]|\cc[2]|*)                                   \\[4pt]
        nearest-neighbor $p$-wave \pion & \sum_{\cc,\cc[2]} (\matt[u]\comp\matt{\xx+1}[d]|\cc[2]|* - \matt{\xx+1}[u]\comp*\matt[d]|\cc[2]|*)                                   \\[4pt]
        \proton                         & \matt[u]|\str|\matt[u]|\stg|\matt[d]|\stb| + \matt[u]|\stg|\matt[u]|\stb|\matt[d]|\str| + \matt[u]|\stb|\matt[u]|\str|\matt[d]|\stg| \\[4pt]
        \deltabaryon                    & \matt[u]|\str|\matt[u]|\stg|\matt[u]|\stb|
        \\[2pt] \bottomrule
    \end{tabular}
    \caption{\label{tab:correlators_operators}
        Operators $\corrfield$ exciting some of the simplest color-neutral candidate particles of the model ($p$- and $s$-wave labels are assigned according to their parity transformation properties \cite{note_spin_1d}).
    }
\end{table*}
\begin{figure}
    \begin{subcaptions}{\includegraphics{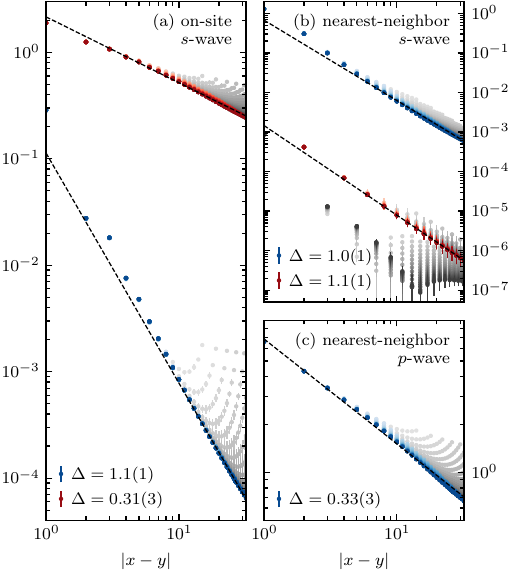}}
        \oversubcaption{0.0, 0.0}{\label{fig:correlator_pion_onsite}}
        \oversubcaption{0.5, 0.0}{\label{fig:correlator_pion_s-wave}}
        \oversubcaption{0.5, 0.5}{\label{fig:correlator_pion_p-wave}}
    \end{subcaptions}
    \caption{\label{fig:correlator_pion}%
        Log-log scale power law decay of \pion{} correlators from \cref{tab:correlators_operators};
        at $\cpl=0.1$ (blue) and $\cpl=10$ (red); $L\in[24, 100]$ (darker tones for longer chains).
        The correlation strength and its uncertainty are obtained averaging over equidistant $(x,y)$ pairs.
        The scaling dimensions $\Delta$ of the associated operators are extrapolated via linear regression.
        Data points $1 \leq\abs{x-y}\leq L/3$ from all system sizes $L$ are included and given a weight proportional to $\abs{x-y}$ (to enhance the asymptotic behaviour);
        points close to the boundary have been excluded and are shown in greyscale.
        The strong coupling $s$-wave pion correlator exhibits an even-odd distance oscillatory pattern, therefore only even distances are fitted.
        The 10\% error estimates come from the comparison with unweighted fits.
    }
\end{figure}

We compute the lattice mass $M$ of
\begin{math}
    \pion\!,\;\proton\!,\;\deltabaryon
\end{math}
particles by means of a finite-size scaling analysis of the respective first bulk gaps.
Indeed, we expect $(\varepsilon_2-\varepsilon_1)\to(\varepsilon_2-\varepsilon_0)\to M$ in the large $L$ limit, provided the edge mode gap and the  bulk-boundary interaction vanish quickly enough (the former is expected to fall exponentially \cite{Vodola2015LongRangeIsing}).
The results are shown in \cref{fig:mass_gaps}.
Protons \proton{}, neutrons \proton*{}, \deltabaryon{} and \deltabaryon*{} baryons have an gap $M \approx 1$ in lattice units at $g=0.1$.
While we cannot exclude that their gaps will close in the $\cpl\to0$ limit, we can safely conclude that is the case for charged pions $\pi^{\pm}$, which are gapless already at finite coupling ($g=0.1$).

\subsubsection{Correlators}\label{sec:excitations:correlators}

Asymptotically, connected correlators
\begin{math}
    G_{yz} =
    \vev*{\corrfield{y}         \corrfield{z}*} -
    \vev*{\corrfield{y}}\!\vev*{\corrfield{z}*}
\end{math}
of massive (massless) fields $\corrfield$ are expected to decay exponentially (algebraically) with space separation \cite{Hastings2006SpectralGapExponential,Sachdev2011QuantumCriticality}.
More precisely,
\begin{math}
    \left.G_{yz} \simeq C e^{-\MASS\dist}\right.
\end{math}
and
\begin{math}
    \left.G_{yz} \simeq C {\dist}^{-2\SDIM}\right.
\end{math}
respectively;
where $\MASS$ is the physical mass (inverse correlation length), $\SDIM$
is the CFT scaling dimension of the field operator, and $C$ can be reabsorbed in the field normalization.

\begin{figure}
    \begin{subcaptions}{\includegraphics{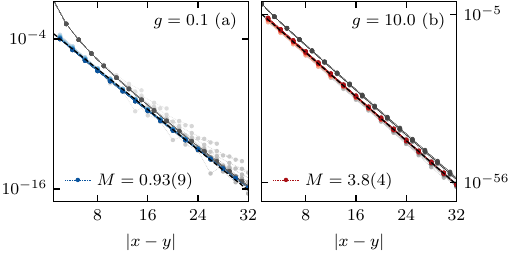}}
        \oversubcaption{0.0, 0.0}{\label{fig:correlator_proton_weak}}
        \oversubcaption{0.5, 0.0}{\label{fig:correlator_proton_strong}}
    \end{subcaptions}
    \caption{\label{fig:correlator_proton}%
        Log scale exponential decay of \proton{} correlator from \cref{tab:correlators_operators}; at $\cpl=0.1$ (blue, \protect\subref{fig:correlator_proton_weak}) and $\cpl=10$ (red, \protect\subref{fig:correlator_proton_strong}); $L\in[24, 100]$ (darker tones for longer chains).
        Methodology as per \cref{fig:correlator_pion}.
        The fit slopes provide the particle's mass $M=1/\corrlen$.
        Only even distances are interpolated due to a clear even-odd distance staggering;
        fitting odd distances yields similar results.
        Up to normalization, \deltabaryon{} correlators are close to \proton{} ones.
    }
\end{figure}

We evaluate the connected vacuum two-point function
\begin{math}
    G_{yz}
\end{math}
for the field operators in \cref{tab:correlators_operators} \cite{note_connected_correlators}.
The $\mass=0$ phase is found to be a liquid of pions with gapped protons and Delta baryons,
in agreement with the energy gaps results from the previous section.
Here we show that this characterization applies to the strong coupling phase as well.
Moreover, correlators allow to distinguish between at least two kind ($s$- and $p$-wave) of \pion{} states.
As we now show, different pions realize superfluid order in the large and small $\cpl$ regimes respectively.
All pion correlators in \cref{fig:correlator_pion} decay as power laws with $\Delta\approx1/3$ or $\Delta\approx1$.
The large-$\cpl$ $p$-wave \pion{} is gapped and it has thus been omitted in \cref{fig:correlator_pion_p-wave}.
As anticipated, \proton{} (\cref{fig:correlator_proton}) and \deltabaryon{} are gapped;
they have similar correlation lengths, compatible with the mass estimates from \cref{fig:mass_gaps}.

\section{Discussion}\label{sec:discussion}
In obtaining the results form the previous section, a series of methods and properties of the model have been derived.
We now summarize them, providing additional context, discussing their implications and giving an outlook on future directions.

We truncated the infinite-dimensional \gaugegroup{} link space of QCD by means of a cutoff in color irrep space \cite{Zohar2015FormulationLatticeGauge}, illustrated in \cref{sec:truncation}.
The truncation can be controlled by tuning the number of included irreps.
This approach is agnostic to the number of spacetime dimensions and extends naturally to any continuous gauge group;
moreover, it is not relevant solely to TN simulation but in general to any computational paradigm based on canonical quantization, such as
quantum simulation.
In \cref{sec:gauss} we showed how it can be combined with a splitting of gauge bosons into rishon \dofs{}
to obtain a TN state ansatz which is gauge invariant by construction.
In \cref{app:basis} we realized explicitly this prescription for (1+1)D QCD.
We considered two quark flavors and the strictest possible truncation but the procedure is completely general and holds even for finite groups.
Indeed, we expect equilibrium TN simulations of less severe truncations to be feasible via the above prescription.
These could shed light on the effects of the truncation and on the untruncated limit, which is key in recovering the continuum physics of true QCD.
Another possible extension is the inclusion of a $\Uone$ electric field.
We conjectured some of its implications, while the details of the implementation are examined in \cref{app:basis}.

Showing that the model has the expected continuum limit is equivalent to proving that it has a critical point at $\cpl,m_u,m_d\to0$.
This was carried out in \cref{sec:results-criticalty}, by inspecting the vacuum entanglement \cite{Ikeda2023DetectingCriticalPoint}.
Incidentally, we found that criticality persists in a whole cylinder around the $m_u=m_d=0$ line in $(g,m_u,m_d)$-space, elongated along the $m_u=m_d$ plane and spanning a weak and a strong coupling phase.
In \cref{sec:excitations} we identified some of the respective gapless bulk modes.
See \cref{fig:phases} for a summary.
At finite lattice spacing, the weak coupling phase is dominated by a (gapless) superfluid of $p$-wave, charged pions $\pi^{\pm}$.
We supported this claim with two independent analyses: finite-size scaling of inter-sector gaps and decay of \pion{} liquid order parameters.
These observations are strong signatures that $\pi^{\pm}$ mesons belong to the physical particle spectrum in the continuum limit \cite{Hernandez20111LatticeField},
motivating future works devoted, \eg{}, to the TN simulation of \pion{}\pion*{} collisions \cite{Rigobello2021EntanglementGeneration11d,Belyansky2023HighEnergyCollision,Florio2023RealTimeNon} in \themodel{}.
Another ambitious extension which is worth pursuing is higher spacetime dimensions \cite{Felser2021EfficientTensorNetwork,Felser2020TwoDimensionalQuantum,Zohar2021QuantumSimulationLattice,Magnifico2021LatticeQuantumElectrodynamics,Pardo2023ResourceEfficientQuantum,Lumia2022TwoDimensionalZ2,Osborne2022LargeScale21d,Emonts2023FindingGroundState,Emonts2023FermionicGaussianPeps}.
There, the existence of a transverse polarization makes the gluon a dynamical field, thus giving access to a richer spectrum containing glueballs \cite{Vadacchino2023ReviewGlueballHunting} and possibly other exotic excitations from the QCD folklore.
A first TN study of a non-Abelian Yang-Mills theory in (2+1)D by some of the authors has just appeared \cite{Cataldi202321dSu2Yang}.
Finally, improved Hamiltonians \cite{Carena2022ImprovedHamiltoniansQuantum,Ciavarella2023QuantumSimulationLattice} and high performance computing will become essential in the strive for precision results.

Let us conclude by suggesting that the studied model is in some sense \enquote{minimal}:
Compared to other truncation schemes \cite{Brower1999QcdAsQuantum}, the one adopted here provides the smallest non-trivial link space dimension while preserving exactly the local \gaugegroup{} symmetry.
Moreover, in the QLM approach, D-theory \cite{Beard1998DTheoryField,Wiese2021QuantumLinkModels} mandates that the untruncated theory is obtained by dimensional reduction from a QLM in one more space dimension.
Despite its elegance,
increasing the space dimension makes the approach somewhat demanding for TN methods.
Conversely, within the chosen truncation scheme our model is the simplest having pions in the continuum spectrum.
Discarding additional irreps completely freezes the gauge field and thus the dynamics.
Restricting to a single quark flavor clearly prevents $\pi^{\pm}$ from existing; furthermore,
the model was studied numerically and no signature of gapless neutral pion $\pi^0$ was found \cite{Silvi2019TensorNetworkSimulation}.
Finally, the splitting of links into rishons and the gluing of the latter in a composite site can be skipped with no consequences on the physics, but then the computational Hilbert space is larger ($64^{L}19^{L-1}$ vs $54^{L}$).
On the other hand, giving up on exact \gaugegroup{} gauge invariance, a number of finite \gaugegroup{} subgroups have been explored in literature \cite{Ludl2010SystematicAnalysisFinite,Flyvbjerg1984GroupSpaceDecimation,Zwicky2009DiscreteMinimalFlavor,Merle2012ExplicitSpontaneousBreaking,Bloch2021EffectiveMathbbz3Model}, some of which of lower order than the link space used here.
Inspiring results were recently obtained from the MC simulation of $S(1080)$ \cite{Alexandru2022SpectrumDigitizedQcd}; yet, the associated link space is much larger than the one constructed here.
Finally, a promising proposal which achieves controllable truncation by deforming the gauge symmetry algebra to a quantum group has been recently put forward \cite{Bimonte1996Suq2LatticeGauge,Zache2023QuantumClassicalSpin,Hayata2023StringNetFormulation,Hayata2023BreakingNewGround}.

\section{Conclusion}
Within a Hamiltonian lattice regularization, we introduced a maximally-truncated (\idest{}~hardcore gluon) model of 2-flavor (1+1)D QCD.
We constructed a gauge invariant computational site and used it to perform TN simulations. We found MPS representations of the model's vacua, single particle and finite density states for a wide range of bare quark mass $m_u,m_d$ and coupling $\cpl$ parameters.
These were instrumental in proving that:
\begin{enumerate*}
    \item the continuum limit of the model is well defined; and
    \item it has charged pions in the particle spectrum\emdash{}in close analogy with (1+3)D QCD.
\end{enumerate*}
We stress that both results are expected for any proper discretization of ordinary QCD, but they were far from obvious with the truncation in place.
Rather, we argued in \cref{sec:discussion} that our model is a minimal realization of a \gaugegroup{} gauge theory displaying such features;
it thus qualifies as an ideal QCD testbed in settings where computational resources are still a bottleneck, such as medium-term quantum computation and real-time TN simulation \cite{Scopa2022EntanglementDynamicsConfining,Vovrosh2022ConfinementInducedImpurity,Banuls2022QuantumInformationPerspective,Knaute2023MesonContentEntanglement}, \eg{} of scattering processes \cite{Rigobello2021EntanglementGeneration11d,Belyansky2023HighEnergyCollision,Florio2023RealTimeNon}.

\begin{acknowledgments}
    We thank M.~Di Liberto, L. Tagliacozzo, T.~V. Zache and E. Zohar for helpful discussions.
    A fork of \software{TeNPy} \cite{Hauschild2018EfficientNumericalSimulations} and \software{simsio} \cite{Rigobello2023Rgbmrc/simsio} were used in calculations, CloudVeneto \cite{Andreetto2019MergingOpenstackBased} is acknowledged for the use of computing facilities.
    We are grateful to the Mainz Institute for Theoretical Physics (MITP) of the DFG Cluster of Excellence PRISMA$^+$ (project 39083149) for its kind hospitality and partial support during the completion of this work.
    MR is also grateful to the Racah Institute of Physics of the Hebrew University of Jerusalem for hospitality and partial support.
    This work is partially funded by MIUR (through PRIN 2017) and fondazione CARIPARO, the INFN project QUANTUM, the EU's Horizon 2020 research and innovation programme (PASQuanS2) and QuantERA (through the T-NISQ and QuantHEP projects), European Union - NextGenerationEU project CN00000013 - Italian Research Center on HPC, Big Data and Quantum Computing, and the Quantum Computing and Simulation Center of Padova University.
\end{acknowledgments}

\appendix

\section{Computational basis}\label{app:basis}
\def\singlet{\irrep{1}}
\def\fundamental{\irrep{3}}
\NewDocumentCommand{\genericmatt}{O{r}}{\field{\mattSym}{#1}{}}
\def\specie{\sigma}
\DeclareExpandableDocumentCommand{\tabhead}{mmm}{\multicolumn{#1}{c}{#2\text{-#3}}}
\NewDocumentCommand{\Zlnk}{st!}{\IfBooleanTF{#2}{0}{\mathllap{\IfBooleanTF{#1}{+}{-}}1}}
\newcounter{bcounter}

In this section we detail the derivation of the gauge invariant local computational basis used in TN simulations.
We start with an arbitrary fermionic matter content and (finite or compact Lie, simple) gauge group.
We review the construction of the local Hilbert spaces of each \dofs{}, their irrep decomposition, the truncation of the gauge variables, their factorization into rishons, and the assembly of the dressed site.
At each step, we take \themodel{} as an example.
See also \cref{fig:building_blocks} for an illustration of the single flavor case.
The generalization to non-simple gauge groups is also briefly hinted for the case of $\SUN(3)_{\text{color}}\times\Uone_{\text{em}}$.
Following the physicist's convention, we regularly confuse an irrep with its representation space.

\subsection{Matter site}
The local Hilbert (Fock) space of a fermion multiplet
\begin{equation}\label{eq:matter_algebra}
    \genericmatt
    \ , \quad
    \{\genericmatt[s],\genericmatt[t]*\} = \delta_{st}
    \ , \quad
    r,s,t \in R
\end{equation}
is the exterior algebra
\begin{math}\textstyle
    \bigwedge(R)
\end{math}.
Fixing an ordering in $R$,
a matrix realization of the anticommutator in \cref{eq:matter_algebra} in the Fock basis
is given by the in-site Jordan-Wigner transformation
\cite{Jordan1928UeberDasPaulische,Susskind1977LatticeFermions}
\begin{equation}\textstyle
    \genericmatt =
    \left(\bigotimes_{{s<r}} \sigma_3\right)
    \otimes\sigma_{-}\otimes
    \left(\bigotimes_{{t>r}} I\right)
    \ .
\end{equation}
Typically $R$ is some representation of the model's symmetry group.
Then the Hilbert space can be decomposed in a direct sum of irreps and basis states are labeled $\ket{j \alpha m}$,
where $j$ is an irrep, $\alpha$ a multiplicity index, and $m$ labels states in $j$.
The expansion in the Fock basis is given in terms of Clebsch-Gordan (CG) coefficients.
Assuming $R$ to be irreducible,
\begin{multline}
    \ov{r_1, \ldots, r_N}{j \alpha m} =
    \mathcal{A}\,
    C^{j_{12}m_{12}\alpha_1}_{Rr_1 Rr_2}
    C^{j_{123}m_{123}\alpha_2}_{j_{12}m_{12} Rr_3}
    \cdots {} \\ {} \cdots
    C^{jm\alpha_{N-1}}_{j_{1 \ldots N-1}m_{1 \ldots N-1} Rr_N}
    \ .
\end{multline}
Here $\ket{r_1, \ldots, r_N}$ is the $N$-particle Fock state obtained starting from the Fock vacuum and consecutively occupying modes $r_1, \dots, r_N$;
$\mathcal{A}$ denotes antisymmetrization over the $r_i$ indices;
$C$ are the CG coefficients;
indices $\alpha_i$ keep track of the multiplicity in a elementary tensor product; and
$\alpha=(\alpha_1, \dots, \alpha_{N-1})$ accounts for the overall degeneracy of irrep $j$.
Generalizing to a reducible representation (\eg{} multiple species) requires additional CG decompositions and follows from the identity
\begin{equation}\textstyle
    \bigwedge\left(\bigoplus_i R_i\right)
    \cong
    \bigotimes_i \bigwedge(R_i)
    \ .
\end{equation}
E.g., for $N_f$ quarks in the fundamental irrep $\fundamental$ of \gaugegroup{},
\begin{equation}\label{eq:hilbert_site_qcd}
    \hilsite
    \cong (\singlet \oplus \fundamental \oplus \fundamental* \oplus \singlet')^{\otimes N_f}
    \ .
\end{equation}
Already at $N_f=1$, the trivial irrep $\singlet$ (singlet) appears twice.
At $N_f=2$, irreps $\irrep{6}$, $\irrep{6}*$ and $\irrep{8}$ enter the final CG decomposition as well, as shown in \cref{tab:computational_basis}.

\subsection{Gauge links}
The link Hilbert space is spanned by states $\ket{g}$, $g \in G$,
where $G$ is the gauge group \cite{Zohar2015FormulationLatticeGauge}.
For continuos groups this is clearly infinite dimensional.
In order to truncate it, we switch to the irrep basis $\ket{jmn}$ via non-Abelian Fourier transform \cite{Burgio2000BasisPhysicalHilbert,Zohar2015FormulationLatticeGauge}.
Recall $m \in j$ while $n$ is an index in the dual irrep $\bar{j}$.
In this basis,
\begin{align}
    \mel{j'm'n'}{E^2}{jmn}
     & = \label{eq:E_mel}
    C_2(j)\; \delta_{jj'}\delta_{mm'}\delta_{nn'}
    \ ,
    \\
    \mel{j'm'n'}{U^{J}_{MN}}{jmn}
     & = \label{eq:U_mel}
    {\sqrt{\tfrac{\dim {j}}{\dim {j'}}}}
    \sum_{\alpha,\beta}
    C^{j'm'\alpha}_{JMjm} \overline{C^{j'n'\beta}_{JNjn}}
    \ ,
\end{align}
where $C_2$ is the quadratic Casimir.
By \cref{eq:U_mel}, the link space is generated acting with $U_{MN}$ in the fundamental irrep on $\ket{000}$; in hopping terms, $J$ has to match the irrep of the matter field (in \cref{sec:model}, the fundamental).
We truncate
\begin{math}\textstyle
    \hillink \cong \bigoplus_{\smash{j}} ( {j} \otimes \overline{j} )
\end{math}
via a cutoff $\Lambda$ on the $E^2$ spectrum,
keeping only irreps $j$ such that $C_2(j)\leq\Lambda$.
For \gaugegroup{}, $4/3\leq\Lambda<3$ gives the 19-dimensional link space
\begin{equation}\label{eq:link_space_qcd}\textstyle
    \hillink \cong
    (\singlet \otimes \singlet)
    \oplus
    (\fundamental \otimes \fundamental*)
    \oplus
    (\fundamental* \otimes \fundamental)
    \ .
\end{equation}

\subsection{Rishon semilinks}\label{app:basis_rishon}
Indices $m,m'$ and $n,n'$ factorize in \cref{eq:E_mel,eq:U_mel}, suggesting that a link can be decomposed in two rishon \dof{} residing on its left and right ends,
\begin{equation}
    \begin{aligned}
        \textstyle
        \hillink
         & \hookrightarrow
        \textstyle
        \hilsemiL \otimes \hilsemiR \cong
        \left(\bigoplus_{\smash{j}} {j} \right)^{\otimes2} \\
        \ket{jmn}
         & \mapsto
        \ket{jm}_L \otimes \ket{\bar{j}n}_R
        \ .
    \end{aligned}
\end{equation}
At the operator level the mapping reads
\begin{align}\label{eq:inclusion_map_operators}
    E^2
     & \ \longmapsto \
    \eta^2 \otimes I + I \otimes \eta^2
    \ ,
    \\
    U^{J}_{MN}
     & \ \longmapsto \
    \zeta^J_M \otimes (\zeta^J_N)^\dagger
    \ ,
    \label{eq:parallel_transporter_to_rishons}
\end{align}
where we defined, on a single rishon space,
\begin{align}
    \mel{kn}{\eta^2}{jm}
     & = \label{eq:E_mel_rishon}
    \tfrac{1}{2}C_2(j) \delta_{jk}\delta_{mn}
    \ ,
    \\
    \mel{kn}{\zeta^J_M}{jm}
     & = \label{eq:U_mel_rishon}
    {\sqrt[4]{\tfrac{\dim {j}}{\dim {k}}}}
    \sum_{\alpha}
    C^{kn\alpha}_{JMjm}
    \ .
\end{align}
\def\Nirreps{\mathcal{N}}
At the TN simulation level, the restriction from $\hilsemiL \otimes \hilsemiR$ to $\hillink$ is enforced by:
\begin{enumerate*}
    \item introducing on each link one Abelian local symmetry with generator $e^{2\pi i\Phi}$,
    \item assigning opposite charges $\pm\phi$ to conjugate irreps $j_L$ and $\bar{j}_R$ at the two ends of the link, and
    \item working in the sector where all link charges are zero.
\end{enumerate*}
Fulfilling the latter requirement will in general entail a decomposition of $\zeta^J_M$ and \cref{eq:parallel_transporter_to_rishons} in a sum of terms, as many as the maximum number of target irreps appearing in a single fusion with $J$ of any link irrep $j$.
Link symmetries can be either $\Uone$ or $\Zn{2\Nirreps+1}$, where $\Nirreps$ is the number of pairs of conjugate nontrivial irreps kept.

The \gaugegroup{} case of \cref{eq:link_space_qcd} reads
\begin{equation}\label{eq:hilbert_semilink_qcd}
    \hilsemi \cong \hilsemiL \cong \hilsemiR \cong \singlet \oplus \fundamental \oplus \fundamental*
\end{equation}
Basis labels (grouped by irrep) are
\begin{math}
    (0),(\str,\stg,\stb),(\stc,\stm,\sty)
\end{math}.
Let $\Pi_j$ be the projector on $j$, \eg{}
\begin{equation}
    \Pi_{\fundamental} = \projector{\str} + \projector{\stg} + \projector{\stb}
    \ ,
\end{equation}
then ($\zeta$ is taken in the fundamental)
\begin{align}\label{eq:rishon_ops}
    e^{2 \pi i \Phi} & = \Pi_{\singlet} + e^{2 \pi i/3} \Pi_{\fundamental} + e^{4 \pi i/3} \Pi_{\fundamental*} \ , \\
    \eta^2           & = \tfrac{4}{3}(\Pi_{\fundamental} + \Pi_{\fundamental*})                              \ ,   \\
    \zeta_{\str}     & = \ketbra{\stc}{0} + \ketbra{0}{\str} - \ketbra{\stb}{\stm} + \ketbra{\stg}{\sty}     \ ,   \\
    \zeta_{\stg}     & = \ketbra{\stm}{0} + \ketbra{0}{\stg} - \ketbra{\str}{\sty} + \ketbra{\stb}{\stc}     \ ,   \\
    \zeta_{\stb}     & = \ketbra{\sty}{0} + \ketbra{0}{\stb} - \ketbra{\stg}{\stc} + \ketbra{\str}{\stm}     \ .
\end{align}
No decomposition of $\zeta$ is needed here because, within the given truncation, each link irrep appears only once in a fusion with the fundamental.

\subsection{Dressed site}
On a cubic lattice in $D$ dimensions, the composite site is forged
fusing a matter site with $2D$ rishon \dof{} and
selecting only physical\emdash{}\idest{} gauge singlet\emdash{}states:
\begin{equation}\label{eq:hilbert_composite}
    \hil = \vspan\{ \ket{j \alpha m} \in \hilsite\otimes\hilsemi^{\otimes2D}:j=0 \} \ ,
\end{equation}
where the usual labeling convention has been adopted.

Combining \cref{eq:hilbert_composite,eq:hilbert_site_qcd,eq:hilbert_semilink_qcd}, the local computational basis of \themodel{} is obtained.
Its $54$ singlets are listed in \cref{tab:computational_basis}, organized by various quantum numbers.
The computational matrix elements of any physical local gauge invariant operator can be evaluated from the CG expansion of computational states in the original \enquote{physical} matter and rishon bases.
The CG expansion is available online, for reproducibility \cite{Rigobello2023QhlgtModels}, together with a script for computing matrix elements and with its output for the operators relevant to our numerical simulations.
The dimension of various local Hilbert spaces for higher truncation cutoffs $\Lambda$ are reported in \cref{tab:truncations}.
The first few truncations are within the reach of present-day TN calculations \cite{Magnifico2021LatticeQuantumElectrodynamics}.

\begin{table}
    \aboverulesep=.35ex
    \belowrulesep=.45ex
    \setcounter{bcounter}{0}
    \newcommand\bind{\stepcounter{bcounter}\thebcounter}
    \begin{equation*}
        \setlength\arraycolsep{1.1em}
        \begin{array}{*{4}{c}}
            \toprule
            \Lambda       &
            \text{link}   &
            \text{rishon} &
            \text{dressed}
            \\ \midrule
            4/3           & 19   & 7   & 54  \\
            3             & 83   & 15  & 92  \\
            10/3          & 155  & 27  & 166 \\
            16/3          & 605  & 57  & 266 \\
            6             & 805  & 77  & 342 \\
            8             & 1534 & 104 & 392 \\
            \bottomrule
        \end{array}
    \end{equation*}
    \caption{\label{tab:truncations}%
        Dimensions of the link, rishon and computational spaces of 2-flavor \qcd{} for truncation cutoffs $\Lambda$ equal to the few lowest $\SUN(3)$ quadratic Casimir eigenvalues.
        The 64-dimensional matter site is unaffected by the truncation.
    }
\end{table}
\begin{table*}
    \def\blackrule{\midrule}
\def\greyrule{\arrayrulecolor{black!40}\specialrule{0.25pt}{1pt}{1.5pt}\arrayrulecolor{black}}
\aboverulesep=.35ex
\belowrulesep=.45ex
\setcounter{bcounter}{0}
\newcommand\bind{\stepcounter{bcounter}\thebcounter}
\begin{equation*}
    \setlength\arraycolsep{1.1em}
    \def\arraystretch{0.75}
    \begin{array}{*{13}{c}}
        \toprule
        \multirow{2}{*}{$\alpha$}     &
        \tabhead{5}{\SUN(3)}{color}   &
        \tabhead{2}{\Zn{3}}{link}     &
        \tabhead{2}{\Uone}{flavor(s)} &
        \tabhead{1}{\Zn{2}}{$\Fsym$}  &
        \tabhead{2}{\SUN(2)}{isospin}
        \\
        \cmidrule(lr){2-6}
        \cmidrule(lr){7-8}
        \cmidrule(lr){9-10}
        \cmidrule(lr){11-11}
        \cmidrule(lr){12-13}
                                      &
        j_{\text{matter}}             & j_{u}      & j_{d}      & j_{R}      & j_{L}      &
        \phi_R                        & \phi_L     &
        N_u                           & N_d        &
        \sigma                        &
        I                             & I_3
        \\ \blackrule
        \bind
                                      & \irrep{1}  & \irrep{1}  & \irrep{1}  & \irrep{1}  & \irrep{1}
                                      & \Zlnk!     & \Zlnk!     & 0          & 0          & +1         & 0   & 0
        \\
        \bind
                                      & \irrep{1}  & \irrep{1}  & \irrep{1}  & \irrep{3}  & \irrep{3}*
                                      & \Zlnk      & \Zlnk*     & 0          & 0          & +1         & 0   & 0
        \\
        \bind
                                      & \irrep{1}  & \irrep{1}  & \irrep{1}  & \irrep{3}* & \irrep{3}
                                      & \Zlnk*     & \Zlnk      & 0          & 0          & +1         & 0   & 0
        \\ \blackrule
        \bind
                                      & \irrep{3}  & \irrep{1}  & \irrep{3}  & \irrep{1}  & \irrep{3}*
                                      & \Zlnk!     & \Zlnk*     & 0          & 1          &            & 1/2 & -1/2
        \\
        \bind
                                      & \irrep{3}  & \irrep{1}  & \irrep{3}  & \irrep{3}  & \irrep{3}
                                      & \Zlnk      & \Zlnk      & 0          & 1          &            & 1/2 & -1/2
        \\
        \bind
                                      & \irrep{3}  & \irrep{1}  & \irrep{3}  & \irrep{3}* & \irrep{1}
                                      & \Zlnk*     & \Zlnk!     & 0          & 1          &            & 1/2 & -1/2
        \\ \greyrule
        \bind
                                      & \irrep{3}  & \irrep{3}  & \irrep{1}  & \irrep{1}  & \irrep{3}*
                                      & \Zlnk!     & \Zlnk*     & 1          & 0          &            & 1/2 & +1/2
        \\
        \bind
                                      & \irrep{3}  & \irrep{3}  & \irrep{1}  & \irrep{3}  & \irrep{3}
                                      & \Zlnk      & \Zlnk      & 1          & 0          &            & 1/2 & +1/2
        \\
        \bind
                                      & \irrep{3}  & \irrep{3}  & \irrep{1}  & \irrep{3}* & \irrep{1}
                                      & \Zlnk*     & \Zlnk!     & 1          & 0          &            & 1/2 & +1/2
        \\ \blackrule
        \bind
                                      & \irrep{6}  & \irrep{3}  & \irrep{3}  & \irrep{3}* & \irrep{3}*
                                      & \Zlnk*     & \Zlnk*     & 1          & 1          & +1         & 0   & 0
        \\ \blackrule
        \bind
                                      & \irrep{3}* & \irrep{1}  & \irrep{3}* & \irrep{1}  & \irrep{3}
                                      & \Zlnk!     & \Zlnk*     & 0          & 2          &            & 1   & -1
        \\
        \bind
                                      & \irrep{3}* & \irrep{1}  & \irrep{3}* & \irrep{3}  & \irrep{1}
                                      & \Zlnk*     & \Zlnk!     & 0          & 2          &            & 1   & -1
        \\
        \bind
                                      & \irrep{3}* & \irrep{1}  & \irrep{3}* & \irrep{3}* & \irrep{3}*
                                      & \Zlnk*     & \Zlnk*     & 0          & 2          &            & 1   & -1
        \\ \greyrule
        \bind
                                      & \irrep{3}* & \irrep{3}  & \irrep{3}  & \irrep{1}  & \irrep{3}
                                      & \Zlnk!     & \Zlnk*     & 1          & 1          & -1         & 1   & 0
        \\
        \bind
                                      & \irrep{3}* & \irrep{3}  & \irrep{3}  & \irrep{3}  & \irrep{1}
                                      & \Zlnk*     & \Zlnk!     & 1          & 1          & -1         & 1   & 0
        \\
        \bind
                                      & \irrep{3}* & \irrep{3}  & \irrep{3}  & \irrep{3}* & \irrep{3}*
                                      & \Zlnk*     & \Zlnk*     & 1          & 1          & -1         & 1   & 0
        \\ \greyrule
        \bind
                                      & \irrep{3}* & \irrep{3}* & \irrep{1}  & \irrep{1}  & \irrep{3}
                                      & \Zlnk!     & \Zlnk*     & 2          & 0          &            & 1   & +1
        \\
        \bind
                                      & \irrep{3}* & \irrep{3}* & \irrep{1}  & \irrep{3}  & \irrep{1}
                                      & \Zlnk*     & \Zlnk!     & 2          & 0          &            & 1   & +1
        \\
        \bind
                                      & \irrep{3}* & \irrep{3}* & \irrep{1}  & \irrep{3}* & \irrep{3}*
                                      & \Zlnk*     & \Zlnk*     & 2          & 0          &            & 1   & +1
        \\ \blackrule
        \bind
                                      & \irrep{8}  & \irrep{3}  & \irrep{3}* & \irrep{3}  & \irrep{3}*
                                      & \Zlnk      & \Zlnk*     & 1          & 2          &            & 1/2 & -1/2
        \\
        \bind
                                      & \irrep{8}  & \irrep{3}  & \irrep{3}* & \irrep{3}* & \irrep{3}
                                      & \Zlnk*     & \Zlnk      & 1          & 2          &            & 1/2 & -1/2
        \\ \greyrule
        \bind
                                      & \irrep{8}  & \irrep{3}* & \irrep{3}  & \irrep{3}  & \irrep{3}*
                                      & \Zlnk      & \Zlnk*     & 2          & 1          &            & 1/2 & +1/2
        \\
        \bind
                                      & \irrep{8}  & \irrep{3}* & \irrep{3}  & \irrep{3}* & \irrep{3}
                                      & \Zlnk*     & \Zlnk      & 2          & 1          &            & 1/2 & +1/2
        \\ \blackrule
        \bind
                                      & \irrep{1}  & \irrep{1}  & \irrep{1}' & \irrep{1}  & \irrep{1}
                                      & \Zlnk!     & \Zlnk!     & 0          & 3          &            & 3/2 & -3/2
        \\
        \bind
                                      & \irrep{1}  & \irrep{1}  & \irrep{1}' & \irrep{3}  & \irrep{3}*
                                      & \Zlnk      & \Zlnk*     & 0          & 3          &            & 3/2 & -3/2
        \\
        \bind
                                      & \irrep{1}  & \irrep{1}  & \irrep{1}' & \irrep{3}* & \irrep{3}
                                      & \Zlnk*     & \Zlnk      & 0          & 3          &            & 3/2 & -3/2

        \\ \greyrule
        \bind
                                      & \irrep{1}  & \irrep{3}  & \irrep{3}* & \irrep{1}  & \irrep{1}
                                      & \Zlnk!     & \Zlnk!     & 1          & 2          &            & 3/2 & -1/2

        \\
        \bind
                                      & \irrep{1}  & \irrep{3}  & \irrep{3}* & \irrep{3}  & \irrep{3}*
                                      & \Zlnk      & \Zlnk*     & 1          & 2          &            & 3/2 & -1/2

        \\
        \bind
                                      & \irrep{1}  & \irrep{3}  & \irrep{3}* & \irrep{3}* & \irrep{3}
                                      & \Zlnk*     & \Zlnk      & 1          & 2          &            & 3/2 & -1/2
        \\ \greyrule
        \bind
                                      & \irrep{1}  & \irrep{3}* & \irrep{3}  & \irrep{1}  & \irrep{1}
                                      & \Zlnk!     & \Zlnk!     & 2          & 1          &            & 3/2 & +1/2
        \\
        \bind
                                      & \irrep{1}  & \irrep{3}* & \irrep{3}  & \irrep{3}  & \irrep{3}*
                                      & \Zlnk      & \Zlnk*     & 2          & 1          &            & 3/2 & +1/2
        \\
        \bind
                                      & \irrep{1}  & \irrep{3}* & \irrep{3}  & \irrep{3}* & \irrep{3}
                                      & \Zlnk*     & \Zlnk      & 2          & 1          &            & 3/2 & +1/2
        \\ \greyrule
        \bind
                                      & \irrep{1}  & \irrep{1}' & \irrep{1}  & \irrep{1}  & \irrep{1}
                                      & \Zlnk!     & \Zlnk!     & 3          & 0          &            & 3/2 & +3/2
        \\
        \bind
                                      & \irrep{1}  & \irrep{1}' & \irrep{1}  & \irrep{3}  & \irrep{3}*
                                      & \Zlnk      & \Zlnk*     & 3          & 0          &            & 3/2 & +3/2
        \\
        \bind
                                      & \irrep{1}  & \irrep{1}' & \irrep{1}  & \irrep{3}* & \irrep{3}
                                      & \Zlnk*     & \Zlnk      & 3          & 0          &            & 3/2 & +3/2
        \\ \blackrule
        \bind
                                      & \irrep{6}* & \irrep{3}* & \irrep{3}* & \irrep{3}  & \irrep{3}
                                      & \Zlnk      & \Zlnk      & 2          & 2          & +1         & 0   & 0
        \\ \blackrule
        \bind
                                      & \irrep{3}  & \irrep{3}  & \irrep{1}' & \irrep{1}  & \irrep{3}*
                                      & \Zlnk!     & \Zlnk*     & 1          & 3          &            & 1   & -1
        \\
        \bind
                                      & \irrep{3}  & \irrep{3}  & \irrep{1}' & \irrep{3}  & \irrep{3}
                                      & \Zlnk      & \Zlnk      & 1          & 3          &            & 1   & -1
        \\
        \bind
                                      & \irrep{3}  & \irrep{3}  & \irrep{1}' & \irrep{3}* & \irrep{1}
                                      & \Zlnk*     & \Zlnk!     & 1          & 3          &            & 1   & -1
        \\ \greyrule
        \bind
                                      & \irrep{3}  & \irrep{3}* & \irrep{3}* & \irrep{1}  & \irrep{3}*
                                      & \Zlnk!     & \Zlnk*     & 2          & 2          & -1         & 1   & 0
        \\
        \bind
                                      & \irrep{3}  & \irrep{3}* & \irrep{3}* & \irrep{3}  & \irrep{3}
                                      & \Zlnk      & \Zlnk      & 2          & 2          & -1         & 1   & 0
        \\
        \bind
                                      & \irrep{3}  & \irrep{3}* & \irrep{3}* & \irrep{3}* & \irrep{1}
                                      & \Zlnk*     & \Zlnk!     & 2          & 2          & -1         & 1   & 0
        \\ \greyrule
        \bind
                                      & \irrep{3}  & \irrep{1}' & \irrep{3}  & \irrep{1}  & \irrep{3}*
                                      & \Zlnk!     & \Zlnk*     & 3          & 1          &            & 1   & +1
        \\
        \bind
                                      & \irrep{3}  & \irrep{1}' & \irrep{3}  & \irrep{3}  & \irrep{3}
                                      & \Zlnk      & \Zlnk      & 3          & 1          &            & 1   & +1
        \\
        \bind
                                      & \irrep{3}  & \irrep{1}' & \irrep{3}  & \irrep{3}* & \irrep{1}
                                      & \Zlnk*     & \Zlnk!     & 3          & 1          &            & 1   & +1
        \\ \blackrule
        \bind
                                      & \irrep{3}* & \irrep{3}* & \irrep{1}' & \irrep{1}  & \irrep{3}
                                      & \Zlnk!     & \Zlnk      & 2          & 3          &            & 1/2 & -1/2
        \\
        \bind
                                      & \irrep{3}* & \irrep{3}* & \irrep{1}' & \irrep{3}  & \irrep{1}
                                      & \Zlnk      & \Zlnk!     & 2          & 3          &            & 1/2 & -1/2
        \\
        \bind
                                      & \irrep{3}* & \irrep{3}* & \irrep{1}' & \irrep{3}* & \irrep{3}*
                                      & \Zlnk*     & \Zlnk      & 2          & 3          &            & 1/2 & -1/2
        \\ \greyrule
        \bind
                                      & \irrep{3}* & \irrep{1}' & \irrep{3}* & \irrep{1}  & \irrep{3}
                                      & \Zlnk!     & \Zlnk      & 3          & 2          &            & 1/2 & +1/2
        \\
        \bind
                                      & \irrep{3}* & \irrep{1}' & \irrep{3}* & \irrep{3}  & \irrep{1}
                                      & \Zlnk      & \Zlnk!     & 3          & 2          &            & 1/2 & +1/2
        \\
        \bind
                                      & \irrep{3}* & \irrep{1}' & \irrep{3}* & \irrep{3}* & \irrep{3}*
                                      & \Zlnk*     & \Zlnk      & 3          & 2          &            & 1/2 & +1/2
        \\ \blackrule
        \bind
                                      & \irrep{1}  & \irrep{1}' & \irrep{1}' & \irrep{1}  & \irrep{1}
                                      & \Zlnk!     & \Zlnk!     & 3          & 3          & +1         & 0   & 0
        \\
        \bind
                                      & \irrep{1}  & \irrep{1}' & \irrep{1}' & \irrep{3}  & \irrep{3}*
                                      & \Zlnk      & \Zlnk*     & 3          & 3          & +1         & 0   & 0
        \\
        \bind
                                      & \irrep{1}  & \irrep{1}' & \irrep{1}' & \irrep{3}* & \irrep{3}
                                      & \Zlnk*     & \Zlnk      & 3          & 3          & +1         & 0   & 0
        \\ \bottomrule
    \end{array}
\end{equation*}
    \caption{\label{tab:computational_basis}%
        Quantum numbers of the basis states $\ket{\alpha}$:
        \gaugegroup{}-color irreps of matter, $j_{\text{matter}} \in j_{u} \otimes j_{d}$;
        \gaugegroup{}-color irrep $j_{R}$ ($j_{L}$)
        and corresponding $\Zn{3}$-link charge $\phi_{R}$ ($\phi_{L}$)
        of the $R$ ($L$) rishon;
        numbers $N_u$, $N_d$ of up and down quarks;
        $\Zn{2}$-$\Fsym$ flavor parity $\sigma$ of $N_u=N_d$ states;
        $\SUN(2)$-isospin irrep $I$ and projection $I_3=(N_u-N_d)/2$.
        Note that $\phi_{R,L}$ contribute to different link charges; moreover, $\Zn{2}$-$\Fsym$ and $\SUN(2)$-isospin provide good quantum numbers only for degenerate quark masses.
    }
\end{table*}

\subsection{Inclusion of an electric field}
\def\qem{Q}
\def\EM{em}
\def\elecEM{\elecSym_{\text{\EM}}}
A physically motivated extension of the model, which would allow studying electric corrections to \qcd{} in the spontaneously broken electroweak phase \cite{Kordov2023WeakDecayConstants}, consists in adding a $\Uone$-electromagnetic (\EM) component to its gauge group.
To this aim, new $\Uone$ gauge \dofs{} have to be implanted on each link and then split into rishons as per \cref{app:basis_rishon}.

All $\Uone$ irreps are one-dimensional and are labeled by $\qem\in\integers$;
$\qem_u=-2\qem_d$ and we can set $\qem_d=-1$,
from which the electric charge
\begin{math}
    \qem_\alpha=2N_u-N_d
\end{math}
of each state in \cref{tab:computational_basis} follows.
Moreover,
\begin{equation}
    \compSym_{\text{\EM}}^Q\ket{Q'} = \ket{Q'+Q}
    \ , \quad
    \elecEM\ket{Q} = \qem\ket{Q}
    \ .
\end{equation}
Insisting that the bare vacuum of all \dofs{} is a physical state, at least 5 $\Uone$ irreps have to be kept.
We adopt once again the maximal truncation.
Then, each row in \cref{tab:computational_basis} is split in
\begin{math}
    5 - \abs{\qem_\alpha}
\end{math}
entries, with the $L$ rishon in irreps $\qem_L$,
\begin{equation}
    -2-\min(0,\qem_\alpha)<\qem_L<+2-\max(0,\qem_\alpha)
    \ ,
\end{equation}
and $\qem_R=-(\qem_L+\qem_\alpha)$.
The resulting computational basis consists of 150 states.
Alternatively, a $\Zn{5}$ subgroup truncation gives an even more bewildering\emdash{}albeit still attainable \cite{Magnifico2021LatticeQuantumElectrodynamics}\emdash{}270-dimensional computational site.

At the level of the Hamiltonian, the extension amounts to the following formal substitutions in \cref{eq:lattice_yang_mills_hamiltonian}
\begin{equation}
    \compSym \to \compSym_{\text{color}} \compSym_{\text{\EM}}^{Q_{\ff}}
    \ , \quad
    \cpl \elecSym \to \cpl_{\text{color}} \elecSym_{\text{color}} + \cpl_{\text{em}} \elecEM
    \ ;
\end{equation}
regardless of the chosen truncation scheme.

\section{Strong coupling expansion}\label{app:sce}
\def\hilunp{\hil_{\text{eff}}}
\def\Hunp{H_0}
\def\Hper{H_1}
\def\Heff{H_{\text{eff}}}

In this section we perform a strong coupling expansion (SCE) for \themodel{}.
We work in the regime $\cpl\gg1$, $\mass=0$ and treat the hopping term in \cref{eq:lattice_yang_mills_hamiltonian} as a perturbation of the chromoelectric energy term:
\begin{align}
    \quad
    \Hunp & = \frac{\cpl^2}{2} \sum_{\xx} \casimir
    \;,
    \label{eq:unperturbed}
    \\
    \Hper & = \frac{i}{2} \sum_{\mathclap{\xx,\ff,\cc[1;2]}}
    \matt* \comp* \matt{\xx+1}|\cc[2]|
    + \hc
    \label{eq:perturbator}
    \;;
\end{align}

We restrict to the unperturbed ground space $\hilunp$\emdash{}the null space $\Hunp$\emdash{}and use second order, degenerate perturbation theory to define an effective Hamiltonian
\begin{equation} \label{eq:ptsecondorder}
    \Heff = \cpl^{-2}\, V^{\dagger} \Hper (-\Hunp)^{p} \Hper V
    \ ,
\end{equation}
which resolves order $O(1/g^2)$ splittings.
Here $A^p$ is the Moore-Penrose pseudoinverse of $A$ and $V$ is the isometry from the full Hilbert space to $\hilunp$.
In the dressed site formulation, $V$ decomposes in a product of local isometries $v_x$ projecting on the gauge-trivial states,
\begin{align}
    v & =
    \ketbra{ddd}{24}
    +\ketbra{udd}{27}
    +\ketbra{uud}{30}
    +\ketbra{uuu}{33}
    +{}
    \nonumber
    \\
      & \hphantom{= {}}
    +\ketbra{\circ}{1}
    +\ketbra{\bullet}{52}
    \ ,
\end{align}
where we labeled states in $\hilunp$ according to their quark content ($\circ=\text{empty}$, $\bullet=\text{full}$).
Furthermore, $\Hunp$ is local on links as well as on dressed sites.
Finally, because each summand in \cref{eq:perturbator} changes the gauge state on exactly one link, both $\Hper$ factors in \cref{eq:ptsecondorder} must to act on the same link.
From the previous observations it follows that $\Heff$ is nearest-neighbor.
Up to an additive constant,
\begin{equation}\label{eq:sce}
    \cpl^2\Heff =
    3\sum_{\xx} S^z_{\frac{1}{2},\xx} S^z_{\frac{1}{2},\xx+1}
    +\sum_{\xx} \vec{S}_{\frac{3}{2},\xx} \cdot \vec{S}_{\frac{3}{2},\xx+1}
    \ ,
\end{equation}
where
\begin{math}
    S^z_{1/2} = (\projector{\circ} - \projector{\bullet})/2
\end{math}
and
\begin{math}
    \vec{S}_{3/2}
\end{math}
are the spin matrices over
\begin{math}
    (
    \ket{ddd},
    \ket{udd},
    \ket{uud},
    \ket{uuu}
    )
\end{math}%
\emdash{}the isospin-$3/2$ quadruplet.
Interestingly, the even and odd baryon number subspaces decouple at leading order in the expansion.
The dynamics of the former is ruled by an antiferromagnetic Ising model whose $\Zn{2}$ symmetry represents charge conjugation; the odd subspace realizes a spin-${3}/{2}$ antiferromagnetic Heisenberg model with $\SUN(2)$-isospin symmetry, which remains unbroken in the massive quarks case, as long as $m_u = m_d$.
Numerical evaluation of the single-site reduced density matrix shows that the Heisenberg model dominates the strong coupling physics: at $g=10.0$ the populations of the isospin-$0$ states are suppressed by more than 3 orders of magnitudes with respect to those of the isospin-$3/2$ quadruplet.

A comparison with the SCE in \cite{Ciavarella2023QuantumSimulationLattice} reveals how the presence of two quarks flavors instead of one greatly enlarges the configuration space of the model, also at the level of the zero chromoelectric energy effective subspace.

\section{Vacuum sector}\label{app:vacuum}
In this section we inspect the vacuum sector of the model\emdash{}\idest{}, the \emph{unflavored} $N_u = N_d = 0$ sector \cite{note_particle_number}\emdash{}with the main goal of assessing whether certain phases undergo spontaneous symmetry breaking (SSB).
SSB is relevant, \eg{}, when comparing expectation values on excited states with their vacuum (unflavoured ground state) expectation value (VEV), as done in \cref{sec:excitations}.
Indeed, certain observables may not admit an unambiguous definition of VEV in the presence of degenerate vacua.

\begin{table}
    \setlength\tabcolsep{1em}
    \begin{tabular}{>{$}r<{$}>{$}l<{$}}
        \toprule
        \multicolumn{1}{c}{$g$} & \multicolumn{1}{c}{$\varepsilon_n - \varepsilon_0\quad(n=0,1,\ldots)$} \\ \midrule
        0.1                     & 0,\ 1,\ 19.5,\ 20.6,\ 63,\ 64,\ 104, \ldots                            \\
        10.0                    & 0,\ 1,\ 3.9,\ 7.5,\ 7.7,\ 9.9,\ 10.0, \ldots                           \\
        \bottomrule
    \end{tabular}
    \caption{\label{tab:vacuum_gaps}%
        Gaps of the first few Hamiltonian eigenvalues $\varepsilon_n$ in the vacuum sector, at weak and strong coupling.
        To emphasize the hierarchy of the splittings, we chose units such that $\varepsilon_1-\varepsilon_0=1$,
        independently for each $g$ value.
    }
\end{table}
\begin{figure}
    \begin{subcaptions}{\includegraphics{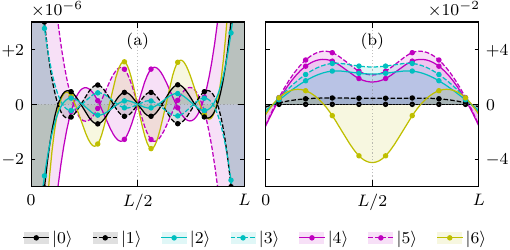}}
        \oversubcaption{0.0, 0.0}{\label{fig:vacuum_profiles_matter}}
        \oversubcaption{0.5, 0.0}{\label{fig:vacuum_profiles_gauge}}
    \end{subcaptions}
    \caption{\label{fig:vacuum_profiles}%
        Up quark number density $\fdens[u]$ \sref{fig:vacuum_profiles_matter} and link energy density $\casimir$ \sref{fig:vacuum_profiles_gauge} of the 7 lowest energy eigenstates $\ket{n}$ in the vacuum sector, as found by DMRG, at $\cpl=0.1$ and $L=8$.
        The down flavored profiles $\fdens[d]$ are identical.
        In \sref{fig:vacuum_profiles_gauge} we subtracted the $\ket{0}$ contribution.
        Points are spline interpolated; the divergencies at the boundaries are an artifact (Runge's phenomenon).
    }
\end{figure}

\subsection{Intra-sector excitations}
We compute the 7 lowest Hamiltonian eigenstates in the unflavored sector, at $\mass = 0$ and both small and large $g$.
Their gaps are reported in \cref{tab:vacuum_gaps}.
At large coupling there is no clear hierarchy among them, bolstering the argument in favour of a unique vacuum
and rendering the subtraction of VEVs straightforward and unambiguous.
The strong coupling vacuum density profiles are \Csym{}, \Psym{}, \Fsym{} antisymmetric:
\begin{math}
    \vev{\fdens[u]}=-\vev{\fdens{-\xx}[u]}=-\vev{\fdens{\xx}[d]}
\end{math}.
At small coupling, the eigenstates organize in quasi-degenerate doublets with density profiles (see \cref{fig:vacuum_profiles_matter}) which are charge \Csym{} and parity \Psym{} conjugate one of the other (each profile is individually \CPsym{} and flavor \Fsym{} symmetric).
Since the Hamiltonian is \Csym{} and \Psym{} symmetric, it would appear reasonable to treat the two lowest states,
$\ket{0}$ and $\ket{1}$,
as degenerate vacua.
On the other hand, even the finer (inter-doublet) splittings in \cref{tab:vacuum_gaps} are well resolved by DMRG and, as shown by \cref{fig:vacuum_profiles_gauge}, they originate from a physical effect in the gauge link configuration.
We thus rule out SSB and attribute the $\Csym$ and $\Psym$ violations in the density profiles to the hybridization of the eigenstates found by DMRG with a low-energy flavourless excitation\emdash{}such as a neutral pion $\pi^0$\emdash{}whose gap we expect to close in the thermodynamic limit.
In order to circumvent problems originating from symmetry violations, we do not subtract number density VEVs in \cref{sec:excitations}.

\subsection{Structure factors}
\begin{figure}
    \begin{subcaptions}{\includegraphics{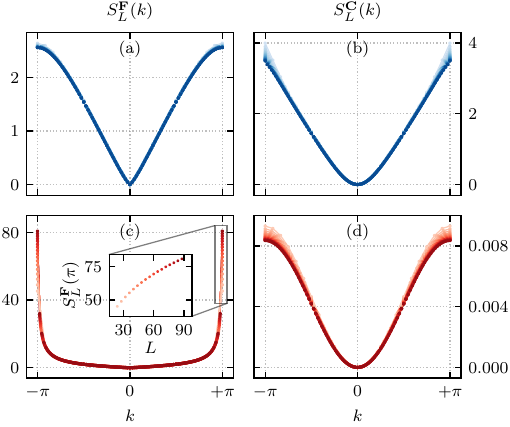}}
        \oversubcaption{0.0, 0.0}{\label{fig:vacuum_SSB_F_weak}}
        \oversubcaption{0.5, 0.0}{\label{fig:vacuum_SSB_C_weak}}
        \oversubcaption{0.0, 0.5}{\label{fig:vacuum_SSB_F_strong}}
        \oversubcaption{0.5, 0.5}{\label{fig:vacuum_SSB_C_strong}}
    \end{subcaptions}
    \caption{\label{fig:vacuum_SSB}%
        Structure factors from \cref{eq:structure_factor} detecting $\Fsym$ (left) and $\Csym$ (right) long-range order at $g=0.1$ (top, blue) and $g=10.0$ (bottom, red).
        Finite size scaling with $L\in[24, 88]$ (darker tones correspond to longer chains).
    }
\end{figure}
We conclude by presenting another diagnostic for (the lack of) SSB.
If the symmetry $\sigma$ suspected of being broken is known, SSB can be detected testing for long-range order via some associated order parameter.
\Cref{fig:vacuum_SSB} shows the finite size scaling of the structure factors,
\begin{equation}\label{eq:structure_factor}
    S_L^{\sigma}(k) = \frac{1}{L} \sum_{y,z} e^{-ik(y-z)} \vev{O^{\sigma}_y O^{\sigma}_z}
    \ ,
\end{equation}
for $\sigma\in\{\Fsym,\Csym\}$, with
\begin{math}
    O^{\Fsym,\Csym}_{\xx}=\fdens[u] \mp \fdens[d]
\end{math},
A peak $S_L^{\sigma}(k) \sim L$ would reveal long-range order in $\sigma$ with $2\pi/k$-periodicity.
The peak at $k=\pi$ in \cref{fig:vacuum_SSB_F_strong} shows the emergence of antiferromagnetic flavor order at strong coupling.
Still, $S_L^{\Fsym}(\pi)$ grows sub-linearly with $L$ (inset plot), suggesting that the order is quasi-long-range.
This result is compatible with the Mermin-Wagner theorem \cite{Mermin1966AbsenceFerromagnetismAntiferromagnetism,Hohenberg1967ExistenceLongRange,Sachdev2011QuantumPhaseTransitions}, forbidding spontaneous breaking of continuos symmetries in 1D quantum models with short range interactions ($\Fsym$ corresponds to a $\Zn{2}$ subgroup of $\SUN(2)$-isospin).
The fact that the isospin-$3/2$ quadruplet
dominates the strong coupling physics in \cref{eq:sce}, while $\ket{\circ}$ and $\ket{\bullet}$ are highly suppressed, explains both the quasi-long-range $\Fsym$ order (arising from the antiferromagnetic XXX$_{3/2}$ model)
and, simultaneously, the lack of $\Csym$ breaking (which would be expected for an antiferromagnetic Ising model).
In conclusion, we find no evidence of SSB.

\section{Convergence}\label{app:convergence}
\begin{figure}
    \vspace{1em}
    \begin{subcaptions}{\includegraphics{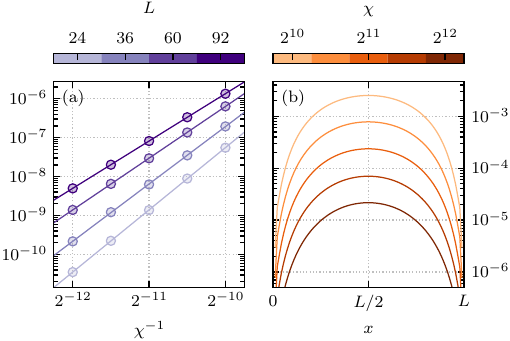}}
        \oversubcaption{0.0, 0.0}{\label{fig:convergence_energy}}
        \oversubcaption{0.5, 0.0}{\label{fig:convergence_entropy}}
    \end{subcaptions}
    \caption{\label{fig:convergence}%
    Convergence of the energy $\varepsilon_{L}$ \sref{fig:convergence_energy} and entanglement entropy $S(x)$ \sref{fig:convergence_entropy}: relative deviation $\abs{O_{\chi}/O_{\chi\to\infty}-1}$ of their value at finite bond dimension $\chi$ from the exact one, where $O = \varepsilon_{L}, S(x)$.
    The exact value is estimated via a power law interpolation $O_{\chi} = A \chi^B + O_{\chi\to\infty}$.
    The precision decreases with the system size, as shown in \sref{fig:convergence_energy} for the energy.
    The same holds for the entropy, although only the largest size $L=92$ is reported in \sref{fig:convergence_entropy}.
    }
\end{figure}
The aim of this work consists in characterizing certain qualitative features of \themodel{} (mainly, existence of the continuum limit and its particles) rather than extracting quantitative numerical estimates.
For this reason, the finite bond dimension extrapolation to $\chi\to\infty$ has not been performed systematically.
We nonetheless verified the convergence of numerical simulations in the more demanding weak coupling phase.
We achieved 6 digits or higher relative precision for ground state energies (\cref{fig:convergence_energy}).
Consistent with expectations, entanglement entropy suffers more severely from the TN approximation, with truncation effects reaching almost the percent order at mid-chain in the longest chains (\cref{fig:convergence_entropy}).


%


\begin{thebibliography}{169}%
\makeatletter
\providecommand \@ifxundefined [1]{%
 \@ifx{#1\undefined}
}%
\providecommand \@ifnum [1]{%
 \ifnum #1\expandafter \@firstoftwo
 \else \expandafter \@secondoftwo
 \fi
}%
\providecommand \@ifx [1]{%
 \ifx #1\expandafter \@firstoftwo
 \else \expandafter \@secondoftwo
 \fi
}%
\providecommand \natexlab [1]{#1}%
\providecommand \enquote  [1]{``#1''}%
\providecommand \bibnamefont  [1]{#1}%
\providecommand \bibfnamefont [1]{#1}%
\providecommand \citenamefont [1]{#1}%
\providecommand \href@noop [0]{\@secondoftwo}%
\providecommand \href [0]{\begingroup \@sanitize@url \@href}%
\providecommand \@href[1]{\@@startlink{#1}\@@href}%
\providecommand \@@href[1]{\endgroup#1\@@endlink}%
\providecommand \@sanitize@url [0]{\catcode `\\12\catcode `\$12\catcode
  `\&12\catcode `\#12\catcode `\^12\catcode `\_12\catcode `\%12\relax}%
\providecommand \@@startlink[1]{}%
\providecommand \@@endlink[0]{}%
\providecommand \url  [0]{\begingroup\@sanitize@url \@url }%
\providecommand \@url [1]{\endgroup\@href {#1}{\urlprefix }}%
\providecommand \urlprefix  [0]{URL }%
\providecommand \Eprint [0]{\href }%
\providecommand \doibase [0]{https://doi.org/}%
\providecommand \selectlanguage [0]{\@gobble}%
\providecommand \bibinfo  [0]{\@secondoftwo}%
\providecommand \bibfield  [0]{\@secondoftwo}%
\providecommand \translation [1]{[#1]}%
\providecommand \BibitemOpen [0]{}%
\providecommand \bibitemStop [0]{}%
\providecommand \bibitemNoStop [0]{.\EOS\space}%
\providecommand \EOS [0]{\spacefactor3000\relax}%
\providecommand \BibitemShut  [1]{\csname bibitem#1\endcsname}%
\let\auto@bib@innerbib\@empty
\bibitem [{\citenamefont {Fritzsch}\ \emph {et~al.}(1973)\citenamefont
  {Fritzsch}, \citenamefont {Gell-Mann},\ and\ \citenamefont
  {Leutwyler}}]{Fritzsch1973AdvantagesColorOctet}%
  \BibitemOpen
  \bibfield  {author} {\bibinfo {author} {\bibfnamefont {H.}~\bibnamefont
  {Fritzsch}}, \bibinfo {author} {\bibfnamefont {M.}~\bibnamefont
  {Gell-Mann}},\ and\ \bibinfo {author} {\bibfnamefont {H.}~\bibnamefont
  {Leutwyler}},\ }\bibfield  {title} {\bibinfo {title} {Advantages of the color
  octet gluon picture},\ }\href {https://doi.org/10.1016/0370-2693(73)90625-4}
  {\bibfield  {journal} {\bibinfo  {journal} {Physics Letters B}\ }\textbf
  {\bibinfo {volume} {47}},\ \bibinfo {pages} {365} (\bibinfo {year}
  {1973})}\BibitemShut {NoStop}%
\bibitem [{\citenamefont {Wilson}(1974)}]{Wilson1974ConfinementQuarks}%
  \BibitemOpen
  \bibfield  {author} {\bibinfo {author} {\bibfnamefont {K.~G.}\ \bibnamefont
  {Wilson}},\ }\bibfield  {title} {\bibinfo {title} {Confinement of quarks},\
  }\href {https://doi.org/10.1103/PhysRevD.10.2445} {\bibfield  {journal}
  {\bibinfo  {journal} {Physical Review D}\ }\textbf {\bibinfo {volume} {10}},\
  \bibinfo {pages} {2445} (\bibinfo {year} {1974})}\BibitemShut {NoStop}%
\bibitem [{\citenamefont {Brambilla}\ \emph {et~al.}(2014)\citenamefont
  {Brambilla} \emph {et~al.}}]{Brambilla2014QcdStronglyCoupled}%
  \BibitemOpen
  \bibfield  {author} {\bibinfo {author} {\bibfnamefont {N.}~\bibnamefont
  {Brambilla}} \emph {et~al.},\ }\bibfield  {title} {\bibinfo {title} {{QCD}
  and strongly coupled gauge theories: challenges and perspectives},\ }\href
  {https://doi.org/10.1140/epjc/s10052-014-2981-5} {\bibfield  {journal}
  {\bibinfo  {journal} {The European Physical Journal C}\ }\textbf {\bibinfo
  {volume} {74}},\ \bibinfo {pages} {2981} (\bibinfo {year}
  {2014})}\BibitemShut {NoStop}%
\bibitem [{\citenamefont {D{{\"u}}rr}\ \emph {et~al.}(2008)\citenamefont
  {D{{\"u}}rr}, \citenamefont {Fodor}, \citenamefont {Frison}, \citenamefont
  {Hoelbling}, \citenamefont {Hoffmann}, \citenamefont {Katz}, \citenamefont
  {Krieg}, \citenamefont {Kurth}, \citenamefont {Lellouch}, \citenamefont
  {Lippert}, \citenamefont {Szabo},\ and\ \citenamefont
  {Vulvert}}]{Duerr2008AbInitioDetermination}%
  \BibitemOpen
  \bibfield  {author} {\bibinfo {author} {\bibfnamefont {S.}~\bibnamefont
  {D{{\"u}}rr}}, \bibinfo {author} {\bibfnamefont {Z.}~\bibnamefont {Fodor}},
  \bibinfo {author} {\bibfnamefont {J.}~\bibnamefont {Frison}}, \bibinfo
  {author} {\bibfnamefont {C.}~\bibnamefont {Hoelbling}}, \bibinfo {author}
  {\bibfnamefont {R.}~\bibnamefont {Hoffmann}}, \bibinfo {author}
  {\bibfnamefont {S.~D.}\ \bibnamefont {Katz}}, \bibinfo {author}
  {\bibfnamefont {S.}~\bibnamefont {Krieg}}, \bibinfo {author} {\bibfnamefont
  {T.}~\bibnamefont {Kurth}}, \bibinfo {author} {\bibfnamefont
  {L.}~\bibnamefont {Lellouch}}, \bibinfo {author} {\bibfnamefont
  {T.}~\bibnamefont {Lippert}}, \bibinfo {author} {\bibfnamefont {K.~K.}\
  \bibnamefont {Szabo}},\ and\ \bibinfo {author} {\bibfnamefont
  {G.}~\bibnamefont {Vulvert}},\ }\bibfield  {title} {\bibinfo {title} {Ab
  {Initio} {Determination} of {Light} {Hadron} {Masses}},\ }\href
  {https://doi.org/10.1126/science.1163233} {\bibfield  {journal} {\bibinfo
  {journal} {Science}\ }\textbf {\bibinfo {volume} {322}},\ \bibinfo {pages}
  {1224} (\bibinfo {year} {2008})}\BibitemShut {NoStop}%
\bibitem [{\citenamefont {Lin}(2022)}]{Lin2022HadronSpectroscopyStructure}%
  \BibitemOpen
  \bibfield  {author} {\bibinfo {author} {\bibfnamefont {H.-W.}\ \bibnamefont
  {Lin}},\ }\bibfield  {title} {\bibinfo {title} {Hadron {Spectroscopy} and
  {Structure} from {Lattice} {QCD}},\ }\href
  {https://doi.org/10.1007/s00601-022-01764-y} {\bibfield  {journal} {\bibinfo
  {journal} {Few-Body Systems}\ }\textbf {\bibinfo {volume} {63}},\ \bibinfo
  {pages} {65} (\bibinfo {year} {2022})}\BibitemShut {NoStop}%
\bibitem [{\citenamefont {Detar}\ and\ \citenamefont
  {Gottlieb}(2004)}]{Detar2004LatticeQuantumChromodynamics}%
  \BibitemOpen
  \bibfield  {author} {\bibinfo {author} {\bibfnamefont {C.}~\bibnamefont
  {Detar}}\ and\ \bibinfo {author} {\bibfnamefont {S.}~\bibnamefont
  {Gottlieb}},\ }\bibfield  {title} {\bibinfo {title} {Lattice quantum
  chromodynamics comes of age},\ }\href {https://doi.org/10.1063/1.1688069}
  {\bibfield  {journal} {\bibinfo  {journal} {Physics Today}\ }\textbf
  {\bibinfo {volume} {57}},\ \bibinfo {pages} {45} (\bibinfo {year}
  {2004})}\BibitemShut {NoStop}%
\bibitem [{\citenamefont {Grabowska}\ \emph {et~al.}(2013)\citenamefont
  {Grabowska}, \citenamefont {Kaplan},\ and\ \citenamefont
  {Nicholson}}]{Grabowska2013SignProblemsNoise}%
  \BibitemOpen
  \bibfield  {author} {\bibinfo {author} {\bibfnamefont {D.}~\bibnamefont
  {Grabowska}}, \bibinfo {author} {\bibfnamefont {D.~B.}\ \bibnamefont
  {Kaplan}},\ and\ \bibinfo {author} {\bibfnamefont {A.~N.}\ \bibnamefont
  {Nicholson}},\ }\bibfield  {title} {\bibinfo {title} {Sign problems, noise,
  and chiral symmetry breaking in a {QCD}-like theory},\ }\href
  {https://doi.org/10.1103/PhysRevD.87.014504} {\bibfield  {journal} {\bibinfo
  {journal} {Physical Review D}\ }\textbf {\bibinfo {volume} {87}},\ \bibinfo
  {pages} {014504} (\bibinfo {year} {2013})}\BibitemShut {NoStop}%
\bibitem [{\citenamefont {Nagata}(2022)}]{Nagata2022FiniteDensityLattice}%
  \BibitemOpen
  \bibfield  {author} {\bibinfo {author} {\bibfnamefont {K.}~\bibnamefont
  {Nagata}},\ }\bibfield  {title} {\bibinfo {title} {Finite-density lattice
  {QCD} and sign problem: {Current} status and open problems},\ }\href
  {https://doi.org/10.1016/j.ppnp.2022.103991} {\bibfield  {journal} {\bibinfo
  {journal} {Progress in Particle and Nuclear Physics}\ }\textbf {\bibinfo
  {volume} {127}},\ \bibinfo {pages} {103991} (\bibinfo {year}
  {2022})}\BibitemShut {NoStop}%
\bibitem [{\citenamefont {Kogut}\ and\ \citenamefont
  {Stephanov}(2003)}]{Kogut2003PhasesQuantumChromodynamicsa}%
  \BibitemOpen
  \bibfield  {author} {\bibinfo {author} {\bibfnamefont {J.~B.}\ \bibnamefont
  {Kogut}}\ and\ \bibinfo {author} {\bibfnamefont {M.~A.}\ \bibnamefont
  {Stephanov}},\ }\href {https://doi.org/10.1017/CBO9780511534980} {\emph
  {\bibinfo {title} {The {Phases} of {Quantum} {Chromodynamics}: {From}
  {Confinement} to {Extreme} {Environments}}}},\ Cambridge {Monographs} on
  {Particle} {Physics}, {Nuclear} {Physics} and {Cosmology}\ (\bibinfo
  {publisher} {Cambridge University Press},\ \bibinfo {address} {Cambridge},\
  \bibinfo {year} {2003})\BibitemShut {NoStop}%
\bibitem [{\citenamefont {Ba{\~n}uls}\ and\ \citenamefont
  {Cichy}(2021)}]{Banuls2021TensorsCastTheir}%
  \BibitemOpen
  \bibfield  {author} {\bibinfo {author} {\bibfnamefont {M.~C.}\ \bibnamefont
  {Ba{\~n}uls}}\ and\ \bibinfo {author} {\bibfnamefont {K.}~\bibnamefont
  {Cichy}},\ }\bibfield  {title} {\bibinfo {title} {Tensors cast their nets for
  quarks},\ }\href {https://doi.org/10.1038/s41567-021-01294-0} {\bibfield
  {journal} {\bibinfo  {journal} {Nature Physics}\ }\textbf {\bibinfo {volume}
  {17}},\ \bibinfo {pages} {762} (\bibinfo {year} {2021})}\BibitemShut
  {NoStop}%
\bibitem [{\citenamefont {Borsanyi}\ \emph {et~al.}(2020)\citenamefont
  {Borsanyi}, \citenamefont {Fodor}, \citenamefont {Guenther}, \citenamefont
  {Kara}, \citenamefont {Katz}, \citenamefont {Parotto}, \citenamefont
  {Pasztor}, \citenamefont {Ratti},\ and\ \citenamefont
  {Szab{\'o}}}]{Borsanyi2020QcdCrossoverFinite}%
  \BibitemOpen
  \bibfield  {author} {\bibinfo {author} {\bibfnamefont {S.}~\bibnamefont
  {Borsanyi}}, \bibinfo {author} {\bibfnamefont {Z.}~\bibnamefont {Fodor}},
  \bibinfo {author} {\bibfnamefont {J.~N.}\ \bibnamefont {Guenther}}, \bibinfo
  {author} {\bibfnamefont {R.}~\bibnamefont {Kara}}, \bibinfo {author}
  {\bibfnamefont {S.~D.}\ \bibnamefont {Katz}}, \bibinfo {author}
  {\bibfnamefont {P.}~\bibnamefont {Parotto}}, \bibinfo {author} {\bibfnamefont
  {A.}~\bibnamefont {Pasztor}}, \bibinfo {author} {\bibfnamefont
  {C.}~\bibnamefont {Ratti}},\ and\ \bibinfo {author} {\bibfnamefont {K.~K.}\
  \bibnamefont {Szab{\'o}}},\ }\bibfield  {title} {\bibinfo {title} {{QCD}
  {Crossover} at {Finite} {Chemical} {Potential} from {Lattice}
  {Simulations}},\ }\href {https://doi.org/10.1103/PhysRevLett.125.052001}
  {\bibfield  {journal} {\bibinfo  {journal} {Physical Review Letters}\
  }\textbf {\bibinfo {volume} {125}},\ \bibinfo {pages} {052001} (\bibinfo
  {year} {2020})}\BibitemShut {NoStop}%
\bibitem [{\citenamefont {Verstraete}\ \emph {et~al.}(2008)\citenamefont
  {Verstraete}, \citenamefont {Murg},\ and\ \citenamefont
  {Cirac}}]{Verstraete2008MatrixProductStates}%
  \BibitemOpen
  \bibfield  {author} {\bibinfo {author} {\bibfnamefont {F.}~\bibnamefont
  {Verstraete}}, \bibinfo {author} {\bibfnamefont {V.}~\bibnamefont {Murg}},\
  and\ \bibinfo {author} {\bibfnamefont {J.}~\bibnamefont {Cirac}},\ }\bibfield
   {title} {\bibinfo {title} {Matrix product states, projected entangled pair
  states, and variational renormalization group methods for quantum spin
  systems},\ }\href {https://doi.org/10.1080/14789940801912366} {\bibfield
  {journal} {\bibinfo  {journal} {Advances in Physics}\ }\textbf {\bibinfo
  {volume} {57}},\ \bibinfo {pages} {143} (\bibinfo {year} {2008})}\BibitemShut
  {NoStop}%
\bibitem [{\citenamefont
  {Or{\'u}s}(2014)}]{Orus2014PracticalIntroductionTensor}%
  \BibitemOpen
  \bibfield  {author} {\bibinfo {author} {\bibfnamefont {R.}~\bibnamefont
  {Or{\'u}s}},\ }\bibfield  {title} {\bibinfo {title} {A practical introduction
  to tensor networks: {Matrix} product states and projected entangled pair
  states},\ }\href {https://doi.org/10.1016/j.aop.2014.06.013} {\bibfield
  {journal} {\bibinfo  {journal} {Annals of Physics}\ }\textbf {\bibinfo
  {volume} {349}},\ \bibinfo {pages} {117} (\bibinfo {year}
  {2014})}\BibitemShut {NoStop}%
\bibitem [{\citenamefont
  {Montangero}(2018)}]{Montangero2018IntroductionTensorNetwork}%
  \BibitemOpen
  \bibfield  {author} {\bibinfo {author} {\bibfnamefont {S.}~\bibnamefont
  {Montangero}},\ }\href {https://doi.org/10.1007/978-3-030-01409-4} {\emph
  {\bibinfo {title} {Introduction to {Tensor} {Network} {Methods}: {Numerical}
  simulations of low-dimensional many-body quantum systems}}}\ (\bibinfo
  {publisher} {Springer International Publishing},\ \bibinfo {year}
  {2018})\BibitemShut {NoStop}%
\bibitem [{\citenamefont {Silvi}\ \emph
  {et~al.}(2019{\natexlab{a}})\citenamefont {Silvi}, \citenamefont
  {Tschirsich}, \citenamefont {Gerster}, \citenamefont {J{\"u}nemann},
  \citenamefont {Jaschke}, \citenamefont {Rizzi},\ and\ \citenamefont
  {Montangero}}]{Silvi2019TensorNetworksAnthology}%
  \BibitemOpen
  \bibfield  {author} {\bibinfo {author} {\bibfnamefont {P.}~\bibnamefont
  {Silvi}}, \bibinfo {author} {\bibfnamefont {F.}~\bibnamefont {Tschirsich}},
  \bibinfo {author} {\bibfnamefont {M.}~\bibnamefont {Gerster}}, \bibinfo
  {author} {\bibfnamefont {J.}~\bibnamefont {J{\"u}nemann}}, \bibinfo {author}
  {\bibfnamefont {D.}~\bibnamefont {Jaschke}}, \bibinfo {author} {\bibfnamefont
  {M.}~\bibnamefont {Rizzi}},\ and\ \bibinfo {author} {\bibfnamefont
  {S.}~\bibnamefont {Montangero}},\ }\bibfield  {title} {\bibinfo {title} {The
  {Tensor} {Networks} {Anthology}: {Simulation} techniques for many-body
  quantum lattice systems},\ }\bibfield  {journal} {\bibinfo  {journal}
  {SciPost Physics Lecture Notes}\ }\textbf {\bibinfo {volume} {8}},\ \href
  {https://doi.org/10.21468/SciPostPhysLectNotes.8}
  {10.21468/SciPostPhysLectNotes.8} (\bibinfo {year}
  {2019}{\natexlab{a}})\BibitemShut {NoStop}%
\bibitem [{\citenamefont {Emonts}\ \emph {et~al.}(2023)\citenamefont {Emonts},
  \citenamefont {Kelman}, \citenamefont {Borla}, \citenamefont {Moroz},
  \citenamefont {Gazit},\ and\ \citenamefont
  {Zohar}}]{Emonts2023FindingGroundState}%
  \BibitemOpen
  \bibfield  {author} {\bibinfo {author} {\bibfnamefont {P.}~\bibnamefont
  {Emonts}}, \bibinfo {author} {\bibfnamefont {A.}~\bibnamefont {Kelman}},
  \bibinfo {author} {\bibfnamefont {U.}~\bibnamefont {Borla}}, \bibinfo
  {author} {\bibfnamefont {S.}~\bibnamefont {Moroz}}, \bibinfo {author}
  {\bibfnamefont {S.}~\bibnamefont {Gazit}},\ and\ \bibinfo {author}
  {\bibfnamefont {E.}~\bibnamefont {Zohar}},\ }\bibfield  {title} {\bibinfo
  {title} {Finding the ground state of a lattice gauge theory with fermionic
  tensor networks: a $2+1d$ $\mathbb{Z}_2$ demonstration},\ }\href
  {https://doi.org/10.1103/PhysRevD.107.014505} {\bibfield  {journal} {\bibinfo
   {journal} {Physical Review D}\ }\textbf {\bibinfo {volume} {107}},\ \bibinfo
  {pages} {014505} (\bibinfo {year} {2023})},\ \Eprint
  {https://arxiv.org/abs/2211.00023} {arxiv:2211.00023} \BibitemShut {NoStop}%
\bibitem [{\citenamefont {Eisert}\ \emph {et~al.}(2010)\citenamefont {Eisert},
  \citenamefont {Cramer},\ and\ \citenamefont
  {Plenio}}]{Eisert2010ColloquiumAreaLaws}%
  \BibitemOpen
  \bibfield  {author} {\bibinfo {author} {\bibfnamefont {J.}~\bibnamefont
  {Eisert}}, \bibinfo {author} {\bibfnamefont {M.}~\bibnamefont {Cramer}},\
  and\ \bibinfo {author} {\bibfnamefont {M.~B.}\ \bibnamefont {Plenio}},\
  }\bibfield  {title} {\bibinfo {title} {Colloquium: {Area} laws for the
  entanglement entropy},\ }\href {https://doi.org/10.1103/RevModPhys.82.277}
  {\bibfield  {journal} {\bibinfo  {journal} {Reviews of Modern Physics}\
  }\textbf {\bibinfo {volume} {82}},\ \bibinfo {pages} {277} (\bibinfo {year}
  {2010})}\BibitemShut {NoStop}%
\bibitem [{\citenamefont
  {Accardi}(1981)}]{Accardi1981TopicsQuantumProbability}%
  \BibitemOpen
  \bibfield  {author} {\bibinfo {author} {\bibfnamefont {L.}~\bibnamefont
  {Accardi}},\ }\bibfield  {title} {\bibinfo {title} {Topics in quantum
  probability},\ }\href {https://doi.org/10.1016/0370-1573(81)90070-3}
  {\bibfield  {journal} {\bibinfo  {journal} {Physics Reports}\ }\textbf
  {\bibinfo {volume} {77}},\ \bibinfo {pages} {169} (\bibinfo {year}
  {1981})}\BibitemShut {NoStop}%
\bibitem [{\citenamefont {Fannes}\ \emph {et~al.}(1989)\citenamefont {Fannes},
  \citenamefont {Nachtergaele},\ and\ \citenamefont
  {Werner}}]{Fannes1989ExactAntiferromagneticGround}%
  \BibitemOpen
  \bibfield  {author} {\bibinfo {author} {\bibfnamefont {M.}~\bibnamefont
  {Fannes}}, \bibinfo {author} {\bibfnamefont {B.}~\bibnamefont
  {Nachtergaele}},\ and\ \bibinfo {author} {\bibfnamefont {R.~F.}\ \bibnamefont
  {Werner}},\ }\bibfield  {title} {\bibinfo {title} {Exact {Antiferromagnetic}
  {Ground} {States} of {Quantum} {Spin} {Chains}},\ }\href
  {https://doi.org/10.1209/0295-5075/10/7/005} {\bibfield  {journal} {\bibinfo
  {journal} {Europhysics Letters}\ }\textbf {\bibinfo {volume} {10}},\ \bibinfo
  {pages} {633} (\bibinfo {year} {1989})}\BibitemShut {NoStop}%
\bibitem [{\citenamefont {{\"O}stlund}\ and\ \citenamefont
  {Rommer}(1995)}]{Oestlund1995ThermodynamicLimitDensity}%
  \BibitemOpen
  \bibfield  {author} {\bibinfo {author} {\bibfnamefont {S.}~\bibnamefont
  {{\"O}stlund}}\ and\ \bibinfo {author} {\bibfnamefont {S.}~\bibnamefont
  {Rommer}},\ }\bibfield  {title} {\bibinfo {title} {Thermodynamic {Limit} of
  {Density} {Matrix} {Renormalization}},\ }\href
  {https://doi.org/10.1103/PhysRevLett.75.3537} {\bibfield  {journal} {\bibinfo
   {journal} {Physical Review Letters}\ }\textbf {\bibinfo {volume} {75}},\
  \bibinfo {pages} {3537} (\bibinfo {year} {1995})}\BibitemShut {NoStop}%
\bibitem [{\citenamefont {Verstraete}\ and\ \citenamefont
  {Cirac}(2004)}]{Verstraete2004RenormalizationAlgorithmsQuantum}%
  \BibitemOpen
  \bibfield  {author} {\bibinfo {author} {\bibfnamefont {F.}~\bibnamefont
  {Verstraete}}\ and\ \bibinfo {author} {\bibfnamefont {J.~I.}\ \bibnamefont
  {Cirac}},\ }\href {https://doi.org/10.48550/arXiv.cond-mat/0407066} {\bibinfo
  {title} {Renormalization algorithms for {Quantum}-{Many} {Body} {Systems} in
  two and higher dimensions}} (\bibinfo {year} {2004}),\ \Eprint
  {https://arxiv.org/abs/cond-mat/0407066} {arXiv:cond-mat/0407066}
  \BibitemShut {NoStop}%
\bibitem [{\citenamefont {Verstraete}\ \emph {et~al.}(2006)\citenamefont
  {Verstraete}, \citenamefont {Wolf}, \citenamefont {Perez-Garcia},\ and\
  \citenamefont {Cirac}}]{Verstraete2006CriticalityAreaLaw}%
  \BibitemOpen
  \bibfield  {author} {\bibinfo {author} {\bibfnamefont {F.}~\bibnamefont
  {Verstraete}}, \bibinfo {author} {\bibfnamefont {M.~M.}\ \bibnamefont
  {Wolf}}, \bibinfo {author} {\bibfnamefont {D.}~\bibnamefont {Perez-Garcia}},\
  and\ \bibinfo {author} {\bibfnamefont {J.~I.}\ \bibnamefont {Cirac}},\
  }\bibfield  {title} {\bibinfo {title} {Criticality, the {A}rea {L}aw, and the
  {C}omputational {P}ower of {P}rojected {E}ntangled {P}air {S}tates},\ }\href
  {https://doi.org/10.1103/PhysRevLett.96.220601} {\bibfield  {journal}
  {\bibinfo  {journal} {Physical Review Letters}\ }\textbf {\bibinfo {volume}
  {96}},\ \bibinfo {pages} {220601} (\bibinfo {year} {2006})}\BibitemShut
  {NoStop}%
\bibitem [{\citenamefont {Fannes}\ \emph
  {et~al.}(1992{\natexlab{a}})\citenamefont {Fannes}, \citenamefont
  {Nachtergaele},\ and\ \citenamefont {Werner}}]{Fannes1992GroundStatesVbs}%
  \BibitemOpen
  \bibfield  {author} {\bibinfo {author} {\bibfnamefont {M.}~\bibnamefont
  {Fannes}}, \bibinfo {author} {\bibfnamefont {B.}~\bibnamefont
  {Nachtergaele}},\ and\ \bibinfo {author} {\bibfnamefont {R.~F.}\ \bibnamefont
  {Werner}},\ }\bibfield  {title} {\bibinfo {title} {Ground states of {VBS}
  models on cayley trees},\ }\href {https://doi.org/10.1007/BF01055710}
  {\bibfield  {journal} {\bibinfo  {journal} {Journal of Statistical Physics}\
  }\textbf {\bibinfo {volume} {66}},\ \bibinfo {pages} {939} (\bibinfo {year}
  {1992}{\natexlab{a}})}\BibitemShut {NoStop}%
\bibitem [{\citenamefont {Felser}\ \emph
  {et~al.}(2021{\natexlab{a}})\citenamefont {Felser}, \citenamefont
  {Notarnicola},\ and\ \citenamefont
  {Montangero}}]{Felser2021EfficientTensorNetwork}%
  \BibitemOpen
  \bibfield  {author} {\bibinfo {author} {\bibfnamefont {T.}~\bibnamefont
  {Felser}}, \bibinfo {author} {\bibfnamefont {S.}~\bibnamefont
  {Notarnicola}},\ and\ \bibinfo {author} {\bibfnamefont {S.}~\bibnamefont
  {Montangero}},\ }\bibfield  {title} {\bibinfo {title} {Efficient {Tensor}
  {Network} {Ansatz} for {High}-{Dimensional} {Quantum} {Many}-{Body}
  {Problems}},\ }\href {https://doi.org/10.1103/PhysRevLett.126.170603}
  {\bibfield  {journal} {\bibinfo  {journal} {Physical Review Letters}\
  }\textbf {\bibinfo {volume} {126}},\ \bibinfo {pages} {170603} (\bibinfo
  {year} {2021}{\natexlab{a}})},\ \Eprint {https://arxiv.org/abs/2011.08200}
  {arXiv:2011.08200} \BibitemShut {NoStop}%
\bibitem [{\citenamefont {Ferrari}\ \emph {et~al.}(2022)\citenamefont
  {Ferrari}, \citenamefont {Magnifico},\ and\ \citenamefont
  {Montangero}}]{Ferrari2022AdaptiveWeightedTree}%
  \BibitemOpen
  \bibfield  {author} {\bibinfo {author} {\bibfnamefont {G.}~\bibnamefont
  {Ferrari}}, \bibinfo {author} {\bibfnamefont {G.}~\bibnamefont {Magnifico}},\
  and\ \bibinfo {author} {\bibfnamefont {S.}~\bibnamefont {Montangero}},\
  }\bibfield  {title} {\bibinfo {title} {Adaptive-weighted tree tensor networks
  for disordered quantum many-body systems},\ }\href
  {https://doi.org/10.1103/PhysRevB.105.214201} {\bibfield  {journal} {\bibinfo
   {journal} {Phys. Rev. B}\ }\textbf {\bibinfo {volume} {105}},\ \bibinfo
  {pages} {214201} (\bibinfo {year} {2022})}\BibitemShut {NoStop}%
\bibitem [{\citenamefont {Vidal}(2007)}]{Vidal2007EntanglementRenormalization}%
  \BibitemOpen
  \bibfield  {author} {\bibinfo {author} {\bibfnamefont {G.}~\bibnamefont
  {Vidal}},\ }\bibfield  {title} {\bibinfo {title} {Entanglement
  {R}enormalization},\ }\href {https://doi.org/10.1103/PhysRevLett.99.220405}
  {\bibfield  {journal} {\bibinfo  {journal} {Physical Review Letters}\
  }\textbf {\bibinfo {volume} {99}},\ \bibinfo {pages} {220405} (\bibinfo
  {year} {2007})}\BibitemShut {NoStop}%
\bibitem [{\citenamefont {Fannes}\ \emph
  {et~al.}(1992{\natexlab{b}})\citenamefont {Fannes}, \citenamefont
  {Nachtergaele},\ and\ \citenamefont
  {Werner}}]{Fannes1992FinitelyCorrelatedStates}%
  \BibitemOpen
  \bibfield  {author} {\bibinfo {author} {\bibfnamefont {M.}~\bibnamefont
  {Fannes}}, \bibinfo {author} {\bibfnamefont {B.}~\bibnamefont
  {Nachtergaele}},\ and\ \bibinfo {author} {\bibfnamefont {R.~F.}\ \bibnamefont
  {Werner}},\ }\bibfield  {title} {\bibinfo {title} {Finitely correlated states
  on quantum spin chains},\ }\href {https://doi.org/10.1007/BF02099178}
  {\bibfield  {journal} {\bibinfo  {journal} {Communications in Mathematical
  Physics}\ }\textbf {\bibinfo {volume} {144}},\ \bibinfo {pages} {443}
  (\bibinfo {year} {1992}{\natexlab{b}})}\BibitemShut {NoStop}%
\bibitem [{\citenamefont {Levin}\ and\ \citenamefont
  {Nave}(2007)}]{Levin2007TensorRenormalizationGroup}%
  \BibitemOpen
  \bibfield  {author} {\bibinfo {author} {\bibfnamefont {M.}~\bibnamefont
  {Levin}}\ and\ \bibinfo {author} {\bibfnamefont {C.~P.}\ \bibnamefont
  {Nave}},\ }\bibfield  {title} {\bibinfo {title} {Tensor {Renormalization}
  {Group} {Approach} to {Two}-{Dimensional} {Classical} {Lattice} {Models}},\
  }\href {https://doi.org/10.1103/PhysRevLett.99.120601} {\bibfield  {journal}
  {\bibinfo  {journal} {Physical Review Letters}\ }\textbf {\bibinfo {volume}
  {99}},\ \bibinfo {pages} {120601} (\bibinfo {year} {2007})}\BibitemShut
  {NoStop}%
\bibitem [{\citenamefont {Stoudenmire}\ and\ \citenamefont
  {Schwab}(2016)}]{Stoudenmire2016SupervisedLearningTensor}%
  \BibitemOpen
  \bibfield  {author} {\bibinfo {author} {\bibfnamefont {E.}~\bibnamefont
  {Stoudenmire}}\ and\ \bibinfo {author} {\bibfnamefont {D.~J.}\ \bibnamefont
  {Schwab}},\ }\bibfield  {title} {\bibinfo {title} {Supervised {Learning} with
  {Tensor} {Networks}},\ }in\ \href
  {https://papers.nips.cc/paper_files/paper/2016/hash/5314b9674c86e3f9d1ba25ef9bb32895-Abstract.html}
  {\emph {\bibinfo {booktitle} {Advances in {Neural} {Information} {Processing}
  {Systems}}}},\ Vol.~\bibinfo {volume} {29}\ (\bibinfo  {publisher} {Curran
  Associates, Inc.},\ \bibinfo {year} {2016})\BibitemShut {NoStop}%
\bibitem [{\citenamefont {Felser}\ \emph
  {et~al.}(2021{\natexlab{b}})\citenamefont {Felser}, \citenamefont {Trenti},
  \citenamefont {Sestini}, \citenamefont {Gianelle}, \citenamefont {Zuliani},
  \citenamefont {Lucchesi},\ and\ \citenamefont
  {Montangero}}]{Felser2021QuantumInspiredMachine}%
  \BibitemOpen
  \bibfield  {author} {\bibinfo {author} {\bibfnamefont {T.}~\bibnamefont
  {Felser}}, \bibinfo {author} {\bibfnamefont {M.}~\bibnamefont {Trenti}},
  \bibinfo {author} {\bibfnamefont {L.}~\bibnamefont {Sestini}}, \bibinfo
  {author} {\bibfnamefont {A.}~\bibnamefont {Gianelle}}, \bibinfo {author}
  {\bibfnamefont {D.}~\bibnamefont {Zuliani}}, \bibinfo {author} {\bibfnamefont
  {D.}~\bibnamefont {Lucchesi}},\ and\ \bibinfo {author} {\bibfnamefont
  {S.}~\bibnamefont {Montangero}},\ }\bibfield  {title} {\bibinfo {title}
  {Quantum-inspired machine learning on high-energy physics data},\ }\href
  {https://doi.org/10.1038/s41534-021-00443-w} {\bibfield  {journal} {\bibinfo
  {journal} {npj Quantum Information}\ }\textbf {\bibinfo {volume} {7}},\
  \bibinfo {pages} {1} (\bibinfo {year} {2021}{\natexlab{b}})}\BibitemShut
  {NoStop}%
\bibitem [{\citenamefont {Rico}\ \emph {et~al.}(2014)\citenamefont {Rico},
  \citenamefont {Pichler}, \citenamefont {Dalmonte}, \citenamefont {Zoller},\
  and\ \citenamefont {Montangero}}]{Rico2014TensorNetworksLattice}%
  \BibitemOpen
  \bibfield  {author} {\bibinfo {author} {\bibfnamefont {E.}~\bibnamefont
  {Rico}}, \bibinfo {author} {\bibfnamefont {T.}~\bibnamefont {Pichler}},
  \bibinfo {author} {\bibfnamefont {M.}~\bibnamefont {Dalmonte}}, \bibinfo
  {author} {\bibfnamefont {P.}~\bibnamefont {Zoller}},\ and\ \bibinfo {author}
  {\bibfnamefont {S.}~\bibnamefont {Montangero}},\ }\bibfield  {title}
  {\bibinfo {title} {Tensor {Networks} for {Lattice} {Gauge} {Theories} and
  {Atomic} {Quantum} {Simulation}},\ }\href
  {https://doi.org/10.1103/PhysRevLett.112.201601} {\bibfield  {journal}
  {\bibinfo  {journal} {Physical Review Letters}\ }\textbf {\bibinfo {volume}
  {112}},\ \bibinfo {pages} {201601} (\bibinfo {year} {2014})}\BibitemShut
  {NoStop}%
\bibitem [{\citenamefont {Tagliacozzo}\ \emph {et~al.}(2014)\citenamefont
  {Tagliacozzo}, \citenamefont {Celi},\ and\ \citenamefont
  {Lewenstein}}]{Tagliacozzo2014TensorNetworksLattice}%
  \BibitemOpen
  \bibfield  {author} {\bibinfo {author} {\bibfnamefont {L.}~\bibnamefont
  {Tagliacozzo}}, \bibinfo {author} {\bibfnamefont {A.}~\bibnamefont {Celi}},\
  and\ \bibinfo {author} {\bibfnamefont {M.}~\bibnamefont {Lewenstein}},\
  }\bibfield  {title} {\bibinfo {title} {Tensor {Networks} for {Lattice}
  {Gauge} {Theories} with {Continuous} {Groups}},\ }\href
  {https://doi.org/10.1103/PhysRevX.4.041024} {\bibfield  {journal} {\bibinfo
  {journal} {Physical Review X}\ }\textbf {\bibinfo {volume} {4}},\ \bibinfo
  {pages} {041024} (\bibinfo {year} {2014})}\BibitemShut {NoStop}%
\bibitem [{\citenamefont {Silvi}\ \emph {et~al.}(2014)\citenamefont {Silvi},
  \citenamefont {Rico}, \citenamefont {Calarco},\ and\ \citenamefont
  {Montangero}}]{Silvi2014LatticeGaugeTensor}%
  \BibitemOpen
  \bibfield  {author} {\bibinfo {author} {\bibfnamefont {P.}~\bibnamefont
  {Silvi}}, \bibinfo {author} {\bibfnamefont {E.}~\bibnamefont {Rico}},
  \bibinfo {author} {\bibfnamefont {T.}~\bibnamefont {Calarco}},\ and\ \bibinfo
  {author} {\bibfnamefont {S.}~\bibnamefont {Montangero}},\ }\bibfield  {title}
  {\bibinfo {title} {Lattice gauge tensor networks},\ }\href
  {https://doi.org/10.1088/1367-2630/16/10/103015} {\bibfield  {journal}
  {\bibinfo  {journal} {New Journal of Physics}\ }\textbf {\bibinfo {volume}
  {16}},\ \bibinfo {pages} {103015} (\bibinfo {year} {2014})}\BibitemShut
  {NoStop}%
\bibitem [{\citenamefont {Pichler}\ \emph {et~al.}(2016)\citenamefont
  {Pichler}, \citenamefont {Dalmonte}, \citenamefont {Rico}, \citenamefont
  {Zoller},\ and\ \citenamefont {Montangero}}]{Pichler2016RealTimeDynamics}%
  \BibitemOpen
  \bibfield  {author} {\bibinfo {author} {\bibfnamefont {T.}~\bibnamefont
  {Pichler}}, \bibinfo {author} {\bibfnamefont {M.}~\bibnamefont {Dalmonte}},
  \bibinfo {author} {\bibfnamefont {E.}~\bibnamefont {Rico}}, \bibinfo {author}
  {\bibfnamefont {P.}~\bibnamefont {Zoller}},\ and\ \bibinfo {author}
  {\bibfnamefont {S.}~\bibnamefont {Montangero}},\ }\bibfield  {title}
  {\bibinfo {title} {Real-{Time} {Dynamics} in {U}(1) {Lattice} {Gauge}
  {Theories} with {Tensor} {Networks}},\ }\href
  {https://doi.org/10.1103/PhysRevX.6.011023} {\bibfield  {journal} {\bibinfo
  {journal} {Physical Review X}\ }\textbf {\bibinfo {volume} {6}},\ \bibinfo
  {pages} {011023} (\bibinfo {year} {2016})}\BibitemShut {NoStop}%
\bibitem [{\citenamefont {Ercolessi}\ \emph {et~al.}(2018)\citenamefont
  {Ercolessi}, \citenamefont {Facchi}, \citenamefont {Magnifico}, \citenamefont
  {Pascazio},\ and\ \citenamefont {Pepe}}]{Ercolessi2018PhaseTransitionsZn}%
  \BibitemOpen
  \bibfield  {author} {\bibinfo {author} {\bibfnamefont {E.}~\bibnamefont
  {Ercolessi}}, \bibinfo {author} {\bibfnamefont {P.}~\bibnamefont {Facchi}},
  \bibinfo {author} {\bibfnamefont {G.}~\bibnamefont {Magnifico}}, \bibinfo
  {author} {\bibfnamefont {S.}~\bibnamefont {Pascazio}},\ and\ \bibinfo
  {author} {\bibfnamefont {F.~V.}\ \bibnamefont {Pepe}},\ }\bibfield  {title}
  {\bibinfo {title} {Phase transitions in {$Z_n$} gauge models: {Towards}
  quantum simulations of the {Schwinger}-{Weyl} {QED}},\ }\href
  {https://doi.org/10.1103/PhysRevD.98.074503} {\bibfield  {journal} {\bibinfo
  {journal} {Physical Review D: Particles and Fields}\ }\textbf {\bibinfo
  {volume} {98}},\ \bibinfo {pages} {074503} (\bibinfo {year}
  {2018})}\BibitemShut {NoStop}%
\bibitem [{\citenamefont {Magnifico}\ \emph {et~al.}(2020)\citenamefont
  {Magnifico}, \citenamefont {Dalmonte}, \citenamefont {Facchi}, \citenamefont
  {Pascazio}, \citenamefont {Pepe},\ and\ \citenamefont
  {Ercolessi}}]{Magnifico2020RealTimeDynamics}%
  \BibitemOpen
  \bibfield  {author} {\bibinfo {author} {\bibfnamefont {G.}~\bibnamefont
  {Magnifico}}, \bibinfo {author} {\bibfnamefont {M.}~\bibnamefont {Dalmonte}},
  \bibinfo {author} {\bibfnamefont {P.}~\bibnamefont {Facchi}}, \bibinfo
  {author} {\bibfnamefont {S.}~\bibnamefont {Pascazio}}, \bibinfo {author}
  {\bibfnamefont {F.~V.}\ \bibnamefont {Pepe}},\ and\ \bibinfo {author}
  {\bibfnamefont {E.}~\bibnamefont {Ercolessi}},\ }\bibfield  {title} {\bibinfo
  {title} {Real {T}ime {D}ynamics and {C}onfinement in the {$\mathbb{Z}_{n}$}
  {S}chwinger-{W}eyl lattice model for $1+1$ {QED}},\ }\href
  {https://doi.org/10.22331/q-2020-06-15-281} {\bibfield  {journal} {\bibinfo
  {journal} {{Quantum}}\ }\textbf {\bibinfo {volume} {4}},\ \bibinfo {pages}
  {281} (\bibinfo {year} {2020})}\BibitemShut {NoStop}%
\bibitem [{\citenamefont {Ba{\~n}uls}\ \emph {et~al.}(2017)\citenamefont
  {Ba{\~n}uls}, \citenamefont {Cichy}, \citenamefont {Cirac}, \citenamefont
  {Jansen},\ and\ \citenamefont
  {K{\"u}hn}}]{Banuls2017EfficientBasisFormulation}%
  \BibitemOpen
  \bibfield  {author} {\bibinfo {author} {\bibfnamefont {M.~C.}\ \bibnamefont
  {Ba{\~n}uls}}, \bibinfo {author} {\bibfnamefont {K.}~\bibnamefont {Cichy}},
  \bibinfo {author} {\bibfnamefont {J.~I.}\ \bibnamefont {Cirac}}, \bibinfo
  {author} {\bibfnamefont {K.}~\bibnamefont {Jansen}},\ and\ \bibinfo {author}
  {\bibfnamefont {S.}~\bibnamefont {K{\"u}hn}},\ }\bibfield  {title} {\bibinfo
  {title} {Efficient {Basis} {Formulation} for ($1+1$)-{Dimensional} {SU}(2)
  {Lattice} {Gauge} {Theory}: {Spectral} {Calculations} with {Matrix} {Product}
  {States}},\ }\href {https://doi.org/10.1103/PhysRevX.7.041046} {\bibfield
  {journal} {\bibinfo  {journal} {Physical Review X}\ }\textbf {\bibinfo
  {volume} {7}},\ \bibinfo {pages} {041046} (\bibinfo {year}
  {2017})}\BibitemShut {NoStop}%
\bibitem [{\citenamefont {Ba{\~n}uls}\ \emph {et~al.}(2019)\citenamefont
  {Ba{\~n}uls}, \citenamefont {Cichy}, \citenamefont {Cirac}, \citenamefont
  {Jansen},\ and\ \citenamefont {K{\"u}hn}}]{Banuls2019TensorNetworksTheir}%
  \BibitemOpen
  \bibfield  {author} {\bibinfo {author} {\bibfnamefont {M.}~\bibnamefont
  {Ba{\~n}uls}}, \bibinfo {author} {\bibfnamefont {K.}~\bibnamefont {Cichy}},
  \bibinfo {author} {\bibfnamefont {J.}~\bibnamefont {Cirac}}, \bibinfo
  {author} {\bibfnamefont {K.}~\bibnamefont {Jansen}},\ and\ \bibinfo {author}
  {\bibfnamefont {S.}~\bibnamefont {K{\"u}hn}},\ }\bibfield  {title} {\bibinfo
  {title} {Tensor networks and their use for lattice gauge theories},\ }in\
  \href {https://doi.org/10.22323/1.334.0022} {\emph {\bibinfo {booktitle}
  {Proceedings of The 36th Annual International Symposium on Lattice Field
  Theory — PoS(LATTICE2018)}}}\ (\bibinfo  {publisher} {Sissa Medialab},\
  \bibinfo {year} {2019})\BibitemShut {NoStop}%
\bibitem [{\citenamefont {Ba{\~n}uls}\ \emph {et~al.}(2020)\citenamefont
  {Ba{\~n}uls} \emph {et~al.}}]{Banuls2020SimulatingLatticeGauge}%
  \BibitemOpen
  \bibfield  {author} {\bibinfo {author} {\bibfnamefont {M.~C.}\ \bibnamefont
  {Ba{\~n}uls}} \emph {et~al.},\ }\bibfield  {title} {\bibinfo {title}
  {Simulating lattice gauge theories within quantum technologies},\ }\href
  {https://doi.org/10.1140/epjd/e2020-100571-8} {\bibfield  {journal} {\bibinfo
   {journal} {The European Physical Journal D: Atomic, Molecular, Optical and
  Plasma Physics}\ }\textbf {\bibinfo {volume} {74}},\ \bibinfo {pages} {165}
  (\bibinfo {year} {2020})}\BibitemShut {NoStop}%
\bibitem [{\citenamefont
  {Zohar}(2021{\natexlab{a}})}]{Zohar2021WilsonLoopsArea}%
  \BibitemOpen
  \bibfield  {author} {\bibinfo {author} {\bibfnamefont {E.}~\bibnamefont
  {Zohar}},\ }\bibfield  {title} {\bibinfo {title} {Wilson loops and area laws
  in lattice gauge theory tensor networks},\ }\href
  {https://doi.org/10.1103/PhysRevResearch.3.033179} {\bibfield  {journal}
  {\bibinfo  {journal} {Physical Review Research}\ }\textbf {\bibinfo {volume}
  {3}},\ \bibinfo {pages} {033179} (\bibinfo {year}
  {2021}{\natexlab{a}})}\BibitemShut {NoStop}%
\bibitem [{\citenamefont {Montangero}\ \emph {et~al.}(2022)\citenamefont
  {Montangero}, \citenamefont {Rico},\ and\ \citenamefont
  {Silvi}}]{Montangero2022LoopFreeTensor}%
  \BibitemOpen
  \bibfield  {author} {\bibinfo {author} {\bibfnamefont {S.}~\bibnamefont
  {Montangero}}, \bibinfo {author} {\bibfnamefont {E.}~\bibnamefont {Rico}},\
  and\ \bibinfo {author} {\bibfnamefont {P.}~\bibnamefont {Silvi}},\ }\bibfield
   {title} {\bibinfo {title} {Loop-free tensor networks for high-energy
  physics},\ }\bibfield  {journal} {\bibinfo  {journal} {Philosophical
  Transactions of the Royal Society A}\ }\href
  {https://doi.org/10.1098/rsta.2021.0065} {10.1098/rsta.2021.0065} (\bibinfo
  {year} {2022})\BibitemShut {NoStop}%
\bibitem [{\citenamefont {Bloch}\ and\ \citenamefont
  {Lohmayer}(2022)}]{Bloch2022GrassmannTensorNetwork}%
  \BibitemOpen
  \bibfield  {author} {\bibinfo {author} {\bibfnamefont {J.}~\bibnamefont
  {Bloch}}\ and\ \bibinfo {author} {\bibfnamefont {R.}~\bibnamefont
  {Lohmayer}},\ }\bibfield  {title} {\bibinfo {title} {Grassmann tensor-network
  method for strong-coupling {QCD}},\ }in\ \href
  {https://doi.org/10.22323/1.430.0004} {\emph {\bibinfo {booktitle}
  {Proceedings of {The} 39th {International} {Symposium} on {Lattice} {Field}
  {Theory} — {PoS}({LATTICE2022})}}}\ (\bibinfo {year} {2022})\ p.\ \bibinfo
  {pages} {004},\ \Eprint {https://arxiv.org/abs/2210.08935} {arxiv:2210.08935}
  \BibitemShut {NoStop}%
\bibitem [{\citenamefont {Di~Meglio}\ \emph {et~al.}(2023)\citenamefont
  {Di~Meglio} \emph {et~al.}}]{DiMeglio2023QuantumComputingHigh}%
  \BibitemOpen
  \bibfield  {author} {\bibinfo {author} {\bibfnamefont {A.}~\bibnamefont
  {Di~Meglio}} \emph {et~al.},\ }\href
  {https://doi.org/10.48550/arXiv.2307.03236} {\emph {\bibinfo {title} {Quantum
  {Computing} for {High}-{Energy} {Physics}: {State} of the {Art} and
  {Challenges}. {Summary} of the {QC4HEP} {Working} {Group}}}},\ \bibinfo
  {type} {Tech. Rep.}\ (\bibinfo {year} {2023})\ \Eprint
  {https://arxiv.org/abs/2307.03236} {arxiv:2307.03236} \BibitemShut {NoStop}%
\bibitem [{\citenamefont {Bauer}\ \emph
  {et~al.}(2023{\natexlab{a}})\citenamefont {Bauer} \emph
  {et~al.}}]{Bauer2023QuantumSimulationHigh}%
  \BibitemOpen
  \bibfield  {author} {\bibinfo {author} {\bibfnamefont {C.~W.}\ \bibnamefont
  {Bauer}} \emph {et~al.},\ }\bibfield  {title} {\bibinfo {title} {Quantum
  {Simulation} for {High}-{Energy} {Physics}},\ }\href
  {https://doi.org/10.1103/PRXQuantum.4.027001} {\bibfield  {journal} {\bibinfo
   {journal} {PRX Quantum}\ }\textbf {\bibinfo {volume} {4}},\ \bibinfo {pages}
  {027001} (\bibinfo {year} {2023}{\natexlab{a}})},\ \Eprint
  {https://arxiv.org/abs/2204.03381} {arxiv:2204.03381} \BibitemShut {NoStop}%
\bibitem [{\citenamefont {Hauke}\ \emph {et~al.}(2013)\citenamefont {Hauke},
  \citenamefont {Marcos}, \citenamefont {Dalmonte},\ and\ \citenamefont
  {Zoller}}]{Hauke2013QuantumSimulationLattice}%
  \BibitemOpen
  \bibfield  {author} {\bibinfo {author} {\bibfnamefont {P.}~\bibnamefont
  {Hauke}}, \bibinfo {author} {\bibfnamefont {D.}~\bibnamefont {Marcos}},
  \bibinfo {author} {\bibfnamefont {M.}~\bibnamefont {Dalmonte}},\ and\
  \bibinfo {author} {\bibfnamefont {P.}~\bibnamefont {Zoller}},\ }\bibfield
  {title} {\bibinfo {title} {Quantum {Simulation} of a {Lattice} {Schwinger}
  {Model} in a {Chain} of {Trapped} {Ions}},\ }\href
  {https://doi.org/10.1103/PhysRevX.3.041018} {\bibfield  {journal} {\bibinfo
  {journal} {Physical Review X}\ }\textbf {\bibinfo {volume} {3}},\ \bibinfo
  {pages} {041018} (\bibinfo {year} {2013})}\BibitemShut {NoStop}%
\bibitem [{\citenamefont {Banerjee}\ \emph {et~al.}(2013)\citenamefont
  {Banerjee}, \citenamefont {B{\"o}gli}, \citenamefont {Dalmonte},
  \citenamefont {Rico}, \citenamefont {Stebler}, \citenamefont {Wiese},\ and\
  \citenamefont {Zoller}}]{Banerjee2013AtomicQuantumSimulation}%
  \BibitemOpen
  \bibfield  {author} {\bibinfo {author} {\bibfnamefont {D.}~\bibnamefont
  {Banerjee}}, \bibinfo {author} {\bibfnamefont {M.}~\bibnamefont {B{\"o}gli}},
  \bibinfo {author} {\bibfnamefont {M.}~\bibnamefont {Dalmonte}}, \bibinfo
  {author} {\bibfnamefont {E.}~\bibnamefont {Rico}}, \bibinfo {author}
  {\bibfnamefont {P.}~\bibnamefont {Stebler}}, \bibinfo {author} {\bibfnamefont
  {U.-J.}\ \bibnamefont {Wiese}},\ and\ \bibinfo {author} {\bibfnamefont
  {P.}~\bibnamefont {Zoller}},\ }\bibfield  {title} {\bibinfo {title} {Atomic
  {Quantum} {Simulation} of {$\mathrm{U}(N)$} and {$\mathrm{SU}(N)$}
  {Non}-{Abelian} {Lattice} {Gauge} {Theories}},\ }\href
  {https://doi.org/10.1103/PhysRevLett.110.125303} {\bibfield  {journal}
  {\bibinfo  {journal} {Physical Review Letters}\ }\textbf {\bibinfo {volume}
  {110}},\ \bibinfo {pages} {125303} (\bibinfo {year} {2013})}\BibitemShut
  {NoStop}%
\bibitem [{\citenamefont
  {Zohar}(2016)}]{Zohar2016QuantumSimulationFundamental}%
  \BibitemOpen
  \bibfield  {author} {\bibinfo {author} {\bibfnamefont {E.}~\bibnamefont
  {Zohar}},\ }\bibfield  {title} {\bibinfo {title} {Quantum simulation of
  fundamental physics},\ }\href {https://doi.org/10.1038/534480a} {\bibfield
  {journal} {\bibinfo  {journal} {Nature}\ }\textbf {\bibinfo {volume} {534}},\
  \bibinfo {pages} {480} (\bibinfo {year} {2016})}\BibitemShut {NoStop}%
\bibitem [{\citenamefont {Gonz{\'a}lez-Cuadra}\ \emph
  {et~al.}(2022)\citenamefont {Gonz{\'a}lez-Cuadra}, \citenamefont {Zache},
  \citenamefont {Carrasco}, \citenamefont {Kraus},\ and\ \citenamefont
  {Zoller}}]{GonzalezCuadra2022HardwareEfficientQuantum}%
  \BibitemOpen
  \bibfield  {author} {\bibinfo {author} {\bibfnamefont {D.}~\bibnamefont
  {Gonz{\'a}lez-Cuadra}}, \bibinfo {author} {\bibfnamefont {T.~V.}\
  \bibnamefont {Zache}}, \bibinfo {author} {\bibfnamefont {J.}~\bibnamefont
  {Carrasco}}, \bibinfo {author} {\bibfnamefont {B.}~\bibnamefont {Kraus}},\
  and\ \bibinfo {author} {\bibfnamefont {P.}~\bibnamefont {Zoller}},\
  }\bibfield  {title} {\bibinfo {title} {Hardware {Efficient} {Quantum}
  {Simulation} of {Non}-{Abelian} {Gauge} {Theories} with {Qudits} on {Rydberg}
  {Platforms}},\ }\href {https://doi.org/10.1103/PhysRevLett.129.160501}
  {\bibfield  {journal} {\bibinfo  {journal} {Physical Review Letters}\
  }\textbf {\bibinfo {volume} {129}},\ \bibinfo {pages} {160501} (\bibinfo
  {year} {2022})}\BibitemShut {NoStop}%
\bibitem [{\citenamefont {Kruckenhauser}\ \emph {et~al.}(2022)\citenamefont
  {Kruckenhauser}, \citenamefont {Bijnen}, \citenamefont {Zache}, \citenamefont
  {Liberto},\ and\ \citenamefont
  {Zoller}}]{Kruckenhauser2022HighDimensionalSo4}%
  \BibitemOpen
  \bibfield  {author} {\bibinfo {author} {\bibfnamefont {A.}~\bibnamefont
  {Kruckenhauser}}, \bibinfo {author} {\bibfnamefont {R.~v.}\ \bibnamefont
  {Bijnen}}, \bibinfo {author} {\bibfnamefont {T.~V.}\ \bibnamefont {Zache}},
  \bibinfo {author} {\bibfnamefont {M.~D.}\ \bibnamefont {Liberto}},\ and\
  \bibinfo {author} {\bibfnamefont {P.}~\bibnamefont {Zoller}},\ }\bibfield
  {title} {\bibinfo {title} {High-dimensional {SO}(4)-symmetric {Rydberg}
  manifolds for quantum simulation},\ }\href
  {https://doi.org/10.1088/2058-9565/aca996} {\bibfield  {journal} {\bibinfo
  {journal} {Quantum Science and Technology}\ }\textbf {\bibinfo {volume}
  {8}},\ \bibinfo {pages} {015020} (\bibinfo {year} {2022})}\BibitemShut
  {NoStop}%
\bibitem [{\citenamefont {Knaute}\ and\ \citenamefont
  {Hauke}(2022)}]{Knaute2022RelativisticMesonSpectra}%
  \BibitemOpen
  \bibfield  {author} {\bibinfo {author} {\bibfnamefont {J.}~\bibnamefont
  {Knaute}}\ and\ \bibinfo {author} {\bibfnamefont {P.}~\bibnamefont {Hauke}},\
  }\bibfield  {title} {\bibinfo {title} {Relativistic meson spectra on ion-trap
  quantum simulators},\ }\href {https://doi.org/10.1103/PhysRevA.105.022616}
  {\bibfield  {journal} {\bibinfo  {journal} {Physical Review A}\ }\textbf
  {\bibinfo {volume} {105}},\ \bibinfo {pages} {022616} (\bibinfo {year}
  {2022})},\ \Eprint {https://arxiv.org/abs/2107.09071} {arXiv:2107.09071}
  \BibitemShut {NoStop}%
\bibitem [{\citenamefont {Nguyen}\ \emph {et~al.}(2022)\citenamefont {Nguyen},
  \citenamefont {Tran}, \citenamefont {Zhu}, \citenamefont {Green},
  \citenamefont {Alderete}, \citenamefont {Davoudi},\ and\ \citenamefont
  {Linke}}]{Nguyen2022DigitalQuantumSimulation}%
  \BibitemOpen
  \bibfield  {author} {\bibinfo {author} {\bibfnamefont {N.~H.}\ \bibnamefont
  {Nguyen}}, \bibinfo {author} {\bibfnamefont {M.~C.}\ \bibnamefont {Tran}},
  \bibinfo {author} {\bibfnamefont {Y.}~\bibnamefont {Zhu}}, \bibinfo {author}
  {\bibfnamefont {A.~M.}\ \bibnamefont {Green}}, \bibinfo {author}
  {\bibfnamefont {C.~H.}\ \bibnamefont {Alderete}}, \bibinfo {author}
  {\bibfnamefont {Z.}~\bibnamefont {Davoudi}},\ and\ \bibinfo {author}
  {\bibfnamefont {N.~M.}\ \bibnamefont {Linke}},\ }\bibfield  {title} {\bibinfo
  {title} {Digital {Quantum} {Simulation} of the {Schwinger} {Model} and
  {Symmetry} {Protection} with {Trapped} {Ions}},\ }\href
  {https://doi.org/10.1103/PRXQuantum.3.020324} {\bibfield  {journal} {\bibinfo
   {journal} {PRX Quantum}\ }\textbf {\bibinfo {volume} {3}},\ \bibinfo {pages}
  {020324} (\bibinfo {year} {2022})}\BibitemShut {NoStop}%
\bibitem [{\citenamefont {Davoudi}\ \emph {et~al.}(2023)\citenamefont
  {Davoudi}, \citenamefont {Shaw},\ and\ \citenamefont
  {Stryker}}]{Davoudi2023GeneralQuantumAlgorithms}%
  \BibitemOpen
  \bibfield  {author} {\bibinfo {author} {\bibfnamefont {Z.}~\bibnamefont
  {Davoudi}}, \bibinfo {author} {\bibfnamefont {A.~F.}\ \bibnamefont {Shaw}},\
  and\ \bibinfo {author} {\bibfnamefont {J.~R.}\ \bibnamefont {Stryker}},\
  }\href {https://doi.org/10.48550/arXiv.2212.14030} {\emph {\bibinfo {title}
  {General quantum algorithms for {Hamiltonian} simulation with applications to
  a non-{Abelian} lattice gauge theory}}},\ \bibinfo {type} {Tech. Rep.}\
  (\bibinfo {year} {2023})\ \Eprint {https://arxiv.org/abs/2212.14030}
  {arxiv:2212.14030} \BibitemShut {NoStop}%
\bibitem [{\citenamefont {B{\u{a}}z{\u{a}}van}\ \emph
  {et~al.}(2023)\citenamefont {B{\u{a}}z{\u{a}}van}, \citenamefont {Saner},
  \citenamefont {Tirrito}, \citenamefont {Araneda}, \citenamefont {Srinivas},\
  and\ \citenamefont {Bermudez}}]{Bazavan2023SyntheticZ2Gauge}%
  \BibitemOpen
  \bibfield  {author} {\bibinfo {author} {\bibfnamefont {O.}~\bibnamefont
  {B{\u{a}}z{\u{a}}van}}, \bibinfo {author} {\bibfnamefont {S.}~\bibnamefont
  {Saner}}, \bibinfo {author} {\bibfnamefont {E.}~\bibnamefont {Tirrito}},
  \bibinfo {author} {\bibfnamefont {G.}~\bibnamefont {Araneda}}, \bibinfo
  {author} {\bibfnamefont {R.}~\bibnamefont {Srinivas}},\ and\ \bibinfo
  {author} {\bibfnamefont {A.}~\bibnamefont {Bermudez}},\ }\href
  {https://doi.org/10.48550/arXiv.2305.08700} {\emph {\bibinfo {title}
  {Synthetic $\mathbb{Z}_2$ gauge theories based on parametric excitations of
  trapped ions}}},\ \bibinfo {type} {Tech. Rep.}\ (\bibinfo {year} {2023})\
  \Eprint {https://arxiv.org/abs/2305.08700} {arxiv:2305.08700} \BibitemShut
  {NoStop}%
\bibitem [{\citenamefont {Turco}\ \emph {et~al.}(2023)\citenamefont {Turco},
  \citenamefont {Quinta}, \citenamefont {Seixas},\ and\ \citenamefont
  {Omar}}]{Turco2023TowardsQuantumSimulation}%
  \BibitemOpen
  \bibfield  {author} {\bibinfo {author} {\bibfnamefont {M.}~\bibnamefont
  {Turco}}, \bibinfo {author} {\bibfnamefont {G.~M.}\ \bibnamefont {Quinta}},
  \bibinfo {author} {\bibfnamefont {J.}~\bibnamefont {Seixas}},\ and\ \bibinfo
  {author} {\bibfnamefont {Y.}~\bibnamefont {Omar}},\ }\href@noop {} {\emph
  {\bibinfo {title} {Towards {Quantum} {Simulation} of {Bound} {States}
  {Scattering}}}},\ \bibinfo {type} {Tech. Rep.}\ (\bibinfo {year} {2023})\
  \Eprint {https://arxiv.org/abs/2305.07692} {arxiv:2305.07692} \BibitemShut
  {NoStop}%
\bibitem [{\citenamefont {Martinez}\ \emph {et~al.}(2016)\citenamefont
  {Martinez}, \citenamefont {Muschik}, \citenamefont {Schindler}, \citenamefont
  {Nigg}, \citenamefont {Erhard}, \citenamefont {Heyl}, \citenamefont {Hauke},
  \citenamefont {Dalmonte}, \citenamefont {Monz}, \citenamefont {Zoller},\ and\
  \citenamefont {Blatt}}]{Martinez2016RealTimeDynamics}%
  \BibitemOpen
  \bibfield  {author} {\bibinfo {author} {\bibfnamefont {E.~A.}\ \bibnamefont
  {Martinez}}, \bibinfo {author} {\bibfnamefont {C.~A.}\ \bibnamefont
  {Muschik}}, \bibinfo {author} {\bibfnamefont {P.}~\bibnamefont {Schindler}},
  \bibinfo {author} {\bibfnamefont {D.}~\bibnamefont {Nigg}}, \bibinfo {author}
  {\bibfnamefont {A.}~\bibnamefont {Erhard}}, \bibinfo {author} {\bibfnamefont
  {M.}~\bibnamefont {Heyl}}, \bibinfo {author} {\bibfnamefont {P.}~\bibnamefont
  {Hauke}}, \bibinfo {author} {\bibfnamefont {M.}~\bibnamefont {Dalmonte}},
  \bibinfo {author} {\bibfnamefont {T.}~\bibnamefont {Monz}}, \bibinfo {author}
  {\bibfnamefont {P.}~\bibnamefont {Zoller}},\ and\ \bibinfo {author}
  {\bibfnamefont {R.}~\bibnamefont {Blatt}},\ }\bibfield  {title} {\bibinfo
  {title} {Real-time dynamics of lattice gauge theories with a few-qubit
  quantum computer},\ }\href {https://doi.org/10.1038/nature18318} {\bibfield
  {journal} {\bibinfo  {journal} {Nature}\ }\textbf {\bibinfo {volume} {534}},\
  \bibinfo {pages} {516} (\bibinfo {year} {2016})}\BibitemShut {NoStop}%
\bibitem [{\citenamefont {Bruzewicz}\ \emph {et~al.}(2019)\citenamefont
  {Bruzewicz}, \citenamefont {Chiaverini}, \citenamefont {McConnell},\ and\
  \citenamefont {Sage}}]{Bruzewicz2019TrappedIonQuantum}%
  \BibitemOpen
  \bibfield  {author} {\bibinfo {author} {\bibfnamefont {C.~D.}\ \bibnamefont
  {Bruzewicz}}, \bibinfo {author} {\bibfnamefont {J.}~\bibnamefont
  {Chiaverini}}, \bibinfo {author} {\bibfnamefont {R.}~\bibnamefont
  {McConnell}},\ and\ \bibinfo {author} {\bibfnamefont {J.~M.}\ \bibnamefont
  {Sage}},\ }\bibfield  {title} {\bibinfo {title} {Trapped-ion quantum
  computing: Progress and challenges},\ }\href
  {https://doi.org/10.1063/1.5088164} {\bibfield  {journal} {\bibinfo
  {journal} {Applied Physics Reviews}\ }\textbf {\bibinfo {volume} {6}},\
  \bibinfo {pages} {021314} (\bibinfo {year} {2019})}\BibitemShut {NoStop}%
\bibitem [{\citenamefont {Wintersperger}\ \emph {et~al.}(2020)\citenamefont
  {Wintersperger}, \citenamefont {Braun}, \citenamefont {{\"U}nal},
  \citenamefont {Eckardt}, \citenamefont {Liberto}, \citenamefont {Goldman},
  \citenamefont {Bloch},\ and\ \citenamefont
  {Aidelsburger}}]{Wintersperger2020RealizationAnomalousFloquet}%
  \BibitemOpen
  \bibfield  {author} {\bibinfo {author} {\bibfnamefont {K.}~\bibnamefont
  {Wintersperger}}, \bibinfo {author} {\bibfnamefont {C.}~\bibnamefont
  {Braun}}, \bibinfo {author} {\bibfnamefont {F.~N.}\ \bibnamefont {{\"U}nal}},
  \bibinfo {author} {\bibfnamefont {A.}~\bibnamefont {Eckardt}}, \bibinfo
  {author} {\bibfnamefont {M.~D.}\ \bibnamefont {Liberto}}, \bibinfo {author}
  {\bibfnamefont {N.}~\bibnamefont {Goldman}}, \bibinfo {author} {\bibfnamefont
  {I.}~\bibnamefont {Bloch}},\ and\ \bibinfo {author} {\bibfnamefont
  {M.}~\bibnamefont {Aidelsburger}},\ }\bibfield  {title} {\bibinfo {title}
  {Realization of an anomalous {Floquet} topological system with ultracold
  atoms},\ }\href {https://doi.org/10.1038/s41567-020-0949-y} {\bibfield
  {journal} {\bibinfo  {journal} {Nature Physics}\ }\textbf {\bibinfo {volume}
  {16}},\ \bibinfo {pages} {1058} (\bibinfo {year} {2020})}\BibitemShut
  {NoStop}%
\bibitem [{\citenamefont {Bluvstein}\ \emph {et~al.}(2021)\citenamefont
  {Bluvstein} \emph {et~al.}}]{Bluvstein2021ControllingQuantumMany}%
  \BibitemOpen
  \bibfield  {author} {\bibinfo {author} {\bibfnamefont {D.}~\bibnamefont
  {Bluvstein}} \emph {et~al.},\ }\bibfield  {title} {\bibinfo {title}
  {Controlling quantum many-body dynamics in driven {Rydberg} atom arrays},\
  }\href {https://doi.org/10.1126/science.abg2530} {\bibfield  {journal}
  {\bibinfo  {journal} {Science}\ }\textbf {\bibinfo {volume} {371}},\ \bibinfo
  {pages} {1355} (\bibinfo {year} {2021})}\BibitemShut {NoStop}%
\bibitem [{\citenamefont {Mueller}\ \emph {et~al.}(2022)\citenamefont
  {Mueller}, \citenamefont {Carolan}, \citenamefont {Connelly}, \citenamefont
  {Davoudi}, \citenamefont {Dumitrescu},\ and\ \citenamefont
  {Yeter-Aydeniz}}]{Mueller2022QuantumComputationDynamical}%
  \BibitemOpen
  \bibfield  {author} {\bibinfo {author} {\bibfnamefont {N.}~\bibnamefont
  {Mueller}}, \bibinfo {author} {\bibfnamefont {J.~A.}\ \bibnamefont
  {Carolan}}, \bibinfo {author} {\bibfnamefont {A.}~\bibnamefont {Connelly}},
  \bibinfo {author} {\bibfnamefont {Z.}~\bibnamefont {Davoudi}}, \bibinfo
  {author} {\bibfnamefont {E.~F.}\ \bibnamefont {Dumitrescu}},\ and\ \bibinfo
  {author} {\bibfnamefont {K.}~\bibnamefont {Yeter-Aydeniz}},\ }\href
  {https://doi.org/10.48550/arXiv.2210.03089} {\emph {\bibinfo {title} {Quantum
  computation of dynamical quantum phase transitions and entanglement
  tomography in a lattice gauge theory}}},\ \bibinfo {type} {Tech. Rep.}\
  (\bibinfo {year} {2022})\ \Eprint {https://arxiv.org/abs/2210.03089}
  {arxiv:2210.03089} \BibitemShut {NoStop}%
\bibitem [{\citenamefont {Pomarico}\ \emph {et~al.}(2023)\citenamefont
  {Pomarico}, \citenamefont {Cosmai}, \citenamefont {Facchi}, \citenamefont
  {Lupo}, \citenamefont {Pascazio},\ and\ \citenamefont
  {Pepe}}]{Pomarico2023DynamicalQuantumPhase}%
  \BibitemOpen
  \bibfield  {author} {\bibinfo {author} {\bibfnamefont {D.}~\bibnamefont
  {Pomarico}}, \bibinfo {author} {\bibfnamefont {L.}~\bibnamefont {Cosmai}},
  \bibinfo {author} {\bibfnamefont {P.}~\bibnamefont {Facchi}}, \bibinfo
  {author} {\bibfnamefont {C.}~\bibnamefont {Lupo}}, \bibinfo {author}
  {\bibfnamefont {S.}~\bibnamefont {Pascazio}},\ and\ \bibinfo {author}
  {\bibfnamefont {F.~V.}\ \bibnamefont {Pepe}},\ }\bibfield  {title} {\bibinfo
  {title} {Dynamical quantum phase transitions of the {Schwinger} model:
  real-time dynamics on {IBM} {Quantum}},\ }\href
  {https://doi.org/10.3390/e25040608} {\bibfield  {journal} {\bibinfo
  {journal} {Entropy}\ }\textbf {\bibinfo {volume} {25}},\ \bibinfo {pages}
  {608} (\bibinfo {year} {2023})},\ \Eprint {https://arxiv.org/abs/2302.01151}
  {arXiv:2302.01151} \BibitemShut {NoStop}%
\bibitem [{\citenamefont {Wiese}(2014)}]{Wiese2014TowardsQuantumSimulating}%
  \BibitemOpen
  \bibfield  {author} {\bibinfo {author} {\bibfnamefont {U.-J.}\ \bibnamefont
  {Wiese}},\ }\bibfield  {title} {\bibinfo {title} {Towards quantum simulating
  {QCD}},\ }\href {https://doi.org/10.1016/j.nuclphysa.2014.09.102} {\bibfield
  {journal} {\bibinfo  {journal} {Nuclear Physics A}\ }\bibinfo {series}
  {{QUARK} {MATTER} 2014},\ \textbf {\bibinfo {volume} {931}},\ \bibinfo
  {pages} {246} (\bibinfo {year} {2014})}\BibitemShut {NoStop}%
\bibitem [{\citenamefont {Funcke}\ \emph {et~al.}(2023)\citenamefont {Funcke},
  \citenamefont {Hartung}, \citenamefont {Jansen},\ and\ \citenamefont
  {K{\"u}hn}}]{Funcke2023ReviewQuantumComputing}%
  \BibitemOpen
  \bibfield  {author} {\bibinfo {author} {\bibfnamefont {L.}~\bibnamefont
  {Funcke}}, \bibinfo {author} {\bibfnamefont {T.}~\bibnamefont {Hartung}},
  \bibinfo {author} {\bibfnamefont {K.}~\bibnamefont {Jansen}},\ and\ \bibinfo
  {author} {\bibfnamefont {S.}~\bibnamefont {K{\"u}hn}},\ }\href
  {https://doi.org/10.48550/arXiv.2302.00467} {\emph {\bibinfo {title} {Review
  on {Quantum} {Computing} for {Lattice} {Field} {Theory}}}},\ \bibinfo {type}
  {Tech. Rep.}\ (\bibinfo {year} {2023})\ \Eprint
  {https://arxiv.org/abs/2302.00467} {arxiv:2302.00467} \BibitemShut {NoStop}%
\bibitem [{\citenamefont {Farrell}\ \emph
  {et~al.}(2023{\natexlab{a}})\citenamefont {Farrell}, \citenamefont
  {Chernyshev}, \citenamefont {Powell}, \citenamefont {Zemlevskiy},
  \citenamefont {Illa},\ and\ \citenamefont
  {Savage}}]{Farrell2023PreparationsQuantumSimulationsa}%
  \BibitemOpen
  \bibfield  {author} {\bibinfo {author} {\bibfnamefont {R.~C.}\ \bibnamefont
  {Farrell}}, \bibinfo {author} {\bibfnamefont {I.~A.}\ \bibnamefont
  {Chernyshev}}, \bibinfo {author} {\bibfnamefont {S.~J.~M.}\ \bibnamefont
  {Powell}}, \bibinfo {author} {\bibfnamefont {N.~A.}\ \bibnamefont
  {Zemlevskiy}}, \bibinfo {author} {\bibfnamefont {M.}~\bibnamefont {Illa}},\
  and\ \bibinfo {author} {\bibfnamefont {M.~J.}\ \bibnamefont {Savage}},\
  }\bibfield  {title} {\bibinfo {title} {Preparations for {Quantum}
  {Simulations} of {Quantum} {Chromodynamics} in 1+1 {Dimensions}: ({I})
  {Axial} {Gauge}},\ }\href {https://doi.org/10.1103/PhysRevD.107.054512}
  {\bibfield  {journal} {\bibinfo  {journal} {Physical Review D}\ }\textbf
  {\bibinfo {volume} {107}},\ \bibinfo {pages} {054512} (\bibinfo {year}
  {2023}{\natexlab{a}})},\ \Eprint {https://arxiv.org/abs/2207.01731}
  {arxiv:2207.01731} \BibitemShut {NoStop}%
\bibitem [{\citenamefont {Farrell}\ \emph
  {et~al.}(2023{\natexlab{b}})\citenamefont {Farrell}, \citenamefont
  {Chernyshev}, \citenamefont {Powell}, \citenamefont {Zemlevskiy},
  \citenamefont {Illa},\ and\ \citenamefont
  {Savage}}]{Farrell2023PreparationsQuantumSimulations}%
  \BibitemOpen
  \bibfield  {author} {\bibinfo {author} {\bibfnamefont {R.~C.}\ \bibnamefont
  {Farrell}}, \bibinfo {author} {\bibfnamefont {I.~A.}\ \bibnamefont
  {Chernyshev}}, \bibinfo {author} {\bibfnamefont {S.~J.~M.}\ \bibnamefont
  {Powell}}, \bibinfo {author} {\bibfnamefont {N.~A.}\ \bibnamefont
  {Zemlevskiy}}, \bibinfo {author} {\bibfnamefont {M.}~\bibnamefont {Illa}},\
  and\ \bibinfo {author} {\bibfnamefont {M.~J.}\ \bibnamefont {Savage}},\
  }\bibfield  {title} {\bibinfo {title} {Preparations for {Quantum}
  {Simulations} of {Quantum} {Chromodynamics} in 1+1 {Dimensions}: ({II})
  {Single}-{Baryon} $\beta$-{Decay} in {Real} {Time}},\ }\href
  {https://doi.org/10.1103/PhysRevD.107.054513} {\bibfield  {journal} {\bibinfo
   {journal} {Physical Review D}\ }\textbf {\bibinfo {volume} {107}},\ \bibinfo
  {pages} {054513} (\bibinfo {year} {2023}{\natexlab{b}})},\ \Eprint
  {https://arxiv.org/abs/2209.10781} {arXiv:2209.10781} \BibitemShut {NoStop}%
\bibitem [{\citenamefont {Chawdhry}\ and\ \citenamefont
  {Pellen}(2023)}]{Chawdhry2023QuantumSimulationColour}%
  \BibitemOpen
  \bibfield  {author} {\bibinfo {author} {\bibfnamefont {H.~A.}\ \bibnamefont
  {Chawdhry}}\ and\ \bibinfo {author} {\bibfnamefont {M.}~\bibnamefont
  {Pellen}},\ }\href {https://doi.org/10.48550/arXiv.2303.04818} {\emph
  {\bibinfo {title} {Quantum simulation of colour in perturbative quantum
  chromodynamics}}},\ \bibinfo {type} {Tech. Rep.}\ (\bibinfo {year} {2023})\
  \Eprint {https://arxiv.org/abs/2303.04818} {arxiv:2303.04818} \BibitemShut
  {NoStop}%
\bibitem [{\citenamefont {Bauer}\ \emph
  {et~al.}(2023{\natexlab{b}})\citenamefont {Bauer}, \citenamefont {Davoudi},
  \citenamefont {Klco},\ and\ \citenamefont
  {Savage}}]{Bauer2023QuantumSimulationFundamental}%
  \BibitemOpen
  \bibfield  {author} {\bibinfo {author} {\bibfnamefont {C.~W.}\ \bibnamefont
  {Bauer}}, \bibinfo {author} {\bibfnamefont {Z.}~\bibnamefont {Davoudi}},
  \bibinfo {author} {\bibfnamefont {N.}~\bibnamefont {Klco}},\ and\ \bibinfo
  {author} {\bibfnamefont {M.~J.}\ \bibnamefont {Savage}},\ }\bibfield  {title}
  {\bibinfo {title} {Quantum simulation of fundamental particles and forces},\
  }\href {https://doi.org/10.1038/s42254-023-00599-8} {\bibfield  {journal}
  {\bibinfo  {journal} {Nature Reviews Physics}\ }\textbf {\bibinfo {volume}
  {5}},\ \bibinfo {pages} {420} (\bibinfo {year}
  {2023}{\natexlab{b}})}\BibitemShut {NoStop}%
\bibitem [{\citenamefont
  {Ciavarella}(2023)}]{Ciavarella2023QuantumSimulationLattice}%
  \BibitemOpen
  \bibfield  {author} {\bibinfo {author} {\bibfnamefont {A.~N.}\ \bibnamefont
  {Ciavarella}},\ }\href {https://doi.org/10.48550/arXiv.2307.05593} {\emph
  {\bibinfo {title} {Quantum {Simulation} of {Lattice} {QCD} with {Improved}
  {Hamiltonians}}}},\ \bibinfo {type} {Tech. Rep.}\ (\bibinfo {year} {2023})\
  \Eprint {https://arxiv.org/abs/2307.05593} {arxiv:2307.05593} \BibitemShut
  {NoStop}%
\bibitem [{\citenamefont {Kadam}\ \emph {et~al.}(2023)\citenamefont {Kadam},
  \citenamefont {Raychowdhury},\ and\ \citenamefont
  {Stryker}}]{Kadam2023LoopStringHadron}%
  \BibitemOpen
  \bibfield  {author} {\bibinfo {author} {\bibfnamefont {S.~V.}\ \bibnamefont
  {Kadam}}, \bibinfo {author} {\bibfnamefont {I.}~\bibnamefont
  {Raychowdhury}},\ and\ \bibinfo {author} {\bibfnamefont {J.~R.}\ \bibnamefont
  {Stryker}},\ }\bibfield  {title} {\bibinfo {title} {Loop-string-hadron
  formulation of an {SU}(3) gauge theory with dynamical quarks},\ }\href
  {https://doi.org/10.1103/PhysRevD.107.094513} {\bibfield  {journal} {\bibinfo
   {journal} {Physical Review D}\ }\textbf {\bibinfo {volume} {107}},\ \bibinfo
  {pages} {094513} (\bibinfo {year} {2023})},\ \Eprint
  {https://arxiv.org/abs/2212.04490} {arxiv:2212.04490} \BibitemShut {NoStop}%
\bibitem [{\citenamefont {Wang}\ \emph {et~al.}(2023)\citenamefont {Wang},
  \citenamefont {Wang}, \citenamefont {Vovrosh}, \citenamefont {Knolle},
  \citenamefont {Mintert},\ and\ \citenamefont
  {Mukherjee}}]{Wang2023QuantumSimulationHadronic}%
  \BibitemOpen
  \bibfield  {author} {\bibinfo {author} {\bibfnamefont {Z.}~\bibnamefont
  {Wang}}, \bibinfo {author} {\bibfnamefont {F.}~\bibnamefont {Wang}}, \bibinfo
  {author} {\bibfnamefont {J.}~\bibnamefont {Vovrosh}}, \bibinfo {author}
  {\bibfnamefont {J.}~\bibnamefont {Knolle}}, \bibinfo {author} {\bibfnamefont
  {F.}~\bibnamefont {Mintert}},\ and\ \bibinfo {author} {\bibfnamefont
  {R.}~\bibnamefont {Mukherjee}},\ }\href
  {https://doi.org/10.48550/arXiv.2304.12623} {\emph {\bibinfo {title} {Quantum
  simulation of hadronic states with {Rydberg}-dressed atoms}}},\ \bibinfo
  {type} {Tech. Rep.}\ (\bibinfo {year} {2023})\ \Eprint
  {https://arxiv.org/abs/2304.12623} {arxiv:2304.12623} \BibitemShut {NoStop}%
\bibitem [{\citenamefont {Qian}\ \emph {et~al.}(2022)\citenamefont {Qian},
  \citenamefont {Basili}, \citenamefont {Pal}, \citenamefont {Luecke},\ and\
  \citenamefont {Vary}}]{Qian2022SolvingHadronStructures}%
  \BibitemOpen
  \bibfield  {author} {\bibinfo {author} {\bibfnamefont {W.}~\bibnamefont
  {Qian}}, \bibinfo {author} {\bibfnamefont {R.}~\bibnamefont {Basili}},
  \bibinfo {author} {\bibfnamefont {S.}~\bibnamefont {Pal}}, \bibinfo {author}
  {\bibfnamefont {G.}~\bibnamefont {Luecke}},\ and\ \bibinfo {author}
  {\bibfnamefont {J.~P.}\ \bibnamefont {Vary}},\ }\bibfield  {title} {\bibinfo
  {title} {Solving hadron structures using the basis light-front quantization
  approach on quantum computers},\ }\href
  {https://doi.org/10.1103/PhysRevResearch.4.043193} {\bibfield  {journal}
  {\bibinfo  {journal} {Physical Review Research}\ }\textbf {\bibinfo {volume}
  {4}},\ \bibinfo {pages} {043193} (\bibinfo {year} {2022})}\BibitemShut
  {NoStop}%
\bibitem [{\citenamefont {Yao}(2022)}]{Yao2022QuantumSimulationLight}%
  \BibitemOpen
  \bibfield  {author} {\bibinfo {author} {\bibfnamefont {X.}~\bibnamefont
  {Yao}},\ }\href {https://doi.org/10.48550/arXiv.2205.07902} {\emph {\bibinfo
  {title} {Quantum {Simulation} of {Light}-{Front} {QCD} for {Jet} {Quenching}
  in {Nuclear} {Environments}}}},\ \bibinfo {type} {Tech. Rep.}\ (\bibinfo
  {year} {2022})\ \Eprint {https://arxiv.org/abs/2205.07902} {arxiv:2205.07902}
  \BibitemShut {NoStop}%
\bibitem [{\citenamefont {Barata}\ \emph {et~al.}(2022)\citenamefont {Barata},
  \citenamefont {Du}, \citenamefont {Li}, \citenamefont {Qian},\ and\
  \citenamefont {Salgado}}]{Barata2022MediumInducedJet}%
  \BibitemOpen
  \bibfield  {author} {\bibinfo {author} {\bibfnamefont {J.}~\bibnamefont
  {Barata}}, \bibinfo {author} {\bibfnamefont {X.}~\bibnamefont {Du}}, \bibinfo
  {author} {\bibfnamefont {M.}~\bibnamefont {Li}}, \bibinfo {author}
  {\bibfnamefont {W.}~\bibnamefont {Qian}},\ and\ \bibinfo {author}
  {\bibfnamefont {C.~A.}\ \bibnamefont {Salgado}},\ }\bibfield  {title}
  {\bibinfo {title} {Medium induced jet broadening in a quantum computer},\
  }\href {https://doi.org/10.1103/PhysRevD.106.074013} {\bibfield  {journal}
  {\bibinfo  {journal} {Physical Review D}\ }\textbf {\bibinfo {volume}
  {106}},\ \bibinfo {pages} {074013} (\bibinfo {year} {2022})},\ \Eprint
  {https://arxiv.org/abs/2208.06750} {arxiv:2208.06750} \BibitemShut {NoStop}%
\bibitem [{\citenamefont {Barata}\ and\ \citenamefont
  {Salgado}(2021)}]{Barata2021QuantumStrategyCompute}%
  \BibitemOpen
  \bibfield  {author} {\bibinfo {author} {\bibfnamefont {J.}~\bibnamefont
  {Barata}}\ and\ \bibinfo {author} {\bibfnamefont {C.~A.}\ \bibnamefont
  {Salgado}},\ }\href {https://doi.org/10.48550/arXiv.2104.04661} {\emph
  {\bibinfo {title} {A quantum strategy to compute the jet quenching parameter
  $\hat{q}$}}},\ \bibinfo {type} {Tech. Rep.}\ (\bibinfo {year} {2021})\
  \Eprint {https://arxiv.org/abs/2104.04661} {arxiv:2104.04661} \BibitemShut
  {NoStop}%
\bibitem [{\citenamefont {Berges}\ \emph {et~al.}(2021)\citenamefont {Berges},
  \citenamefont {Heller}, \citenamefont {Mazeliauskas},\ and\ \citenamefont
  {Venugopalan}}]{Berges2021QcdThermalizationAb}%
  \BibitemOpen
  \bibfield  {author} {\bibinfo {author} {\bibfnamefont {J.}~\bibnamefont
  {Berges}}, \bibinfo {author} {\bibfnamefont {M.~P.}\ \bibnamefont {Heller}},
  \bibinfo {author} {\bibfnamefont {A.}~\bibnamefont {Mazeliauskas}},\ and\
  \bibinfo {author} {\bibfnamefont {R.}~\bibnamefont {Venugopalan}},\
  }\bibfield  {title} {\bibinfo {title} {{QCD} thermalization: \textit{{Ab}
  initio} approaches and interdisciplinary connections},\ }\href
  {https://doi.org/10.1103/RevModPhys.93.035003} {\bibfield  {journal}
  {\bibinfo  {journal} {Reviews of Modern Physics}\ }\textbf {\bibinfo {volume}
  {93}},\ \bibinfo {pages} {035003} (\bibinfo {year} {2021})}\BibitemShut
  {NoStop}%
\bibitem [{\citenamefont {{NuQS Collaboration}}\ \emph
  {et~al.}(2020)\citenamefont {{NuQS Collaboration}}, \citenamefont {Lamm},
  \citenamefont {Lawrence},\ and\ \citenamefont
  {Yamauchi}}]{NuQSCollaboration2020PartonPhysicsQuantum}%
  \BibitemOpen
  \bibfield  {author} {\bibinfo {author} {\bibnamefont {{NuQS Collaboration}}},
  \bibinfo {author} {\bibfnamefont {H.}~\bibnamefont {Lamm}}, \bibinfo {author}
  {\bibfnamefont {S.}~\bibnamefont {Lawrence}},\ and\ \bibinfo {author}
  {\bibfnamefont {Y.}~\bibnamefont {Yamauchi}},\ }\bibfield  {title} {\bibinfo
  {title} {Parton physics on a quantum computer},\ }\href
  {https://doi.org/10.1103/PhysRevResearch.2.013272} {\bibfield  {journal}
  {\bibinfo  {journal} {Physical Review Research}\ }\textbf {\bibinfo {volume}
  {2}},\ \bibinfo {pages} {013272} (\bibinfo {year} {2020})}\BibitemShut
  {NoStop}%
\bibitem [{\citenamefont {Farrelly}\ and\ \citenamefont
  {Streich}(2020)}]{Farrelly2020DiscretizingQuantumField}%
  \BibitemOpen
  \bibfield  {author} {\bibinfo {author} {\bibfnamefont {T.}~\bibnamefont
  {Farrelly}}\ and\ \bibinfo {author} {\bibfnamefont {J.}~\bibnamefont
  {Streich}},\ }\href {https://doi.org/10.48550/arXiv.2002.02643} {\emph
  {\bibinfo {title} {Discretizing quantum field theories for quantum
  simulation}}},\ \bibinfo {type} {Tech. Rep.}\ (\bibinfo {year} {2020})\
  \Eprint {https://arxiv.org/abs/2002.02643} {arxiv:2002.02643} \BibitemShut
  {NoStop}%
\bibitem [{\citenamefont {Halimeh}\ \emph {et~al.}(2022)\citenamefont
  {Halimeh}, \citenamefont {Lang},\ and\ \citenamefont
  {Hauke}}]{Halimeh2022GaugeProtectionNon}%
  \BibitemOpen
  \bibfield  {author} {\bibinfo {author} {\bibfnamefont {J.~C.}\ \bibnamefont
  {Halimeh}}, \bibinfo {author} {\bibfnamefont {H.}~\bibnamefont {Lang}},\ and\
  \bibinfo {author} {\bibfnamefont {P.}~\bibnamefont {Hauke}},\ }\bibfield
  {title} {\bibinfo {title} {Gauge protection in non-abelian lattice gauge
  theories},\ }\href {https://doi.org/10.1088/1367-2630/ac5564} {\bibfield
  {journal} {\bibinfo  {journal} {New Journal of Physics}\ }\textbf {\bibinfo
  {volume} {24}},\ \bibinfo {pages} {033015} (\bibinfo {year}
  {2022})}\BibitemShut {NoStop}%
\bibitem [{\citenamefont {Raychowdhury}\ and\ \citenamefont
  {Stryker}(2020)}]{Raychowdhury2020SolvingGausssLaw}%
  \BibitemOpen
  \bibfield  {author} {\bibinfo {author} {\bibfnamefont {I.}~\bibnamefont
  {Raychowdhury}}\ and\ \bibinfo {author} {\bibfnamefont {J.~R.}\ \bibnamefont
  {Stryker}},\ }\bibfield  {title} {\bibinfo {title} {Solving {Gauss}'s law on
  digital quantum computers with loop-string-hadron digitization},\ }\href
  {https://doi.org/10.1103/PhysRevResearch.2.033039} {\bibfield  {journal}
  {\bibinfo  {journal} {Physical Review Research}\ }\textbf {\bibinfo {volume}
  {2}},\ \bibinfo {pages} {033039} (\bibinfo {year} {2020})}\BibitemShut
  {NoStop}%
\bibitem [{\citenamefont {Mathew}\ and\ \citenamefont
  {Raychowdhury}(2022)}]{Mathew2022ProtectingLocalGlobal}%
  \BibitemOpen
  \bibfield  {author} {\bibinfo {author} {\bibfnamefont {E.}~\bibnamefont
  {Mathew}}\ and\ \bibinfo {author} {\bibfnamefont {I.}~\bibnamefont
  {Raychowdhury}},\ }\bibfield  {title} {\bibinfo {title} {Protecting local and
  global symmetries in simulating 1+1-{D} non-abelian gauge theories},\ }\href
  {https://doi.org/10.1103/PhysRevD.106.054510} {\bibfield  {journal} {\bibinfo
   {journal} {Physical Review D}\ }\textbf {\bibinfo {volume} {106}},\ \bibinfo
  {pages} {054510} (\bibinfo {year} {2022})},\ \Eprint
  {https://arxiv.org/abs/2206.07444} {arXiv:2206.07444} \BibitemShut {NoStop}%
\bibitem [{\citenamefont {Jakobs}\ \emph {et~al.}(2023)\citenamefont {Jakobs},
  \citenamefont {Garofalo}, \citenamefont {Hartung}, \citenamefont {Jansen},
  \citenamefont {Ostmeyer}, \citenamefont {Rolfes}, \citenamefont {Romiti},\
  and\ \citenamefont {Urbach}}]{Jakobs2023CanonicalMomentaDigitized}%
  \BibitemOpen
  \bibfield  {author} {\bibinfo {author} {\bibfnamefont {T.}~\bibnamefont
  {Jakobs}}, \bibinfo {author} {\bibfnamefont {M.}~\bibnamefont {Garofalo}},
  \bibinfo {author} {\bibfnamefont {T.}~\bibnamefont {Hartung}}, \bibinfo
  {author} {\bibfnamefont {K.}~\bibnamefont {Jansen}}, \bibinfo {author}
  {\bibfnamefont {J.}~\bibnamefont {Ostmeyer}}, \bibinfo {author}
  {\bibfnamefont {D.}~\bibnamefont {Rolfes}}, \bibinfo {author} {\bibfnamefont
  {S.}~\bibnamefont {Romiti}},\ and\ \bibinfo {author} {\bibfnamefont
  {C.}~\bibnamefont {Urbach}},\ }\bibfield  {title} {\bibinfo {title}
  {Canonical {Momenta} in {Digitized} {SU}(2) {Lattice} {Gauge} {Theory}:
  {Definition} and {Free} {Theory}},\ }\href
  {https://doi.org/10.1140/epjc/s10052-023-11829-9} {\bibfield  {journal}
  {\bibinfo  {journal} {The European Physical Journal C}\ }\textbf {\bibinfo
  {volume} {83}},\ \bibinfo {pages} {669} (\bibinfo {year} {2023})},\ \Eprint
  {https://arxiv.org/abs/2304.02322} {arXiv:2304.02322} \BibitemShut {NoStop}%
\bibitem [{\citenamefont {Bauer}\ \emph
  {et~al.}(2023{\natexlab{c}})\citenamefont {Bauer}, \citenamefont {D'Andrea},
  \citenamefont {Freytsis},\ and\ \citenamefont
  {Grabowska}}]{Bauer2023NewBasisHamiltonian}%
  \BibitemOpen
  \bibfield  {author} {\bibinfo {author} {\bibfnamefont {C.~W.}\ \bibnamefont
  {Bauer}}, \bibinfo {author} {\bibfnamefont {I.}~\bibnamefont {D'Andrea}},
  \bibinfo {author} {\bibfnamefont {M.}~\bibnamefont {Freytsis}},\ and\
  \bibinfo {author} {\bibfnamefont {D.~M.}\ \bibnamefont {Grabowska}},\ }\href
  {https://doi.org/10.48550/arXiv.2307.11829} {\emph {\bibinfo {title} {A new
  basis for {Hamiltonian} {SU}(2) simulations}}},\ \bibinfo {type} {Tech.
  Rep.}\ (\bibinfo {year} {2023})\ \Eprint {https://arxiv.org/abs/2307.11829}
  {arxiv:2307.11829} \BibitemShut {NoStop}%
\bibitem [{\citenamefont {Alexandru}\ \emph {et~al.}(2023)\citenamefont
  {Alexandru}, \citenamefont {Bedaque}, \citenamefont {Carosso}, \citenamefont
  {Cervia}, \citenamefont {Murairi},\ and\ \citenamefont
  {Sheng}}]{Alexandru2023FuzzyGaugeTheory}%
  \BibitemOpen
  \bibfield  {author} {\bibinfo {author} {\bibfnamefont {A.}~\bibnamefont
  {Alexandru}}, \bibinfo {author} {\bibfnamefont {P.~F.}\ \bibnamefont
  {Bedaque}}, \bibinfo {author} {\bibfnamefont {A.}~\bibnamefont {Carosso}},
  \bibinfo {author} {\bibfnamefont {M.~J.}\ \bibnamefont {Cervia}}, \bibinfo
  {author} {\bibfnamefont {E.~M.}\ \bibnamefont {Murairi}},\ and\ \bibinfo
  {author} {\bibfnamefont {A.}~\bibnamefont {Sheng}},\ }\href
  {https://doi.org/10.48550/arXiv.2308.05253} {\emph {\bibinfo {title} {Fuzzy
  {Gauge} {Theory} for {Quantum} {Computers}}}},\ \bibinfo {type} {Tech. Rep.}\
  (\bibinfo {year} {2023})\ \Eprint {https://arxiv.org/abs/2308.05253}
  {arxiv:2308.05253} \BibitemShut {NoStop}%
\bibitem [{\citenamefont {Horn}(1981)}]{Horn1981FiniteMatrixModels}%
  \BibitemOpen
  \bibfield  {author} {\bibinfo {author} {\bibfnamefont {D.}~\bibnamefont
  {Horn}},\ }\bibfield  {title} {\bibinfo {title} {Finite matrix models with
  continuous local gauge invariance},\ }\href
  {https://doi.org/10.1016/0370-2693(81)90763-2} {\bibfield  {journal}
  {\bibinfo  {journal} {Physics Letters B}\ }\textbf {\bibinfo {volume}
  {100}},\ \bibinfo {pages} {149} (\bibinfo {year} {1981})}\BibitemShut
  {NoStop}%
\bibitem [{\citenamefont {Chandrasekharan}\ and\ \citenamefont
  {Wiese}(1997)}]{Chandrasekharan1997QuantumLinkModels}%
  \BibitemOpen
  \bibfield  {author} {\bibinfo {author} {\bibfnamefont {S.}~\bibnamefont
  {Chandrasekharan}}\ and\ \bibinfo {author} {\bibfnamefont {U.~J.}\
  \bibnamefont {Wiese}},\ }\bibfield  {title} {\bibinfo {title} {Quantum link
  models: {A} discrete approach to gauge theories},\ }\href
  {https://doi.org/10.1016/S0550-3213(97)80041-7} {\bibfield  {journal}
  {\bibinfo  {journal} {Nuclear Physics B}\ }\textbf {\bibinfo {volume}
  {492}},\ \bibinfo {pages} {455} (\bibinfo {year} {1997})}\BibitemShut
  {NoStop}%
\bibitem [{\citenamefont {Wiese}(2021)}]{Wiese2021QuantumLinkModels}%
  \BibitemOpen
  \bibfield  {author} {\bibinfo {author} {\bibfnamefont {U.-J.}\ \bibnamefont
  {Wiese}},\ }\bibfield  {title} {\bibinfo {title} {From quantum link models to
  {D}-theory: a resource efficient framework for the quantum simulation and
  computation of gauge theories},\ }\bibfield  {journal} {\bibinfo  {journal}
  {Philosophical Transactions of the Royal Society A: Mathematical, Physical
  and Engineering Sciences}\ }\textbf {\bibinfo {volume} {380}},\ \href
  {https://doi.org/10.1098/rsta.2021.0068} {10.1098/rsta.2021.0068} (\bibinfo
  {year} {2021})\BibitemShut {NoStop}%
\bibitem [{\citenamefont {Horn}\ \emph {et~al.}(1979)\citenamefont {Horn},
  \citenamefont {Weinstein},\ and\ \citenamefont
  {Yankielowicz}}]{Horn1979HamiltonianApproachZn}%
  \BibitemOpen
  \bibfield  {author} {\bibinfo {author} {\bibfnamefont {D.}~\bibnamefont
  {Horn}}, \bibinfo {author} {\bibfnamefont {M.}~\bibnamefont {Weinstein}},\
  and\ \bibinfo {author} {\bibfnamefont {S.}~\bibnamefont {Yankielowicz}},\
  }\bibfield  {title} {\bibinfo {title} {Hamiltonian approach to {$Z(N)$}
  lattice gauge theories},\ }\href {https://doi.org/10.1103/PhysRevD.19.3715}
  {\bibfield  {journal} {\bibinfo  {journal} {Physical Review D: Particles and
  Fields}\ }\textbf {\bibinfo {volume} {19}},\ \bibinfo {pages} {3715}
  (\bibinfo {year} {1979})}\BibitemShut {NoStop}%
\bibitem [{\citenamefont {Notarnicola}\ \emph {et~al.}(2015)\citenamefont
  {Notarnicola}, \citenamefont {Ercolessi}, \citenamefont {Facchi},
  \citenamefont {Marmo}, \citenamefont {Pascazio},\ and\ \citenamefont
  {Pepe}}]{Notarnicola2015DiscreteAbelianGauge}%
  \BibitemOpen
  \bibfield  {author} {\bibinfo {author} {\bibfnamefont {S.}~\bibnamefont
  {Notarnicola}}, \bibinfo {author} {\bibfnamefont {E.}~\bibnamefont
  {Ercolessi}}, \bibinfo {author} {\bibfnamefont {P.}~\bibnamefont {Facchi}},
  \bibinfo {author} {\bibfnamefont {G.}~\bibnamefont {Marmo}}, \bibinfo
  {author} {\bibfnamefont {S.}~\bibnamefont {Pascazio}},\ and\ \bibinfo
  {author} {\bibfnamefont {F.~V.}\ \bibnamefont {Pepe}},\ }\bibfield  {title}
  {\bibinfo {title} {Discrete {Abelian} gauge theories for quantum simulations
  of {QED}},\ }\href {https://doi.org/10.1088/1751-8113/48/30/30FT01}
  {\bibfield  {journal} {\bibinfo  {journal} {Journal of Physics. A.
  Mathematical and Theoretical}\ }\textbf {\bibinfo {volume} {48}},\ \bibinfo
  {pages} {30FT01} (\bibinfo {year} {2015})}\BibitemShut {NoStop}%
\bibitem [{\citenamefont {Alexandru}\ \emph {et~al.}(2022)\citenamefont
  {Alexandru}, \citenamefont {Bedaque}, \citenamefont {Brett},\ and\
  \citenamefont {Lamm}}]{Alexandru2022SpectrumDigitizedQcd}%
  \BibitemOpen
  \bibfield  {author} {\bibinfo {author} {\bibfnamefont {A.}~\bibnamefont
  {Alexandru}}, \bibinfo {author} {\bibfnamefont {P.~F.}\ \bibnamefont
  {Bedaque}}, \bibinfo {author} {\bibfnamefont {R.}~\bibnamefont {Brett}},\
  and\ \bibinfo {author} {\bibfnamefont {H.}~\bibnamefont {Lamm}},\ }\bibfield
  {title} {\bibinfo {title} {Spectrum of digitized {QCD}: {Glueballs} in a
  ${S}(1080)$ gauge theory},\ }\href
  {https://doi.org/10.1103/PhysRevD.105.114508} {\bibfield  {journal} {\bibinfo
   {journal} {Physical Review D}\ }\textbf {\bibinfo {volume} {105}},\ \bibinfo
  {pages} {114508} (\bibinfo {year} {2022})}\BibitemShut {NoStop}%
\bibitem [{\citenamefont {Bimonte}\ \emph {et~al.}(1996)\citenamefont
  {Bimonte}, \citenamefont {Stern},\ and\ \citenamefont
  {Vitale}}]{Bimonte1996Suq2LatticeGauge}%
  \BibitemOpen
  \bibfield  {author} {\bibinfo {author} {\bibfnamefont {G.}~\bibnamefont
  {Bimonte}}, \bibinfo {author} {\bibfnamefont {A.}~\bibnamefont {Stern}},\
  and\ \bibinfo {author} {\bibfnamefont {P.}~\bibnamefont {Vitale}},\
  }\bibfield  {title} {\bibinfo {title} {$\mathrm{SU}_{q}(2)$ lattice gauge
  theory},\ }\href {https://doi.org/10.1103/PhysRevD.54.1054} {\bibfield
  {journal} {\bibinfo  {journal} {Physical Review D}\ }\textbf {\bibinfo
  {volume} {54}},\ \bibinfo {pages} {1054} (\bibinfo {year}
  {1996})}\BibitemShut {NoStop}%
\bibitem [{\citenamefont {Zache}\ \emph {et~al.}(2023)\citenamefont {Zache},
  \citenamefont {Gonz{\'a}lez-Cuadra},\ and\ \citenamefont
  {Zoller}}]{Zache2023QuantumClassicalSpin}%
  \BibitemOpen
  \bibfield  {author} {\bibinfo {author} {\bibfnamefont {T.~V.}\ \bibnamefont
  {Zache}}, \bibinfo {author} {\bibfnamefont {D.}~\bibnamefont
  {Gonz{\'a}lez-Cuadra}},\ and\ \bibinfo {author} {\bibfnamefont
  {P.}~\bibnamefont {Zoller}},\ }\href
  {https://doi.org/10.48550/arXiv.2304.02527} {\emph {\bibinfo {title} {Quantum
  and classical spin network algorithms for $q$-deformed {Kogut}-{Susskind}
  gauge theories}}},\ \bibinfo {type} {Tech. Rep.}\ (\bibinfo {year} {2023})\
  \Eprint {https://arxiv.org/abs/2304.02527} {arxiv:2304.02527} \BibitemShut
  {NoStop}%
\bibitem [{\citenamefont {Hayata}\ and\ \citenamefont
  {Hidaka}(2023{\natexlab{a}})}]{Hayata2023StringNetFormulation}%
  \BibitemOpen
  \bibfield  {author} {\bibinfo {author} {\bibfnamefont {T.}~\bibnamefont
  {Hayata}}\ and\ \bibinfo {author} {\bibfnamefont {Y.}~\bibnamefont
  {Hidaka}},\ }\href@noop {} {\emph {\bibinfo {title} {String-net formulation
  of {Hamiltonian} lattice {Yang}-{Mills} theories and quantum many-body scars
  in a nonabelian gauge theory}}},\ \bibinfo {type} {Tech. Rep.}\ (\bibinfo
  {year} {2023})\ \Eprint {https://arxiv.org/abs/2305.05950} {arxiv:2305.05950}
  \BibitemShut {NoStop}%
\bibitem [{\citenamefont {Hayata}\ and\ \citenamefont
  {Hidaka}(2023{\natexlab{b}})}]{Hayata2023BreakingNewGround}%
  \BibitemOpen
  \bibfield  {author} {\bibinfo {author} {\bibfnamefont {T.}~\bibnamefont
  {Hayata}}\ and\ \bibinfo {author} {\bibfnamefont {Y.}~\bibnamefont
  {Hidaka}},\ }\href {https://doi.org/10.48550/arXiv.2306.12324} {\emph
  {\bibinfo {title} {Breaking new ground for quantum and classical simulations
  of $\mathrm{SU}(3)$ {Yang}-{Mills} theory}}},\ \bibinfo {type} {Tech. Rep.}\
  (\bibinfo {year} {2023})\ \Eprint {https://arxiv.org/abs/2306.12324}
  {arxiv:2306.12324} \BibitemShut {NoStop}%
\bibitem [{\citenamefont {Zohar}\ and\ \citenamefont
  {Burrello}(2015)}]{Zohar2015FormulationLatticeGauge}%
  \BibitemOpen
  \bibfield  {author} {\bibinfo {author} {\bibfnamefont {E.}~\bibnamefont
  {Zohar}}\ and\ \bibinfo {author} {\bibfnamefont {M.}~\bibnamefont
  {Burrello}},\ }\bibfield  {title} {\bibinfo {title} {Formulation of lattice
  gauge theories for quantum simulations},\ }\href
  {https://doi.org/10.1103/PhysRevD.91.054506} {\bibfield  {journal} {\bibinfo
  {journal} {Physical Review D}\ }\textbf {\bibinfo {volume} {91}},\ \bibinfo
  {pages} {054506} (\bibinfo {year} {2015})}\BibitemShut {NoStop}%
\bibitem [{\citenamefont {White}(1992)}]{White1992DensityMatrixFormulation}%
  \BibitemOpen
  \bibfield  {author} {\bibinfo {author} {\bibfnamefont {S.~R.}\ \bibnamefont
  {White}},\ }\bibfield  {title} {\bibinfo {title} {Density {Matrix}
  {Formulation} for {Quantum} {Renormalization} {Groups}},\ }\href
  {https://doi.org/10.1103/PhysRevLett.69.2863} {\bibfield  {journal} {\bibinfo
   {journal} {Physical Review Letters}\ }\textbf {\bibinfo {volume} {69}},\
  \bibinfo {pages} {2863} (\bibinfo {year} {1992})}\BibitemShut {NoStop}%
\bibitem [{\citenamefont
  {Schollw{\"o}ck}(2011)}]{Schollwoeck2011DensityMatrixRenormalization}%
  \BibitemOpen
  \bibfield  {author} {\bibinfo {author} {\bibfnamefont {U.}~\bibnamefont
  {Schollw{\"o}ck}},\ }\bibfield  {title} {\bibinfo {title} {The density-matrix
  renormalization group in the age of matrix product states},\ }\href
  {https://doi.org/10.1016/j.aop.2010.09.012} {\bibfield  {journal} {\bibinfo
  {journal} {Annals of Physics}\ }\textbf {\bibinfo {volume} {326}},\ \bibinfo
  {pages} {96} (\bibinfo {year} {2011})}\BibitemShut {NoStop}%
\bibitem [{\citenamefont {Hauschild}\ and\ \citenamefont
  {Pollmann}(2018)}]{Hauschild2018EfficientNumericalSimulations}%
  \BibitemOpen
  \bibfield  {author} {\bibinfo {author} {\bibfnamefont {J.}~\bibnamefont
  {Hauschild}}\ and\ \bibinfo {author} {\bibfnamefont {F.}~\bibnamefont
  {Pollmann}},\ }\bibfield  {title} {\bibinfo {title} {Efficient numerical
  simulations with {Tensor} {Networks}: {Tensor} {Network} {Python}
  ({TeNPy})},\ }\bibfield  {journal} {\bibinfo  {journal} {{SciPost} Physics
  Lecture Notes}\ }\textbf {\bibinfo {volume} {5}},\ \href
  {https://doi.org/10.21468/SciPostPhysLectNotes.5}
  {10.21468/SciPostPhysLectNotes.5} (\bibinfo {year} {2018})\BibitemShut
  {NoStop}%
\bibitem [{\citenamefont {Kogut}\ and\ \citenamefont
  {Susskind}(1975)}]{Kogut1975HamiltonianFormulationWilsons}%
  \BibitemOpen
  \bibfield  {author} {\bibinfo {author} {\bibfnamefont {J.}~\bibnamefont
  {Kogut}}\ and\ \bibinfo {author} {\bibfnamefont {L.}~\bibnamefont
  {Susskind}},\ }\bibfield  {title} {\bibinfo {title} {Hamiltonian formulation
  of {Wilson}'s lattice gauge theories},\ }\href
  {https://doi.org/10.1103/PhysRevD.11.395} {\bibfield  {journal} {\bibinfo
  {journal} {Physical Review D}\ }\textbf {\bibinfo {volume} {11}},\ \bibinfo
  {pages} {395} (\bibinfo {year} {1975})}\BibitemShut {NoStop}%
\bibitem [{\citenamefont {Banks}\ \emph {et~al.}(1976)\citenamefont {Banks},
  \citenamefont {Susskind},\ and\ \citenamefont
  {Kogut}}]{Banks1976StrongCouplingCalculations}%
  \BibitemOpen
  \bibfield  {author} {\bibinfo {author} {\bibfnamefont {T.}~\bibnamefont
  {Banks}}, \bibinfo {author} {\bibfnamefont {L.}~\bibnamefont {Susskind}},\
  and\ \bibinfo {author} {\bibfnamefont {J.}~\bibnamefont {Kogut}},\ }\bibfield
   {title} {\bibinfo {title} {Strong-coupling calculations of lattice gauge
  theories: (1 + 1)-dimensional exercises},\ }\href
  {https://doi.org/10.1103/PhysRevD.13.1043} {\bibfield  {journal} {\bibinfo
  {journal} {Physical Review D}\ }\textbf {\bibinfo {volume} {13}},\ \bibinfo
  {pages} {1043} (\bibinfo {year} {1976})}\BibitemShut {NoStop}%
\bibitem [{\citenamefont {Susskind}(1977)}]{Susskind1977LatticeFermions}%
  \BibitemOpen
  \bibfield  {author} {\bibinfo {author} {\bibfnamefont {L.}~\bibnamefont
  {Susskind}},\ }\bibfield  {title} {\bibinfo {title} {Lattice fermions},\
  }\href {https://doi.org/10.1103/PhysRevD.16.3031} {\bibfield  {journal}
  {\bibinfo  {journal} {Physical Review D}\ }\textbf {\bibinfo {volume} {16}},\
  \bibinfo {pages} {3031} (\bibinfo {year} {1977})}\BibitemShut {NoStop}%
\bibitem [{\citenamefont {Yang}\ and\ \citenamefont
  {Mills}(1954)}]{Yang1954ConservationIsotopicSpin}%
  \BibitemOpen
  \bibfield  {author} {\bibinfo {author} {\bibfnamefont {C.~N.}\ \bibnamefont
  {Yang}}\ and\ \bibinfo {author} {\bibfnamefont {R.~L.}\ \bibnamefont
  {Mills}},\ }\bibfield  {title} {\bibinfo {title} {Conservation of {Isotopic}
  {Spin} and {Isotopic} {Gauge} {Invariance}},\ }\href
  {https://doi.org/10.1103/PhysRev.96.191} {\bibfield  {journal} {\bibinfo
  {journal} {Physical Review}\ }\textbf {\bibinfo {volume} {96}},\ \bibinfo
  {pages} {191} (\bibinfo {year} {1954})}\BibitemShut {NoStop}%
\bibitem [{\citenamefont {Hamer}\ \emph {et~al.}(1982)\citenamefont {Hamer},
  \citenamefont {Kogut}, \citenamefont {Crewther},\ and\ \citenamefont
  {Mazzolini}}]{Hamer1982MassiveSchwingerModel}%
  \BibitemOpen
  \bibfield  {author} {\bibinfo {author} {\bibfnamefont {C.~J.}\ \bibnamefont
  {Hamer}}, \bibinfo {author} {\bibfnamefont {J.}~\bibnamefont {Kogut}},
  \bibinfo {author} {\bibfnamefont {D.~P.}\ \bibnamefont {Crewther}},\ and\
  \bibinfo {author} {\bibfnamefont {M.~M.}\ \bibnamefont {Mazzolini}},\
  }\bibfield  {title} {\bibinfo {title} {The massive {Schwinger} model on a
  lattice: {Background} field, chiral symmetry and the string tension},\ }\href
  {https://doi.org/10.1016/0550-3213(82)90229-2} {\bibfield  {journal}
  {\bibinfo  {journal} {Nuclear Physics B}\ }\textbf {\bibinfo {volume}
  {208}},\ \bibinfo {pages} {413} (\bibinfo {year} {1982})}\BibitemShut
  {NoStop}%
\bibitem [{\citenamefont {Henneaux}\ and\ \citenamefont
  {Teitelboim}(1992)}]{Henneaux1992QuantizationGaugeSystems}%
  \BibitemOpen
  \bibfield  {author} {\bibinfo {author} {\bibfnamefont {M.}~\bibnamefont
  {Henneaux}}\ and\ \bibinfo {author} {\bibfnamefont {C.}~\bibnamefont
  {Teitelboim}},\ }\href {https://doi.org/10.2307/j.ctv10crg0r} {\emph
  {\bibinfo {title} {Quantization of {Gauge} {Systems}}}}\ (\bibinfo
  {publisher} {Princeton University Press},\ \bibinfo {year}
  {1992})\BibitemShut {NoStop}%
\bibitem [{\citenamefont {Silvi}\ \emph
  {et~al.}(2019{\natexlab{b}})\citenamefont {Silvi}, \citenamefont {Sauer},
  \citenamefont {Tschirsich},\ and\ \citenamefont
  {Montangero}}]{Silvi2019TensorNetworkSimulation}%
  \BibitemOpen
  \bibfield  {author} {\bibinfo {author} {\bibfnamefont {P.}~\bibnamefont
  {Silvi}}, \bibinfo {author} {\bibfnamefont {Y.}~\bibnamefont {Sauer}},
  \bibinfo {author} {\bibfnamefont {F.}~\bibnamefont {Tschirsich}},\ and\
  \bibinfo {author} {\bibfnamefont {S.}~\bibnamefont {Montangero}},\ }\bibfield
   {title} {\bibinfo {title} {Tensor network simulation of an {SU}(3) lattice
  gauge theory in {1D}},\ }\href {https://doi.org/10.1103/PhysRevD.100.074512}
  {\bibfield  {journal} {\bibinfo  {journal} {Physical Review D}\ }\textbf
  {\bibinfo {volume} {100}},\ \bibinfo {pages} {074512} (\bibinfo {year}
  {2019}{\natexlab{b}})}\BibitemShut {NoStop}%
\bibitem [{\citenamefont {Orland}\ and\ \citenamefont
  {Rohrlich}(1990)}]{Orland1990LatticeGaugeMagnets}%
  \BibitemOpen
  \bibfield  {author} {\bibinfo {author} {\bibfnamefont {P.}~\bibnamefont
  {Orland}}\ and\ \bibinfo {author} {\bibfnamefont {D.}~\bibnamefont
  {Rohrlich}},\ }\bibfield  {title} {\bibinfo {title} {Lattice gauge magnets:
  {Local} isospin from spin},\ }\href
  {https://doi.org/10.1016/0550-3213(90)90646-U} {\bibfield  {journal}
  {\bibinfo  {journal} {Nuclear Physics B}\ }\textbf {\bibinfo {volume}
  {338}},\ \bibinfo {pages} {647} (\bibinfo {year} {1990})}\BibitemShut
  {NoStop}%
\bibitem [{\citenamefont {Zohar}\ and\ \citenamefont
  {Cirac}(2019)}]{Zohar2019RemovingStaggeredFermionic}%
  \BibitemOpen
  \bibfield  {author} {\bibinfo {author} {\bibfnamefont {E.}~\bibnamefont
  {Zohar}}\ and\ \bibinfo {author} {\bibfnamefont {J.~I.}\ \bibnamefont
  {Cirac}},\ }\bibfield  {title} {\bibinfo {title} {Removing staggered
  fermionic matter in ${U}({N})$ and ${SU}({N})$ lattice gauge theories},\
  }\href {https://doi.org/10.1103/PhysRevD.99.114511} {\bibfield  {journal}
  {\bibinfo  {journal} {Physical Review D}\ }\textbf {\bibinfo {volume} {99}},\
  \bibinfo {pages} {114511} (\bibinfo {year} {2019})}\BibitemShut {NoStop}%
\bibitem [{\citenamefont {Brower}\ \emph {et~al.}(1999)\citenamefont {Brower},
  \citenamefont {Chandrasekharan},\ and\ \citenamefont
  {Wiese}}]{Brower1999QcdAsQuantum}%
  \BibitemOpen
  \bibfield  {author} {\bibinfo {author} {\bibfnamefont {R.}~\bibnamefont
  {Brower}}, \bibinfo {author} {\bibfnamefont {S.}~\bibnamefont
  {Chandrasekharan}},\ and\ \bibinfo {author} {\bibfnamefont {U.-J.}\
  \bibnamefont {Wiese}},\ }\bibfield  {title} {\bibinfo {title} {{QCD} as a
  quantum link model},\ }\href {https://doi.org/10.1103/PhysRevD.60.094502}
  {\bibfield  {journal} {\bibinfo  {journal} {Physical Review D}\ }\textbf
  {\bibinfo {volume} {60}},\ \bibinfo {pages} {094502} (\bibinfo {year}
  {1999})}\BibitemShut {NoStop}%
\bibitem [{\citenamefont {Rommer}\ and\ \citenamefont
  {{\"O}stlund}(1997)}]{Rommer1997ClassAnsatzWave}%
  \BibitemOpen
  \bibfield  {author} {\bibinfo {author} {\bibfnamefont {S.}~\bibnamefont
  {Rommer}}\ and\ \bibinfo {author} {\bibfnamefont {S.}~\bibnamefont
  {{\"O}stlund}},\ }\bibfield  {title} {\bibinfo {title} {Class of ansatz wave
  functions for one-dimensional spin systems and their relation to the density
  matrix renormalization group},\ }\href
  {https://doi.org/10.1103/PhysRevB.55.2164} {\bibfield  {journal} {\bibinfo
  {journal} {Physical Review B}\ }\textbf {\bibinfo {volume} {55}},\ \bibinfo
  {pages} {2164} (\bibinfo {year} {1997})}\BibitemShut {NoStop}%
\bibitem [{\citenamefont {Tschirsich}\ \emph {et~al.}(2019)\citenamefont
  {Tschirsich}, \citenamefont {Montangero},\ and\ \citenamefont
  {Dalmonte}}]{Tschirsich2019PhaseDiagramConformal}%
  \BibitemOpen
  \bibfield  {author} {\bibinfo {author} {\bibfnamefont {F.}~\bibnamefont
  {Tschirsich}}, \bibinfo {author} {\bibfnamefont {S.}~\bibnamefont
  {Montangero}},\ and\ \bibinfo {author} {\bibfnamefont {M.}~\bibnamefont
  {Dalmonte}},\ }\bibfield  {title} {\bibinfo {title} {Phase diagram and
  conformal string excitations of square ice using gauge invariant matrix
  product states},\ }\href {https://doi.org/10.21468/SciPostPhys.6.3.028}
  {\bibfield  {journal} {\bibinfo  {journal} {SciPost Physics}\ }\textbf
  {\bibinfo {volume} {6}},\ \bibinfo {pages} {028} (\bibinfo {year}
  {2019})}\BibitemShut {NoStop}%
\bibitem [{\citenamefont {Ba{\~n}uls}\ \emph {et~al.}(2013)\citenamefont
  {Ba{\~n}uls}, \citenamefont {Cichy}, \citenamefont {Cirac},\ and\
  \citenamefont {Jansen}}]{Banuls2013MassSpectrumSchwinger}%
  \BibitemOpen
  \bibfield  {author} {\bibinfo {author} {\bibfnamefont {M.~C.}\ \bibnamefont
  {Ba{\~n}uls}}, \bibinfo {author} {\bibfnamefont {K.}~\bibnamefont {Cichy}},
  \bibinfo {author} {\bibfnamefont {J.~I.}\ \bibnamefont {Cirac}},\ and\
  \bibinfo {author} {\bibfnamefont {K.}~\bibnamefont {Jansen}},\ }\bibfield
  {title} {\bibinfo {title} {The mass spectrum of the {Schwinger} model with
  matrix product states},\ }\href {https://doi.org/10.1007/JHEP11(2013)158}
  {\bibfield  {journal} {\bibinfo  {journal} {Journal of High Energy Physics}\
  }\textbf {\bibinfo {volume} {2013}},\ \bibinfo {pages} {158} (\bibinfo {year}
  {2013})}\BibitemShut {NoStop}%
\bibitem [{\citenamefont {Pollmann}\ \emph {et~al.}(2009)\citenamefont
  {Pollmann}, \citenamefont {Mukerjee}, \citenamefont {Turner},\ and\
  \citenamefont {Moore}}]{Pollmann2009TheoryFiniteEntanglement}%
  \BibitemOpen
  \bibfield  {author} {\bibinfo {author} {\bibfnamefont {F.}~\bibnamefont
  {Pollmann}}, \bibinfo {author} {\bibfnamefont {S.}~\bibnamefont {Mukerjee}},
  \bibinfo {author} {\bibfnamefont {A.~M.}\ \bibnamefont {Turner}},\ and\
  \bibinfo {author} {\bibfnamefont {J.~E.}\ \bibnamefont {Moore}},\ }\bibfield
  {title} {\bibinfo {title} {Theory of {Finite}-{Entanglement} {Scaling} at
  {One}-{Dimensional} {Quantum} {Critical} {Points}},\ }\href
  {https://doi.org/10.1103/PhysRevLett.102.255701} {\bibfield  {journal}
  {\bibinfo  {journal} {Physical Review Letters}\ }\textbf {\bibinfo {volume}
  {102}},\ \bibinfo {pages} {255701} (\bibinfo {year} {2009})}\BibitemShut
  {NoStop}%
\bibitem [{\citenamefont {Pirvu}\ \emph {et~al.}(2012)\citenamefont {Pirvu},
  \citenamefont {Vidal}, \citenamefont {Verstraete},\ and\ \citenamefont
  {Tagliacozzo}}]{Pirvu2012MatrixProductStates}%
  \BibitemOpen
  \bibfield  {author} {\bibinfo {author} {\bibfnamefont {B.}~\bibnamefont
  {Pirvu}}, \bibinfo {author} {\bibfnamefont {G.}~\bibnamefont {Vidal}},
  \bibinfo {author} {\bibfnamefont {F.}~\bibnamefont {Verstraete}},\ and\
  \bibinfo {author} {\bibfnamefont {L.}~\bibnamefont {Tagliacozzo}},\
  }\bibfield  {title} {\bibinfo {title} {Matrix product states for critical
  spin chains: {Finite}-size versus finite-entanglement scaling},\ }\href
  {https://doi.org/10.1103/PhysRevB.86.075117} {\bibfield  {journal} {\bibinfo
  {journal} {Physical Review B}\ }\textbf {\bibinfo {volume} {86}},\ \bibinfo
  {pages} {075117} (\bibinfo {year} {2012})}\BibitemShut {NoStop}%
\bibitem [{\citenamefont {Stojevic}\ \emph {et~al.}(2015)\citenamefont
  {Stojevic}, \citenamefont {Haegeman}, \citenamefont {McCulloch},
  \citenamefont {Tagliacozzo},\ and\ \citenamefont
  {Verstraete}}]{Stojevic2015ConformalDataFinite}%
  \BibitemOpen
  \bibfield  {author} {\bibinfo {author} {\bibfnamefont {V.}~\bibnamefont
  {Stojevic}}, \bibinfo {author} {\bibfnamefont {J.}~\bibnamefont {Haegeman}},
  \bibinfo {author} {\bibfnamefont {I.~P.}\ \bibnamefont {McCulloch}}, \bibinfo
  {author} {\bibfnamefont {L.}~\bibnamefont {Tagliacozzo}},\ and\ \bibinfo
  {author} {\bibfnamefont {F.}~\bibnamefont {Verstraete}},\ }\bibfield  {title}
  {\bibinfo {title} {Conformal data from finite entanglement scaling},\ }\href
  {https://doi.org/10.1103/PhysRevB.91.035120} {\bibfield  {journal} {\bibinfo
  {journal} {Physical Review B}\ }\textbf {\bibinfo {volume} {91}},\ \bibinfo
  {pages} {035120} (\bibinfo {year} {2015})}\BibitemShut {NoStop}%
\bibitem [{\citenamefont {Hern{\'a}ndez}(2011)}]{Hernandez20111LatticeField}%
  \BibitemOpen
  \bibfield  {author} {\bibinfo {author} {\bibfnamefont {M.~P.}\ \bibnamefont
  {Hern{\'a}ndez}},\ }\bibfield  {title} {\bibinfo {title} {1 {Lattice} field
  theory fundamentals}\ }(\bibinfo  {publisher} {Oxford University Press},\
  \bibinfo {year} {2011})\ p.~\bibinfo {pages} {20}\BibitemShut {NoStop}%
\bibitem [{\citenamefont {Holzhey}\ \emph {et~al.}(1994)\citenamefont
  {Holzhey}, \citenamefont {Larsen},\ and\ \citenamefont
  {Wilczek}}]{Holzhey1994GeometricRenormalizedEntropy}%
  \BibitemOpen
  \bibfield  {author} {\bibinfo {author} {\bibfnamefont {C.}~\bibnamefont
  {Holzhey}}, \bibinfo {author} {\bibfnamefont {F.}~\bibnamefont {Larsen}},\
  and\ \bibinfo {author} {\bibfnamefont {F.}~\bibnamefont {Wilczek}},\
  }\bibfield  {title} {\bibinfo {title} {Geometric and {Renormalized} {Entropy}
  in {Conformal} {Field} {Theory}},\ }\href
  {https://doi.org/10.1016/0550-3213(94)90402-2} {\bibfield  {journal}
  {\bibinfo  {journal} {Nuclear Physics B}\ }\textbf {\bibinfo {volume}
  {424}},\ \bibinfo {pages} {443} (\bibinfo {year} {1994})},\ \Eprint
  {https://arxiv.org/abs/hep-th/9403108} {arxiv:hep-th/9403108} \BibitemShut
  {NoStop}%
\bibitem [{\citenamefont {Calabrese}\ and\ \citenamefont
  {Cardy}(2004)}]{Calabrese2004EntanglementEntropyQuantum}%
  \BibitemOpen
  \bibfield  {author} {\bibinfo {author} {\bibfnamefont {P.}~\bibnamefont
  {Calabrese}}\ and\ \bibinfo {author} {\bibfnamefont {J.}~\bibnamefont
  {Cardy}},\ }\bibfield  {title} {\bibinfo {title} {Entanglement {Entropy} and
  {Quantum} {Field} {Theory}},\ }\href
  {https://doi.org/10.1088/1742-5468/2004/06/P06002} {\bibfield  {journal}
  {\bibinfo  {journal} {Journal of Statistical Mechanics: Theory and
  Experiment}\ }\textbf {\bibinfo {volume} {2004}},\ \bibinfo {pages} {P06002}
  (\bibinfo {year} {2004})},\ \Eprint {https://arxiv.org/abs/hep-th/0405152}
  {arxiv:hep-th/0405152} \BibitemShut {NoStop}%
\bibitem [{\citenamefont {Srednicki}(1993)}]{Srednicki1993EntropyArea}%
  \BibitemOpen
  \bibfield  {author} {\bibinfo {author} {\bibfnamefont {M.}~\bibnamefont
  {Srednicki}},\ }\bibfield  {title} {\bibinfo {title} {Entropy and {Area}},\
  }\href {https://doi.org/10.1103/PhysRevLett.71.666} {\bibfield  {journal}
  {\bibinfo  {journal} {Physical Review Letters}\ }\textbf {\bibinfo {volume}
  {71}},\ \bibinfo {pages} {666} (\bibinfo {year} {1993})},\ \Eprint
  {https://arxiv.org/abs/hep-th/9303048} {arxiv:hep-th/9303048} \BibitemShut
  {NoStop}%
\bibitem [{\citenamefont {Laflorencie}\ \emph {et~al.}(2006)\citenamefont
  {Laflorencie}, \citenamefont {S{{\o}}rensen}, \citenamefont {Chang},\ and\
  \citenamefont {Affleck}}]{Laflorencie2006BoundaryEffectsCritical}%
  \BibitemOpen
  \bibfield  {author} {\bibinfo {author} {\bibfnamefont {N.}~\bibnamefont
  {Laflorencie}}, \bibinfo {author} {\bibfnamefont {E.~S.}\ \bibnamefont
  {S{{\o}}rensen}}, \bibinfo {author} {\bibfnamefont {M.-S.}\ \bibnamefont
  {Chang}},\ and\ \bibinfo {author} {\bibfnamefont {I.}~\bibnamefont
  {Affleck}},\ }\bibfield  {title} {\bibinfo {title} {Boundary {Effects} in the
  {Critical} {Scaling} of {Entanglement} {Entropy} in {1D} {Systems}},\ }\href
  {https://doi.org/10.1103/PhysRevLett.96.100603} {\bibfield  {journal}
  {\bibinfo  {journal} {Physical Review Letters}\ }\textbf {\bibinfo {volume}
  {96}},\ \bibinfo {pages} {100603} (\bibinfo {year} {2006})}\BibitemShut
  {NoStop}%
\bibitem [{\citenamefont {Calabrese}\ \emph {et~al.}(2010)\citenamefont
  {Calabrese}, \citenamefont {Campostrini}, \citenamefont {Essler},\ and\
  \citenamefont {Nienhuis}}]{Calabrese2010ParityEffectsScaling}%
  \BibitemOpen
  \bibfield  {author} {\bibinfo {author} {\bibfnamefont {P.}~\bibnamefont
  {Calabrese}}, \bibinfo {author} {\bibfnamefont {M.}~\bibnamefont
  {Campostrini}}, \bibinfo {author} {\bibfnamefont {F.}~\bibnamefont
  {Essler}},\ and\ \bibinfo {author} {\bibfnamefont {B.}~\bibnamefont
  {Nienhuis}},\ }\bibfield  {title} {\bibinfo {title} {Parity {Effects} in the
  {Scaling} of {Block} {Entanglement} in {Gapless} {Spin} {Chains}},\ }\href
  {https://doi.org/10.1103/PhysRevLett.104.095701} {\bibfield  {journal}
  {\bibinfo  {journal} {Physical Review Letters}\ }\textbf {\bibinfo {volume}
  {104}},\ \bibinfo {pages} {095701} (\bibinfo {year} {2010})}\BibitemShut
  {NoStop}%
\bibitem [{\citenamefont {Xavier}\ and\ \citenamefont
  {Alcaraz}(2012)}]{Xavier2012FiniteSizeCorrections}%
  \BibitemOpen
  \bibfield  {author} {\bibinfo {author} {\bibfnamefont {J.~C.}\ \bibnamefont
  {Xavier}}\ and\ \bibinfo {author} {\bibfnamefont {F.~C.}\ \bibnamefont
  {Alcaraz}},\ }\bibfield  {title} {\bibinfo {title} {Finite-size corrections
  of the entanglement entropy of critical quantum chains},\ }\href
  {https://doi.org/10.1103/PhysRevB.85.024418} {\bibfield  {journal} {\bibinfo
  {journal} {Physical Review B}\ }\textbf {\bibinfo {volume} {85}},\ \bibinfo
  {pages} {024418} (\bibinfo {year} {2012})}\BibitemShut {NoStop}%
\bibitem [{\citenamefont {Ecker}(1995)}]{Ecker1995ChiralPerturbationTheory}%
  \BibitemOpen
  \bibfield  {author} {\bibinfo {author} {\bibfnamefont {G.}~\bibnamefont
  {Ecker}},\ }\bibfield  {title} {\bibinfo {title} {Chiral perturbation
  theory},\ }\href {https://doi.org/10.1016/0146-6410(95)00041-G} {\bibfield
  {journal} {\bibinfo  {journal} {Progress in Particle and Nuclear Physics}\
  }\textbf {\bibinfo {volume} {35}},\ \bibinfo {pages} {1} (\bibinfo {year}
  {1995})}\BibitemShut {NoStop}%
\bibitem [{\citenamefont {Das}\ \emph {et~al.}(1967)\citenamefont {Das},
  \citenamefont {Guralnik}, \citenamefont {Mathur}, \citenamefont {Low},\ and\
  \citenamefont {Young}}]{Das1967ElectromagneticMassDifference}%
  \BibitemOpen
  \bibfield  {author} {\bibinfo {author} {\bibfnamefont {T.}~\bibnamefont
  {Das}}, \bibinfo {author} {\bibfnamefont {G.~S.}\ \bibnamefont {Guralnik}},
  \bibinfo {author} {\bibfnamefont {V.~S.}\ \bibnamefont {Mathur}}, \bibinfo
  {author} {\bibfnamefont {F.~E.}\ \bibnamefont {Low}},\ and\ \bibinfo {author}
  {\bibfnamefont {J.~E.}\ \bibnamefont {Young}},\ }\bibfield  {title} {\bibinfo
  {title} {Electromagnetic {Mass} {Difference} of {Pions}},\ }\href
  {https://doi.org/10.1103/PhysRevLett.18.759} {\bibfield  {journal} {\bibinfo
  {journal} {Physical Review Letters}\ }\textbf {\bibinfo {volume} {18}},\
  \bibinfo {pages} {759} (\bibinfo {year} {1967})}\BibitemShut {NoStop}%
\bibitem [{\citenamefont
  {Zamolodchikov}(1986)}]{Zamolodchikov1986IrreversibilityFluxRenormalization}%
  \BibitemOpen
  \bibfield  {author} {\bibinfo {author} {\bibfnamefont {A.~B.}\ \bibnamefont
  {Zamolodchikov}},\ }\bibfield  {title} {\bibinfo {title} {Irreversibility of
  the {Flux} of the {Renormalization} {Group} in a {2D} {Field} {Theory}},\
  }\href@noop {} {\bibfield  {journal} {\bibinfo  {journal} {JETP lett}\
  }\textbf {\bibinfo {volume} {43}},\ \bibinfo {pages} {730} (\bibinfo {year}
  {1986})}\BibitemShut {NoStop}%
\bibitem [{\citenamefont {Creutz}(1980)}]{Creutz1980MonteCarloStudy}%
  \BibitemOpen
  \bibfield  {author} {\bibinfo {author} {\bibfnamefont {M.}~\bibnamefont
  {Creutz}},\ }\bibfield  {title} {\bibinfo {title} {Monte {Carlo} study of
  quantized {SU}(2) gauge theory},\ }\href
  {https://doi.org/10.1103/PhysRevD.21.2308} {\bibfield  {journal} {\bibinfo
  {journal} {Physical Review D}\ }\textbf {\bibinfo {volume} {21}},\ \bibinfo
  {pages} {2308} (\bibinfo {year} {1980})}\BibitemShut {NoStop}%
\bibitem [{\citenamefont {Abdalla}\ \emph {et~al.}(2001)\citenamefont
  {Abdalla}, \citenamefont {Abdalla},\ and\ \citenamefont
  {Rothe}}]{Abdalla2001NonPerturbativeMethods}%
  \BibitemOpen
  \bibfield  {author} {\bibinfo {author} {\bibfnamefont {E.}~\bibnamefont
  {Abdalla}}, \bibinfo {author} {\bibfnamefont {M.~C.~B.}\ \bibnamefont
  {Abdalla}},\ and\ \bibinfo {author} {\bibfnamefont {K.~D.}\ \bibnamefont
  {Rothe}},\ }\href {https://doi.org/10.1142/4678} {\emph {\bibinfo {title}
  {Non-{Perturbative} {Methods} in 2 {Dimensional} {Quantum} {Field}
  {Theory}}}},\ \bibinfo {edition} {2nd}\ ed.\ (\bibinfo  {publisher} {{WORLD}
  {SCIENTIFIC}},\ \bibinfo {year} {2001})\BibitemShut {NoStop}%
\bibitem [{\citenamefont {Borla}\ \emph {et~al.}(2020)\citenamefont {Borla},
  \citenamefont {Verresen}, \citenamefont {Grusdt},\ and\ \citenamefont
  {Moroz}}]{Borla2020ConfinedPhasesOne}%
  \BibitemOpen
  \bibfield  {author} {\bibinfo {author} {\bibfnamefont {U.}~\bibnamefont
  {Borla}}, \bibinfo {author} {\bibfnamefont {R.}~\bibnamefont {Verresen}},
  \bibinfo {author} {\bibfnamefont {F.}~\bibnamefont {Grusdt}},\ and\ \bibinfo
  {author} {\bibfnamefont {S.}~\bibnamefont {Moroz}},\ }\bibfield  {title}
  {\bibinfo {title} {Confined {Phases} of {One}-{Dimensional} {Spinless}
  {Fermions} {Coupled} to ${Z}_2$ {Gauge} {Theory}},\ }\href
  {https://doi.org/10.1103/PhysRevLett.124.120503} {\bibfield  {journal}
  {\bibinfo  {journal} {Physical Review Letters}\ }\textbf {\bibinfo {volume}
  {124}},\ \bibinfo {pages} {120503} (\bibinfo {year} {2020})}\BibitemShut
  {NoStop}%
\bibitem [{\citenamefont {Jaffe}(1977)}]{Jaffe1977MultiQuarkHadrons}%
  \BibitemOpen
  \bibfield  {author} {\bibinfo {author} {\bibfnamefont {R.~J.}\ \bibnamefont
  {Jaffe}},\ }\bibfield  {title} {\bibinfo {title} {Multi-{Quark} {Hadrons}. 1.
  {The} {Phenomenology} of (2 {Quark} 2 anti-{Quark}) {Mesons}},\ }\href
  {https://doi.org/10.1103/PhysRevD.15.267} {\bibfield  {journal} {\bibinfo
  {journal} {Phys. Rev. D}\ }\textbf {\bibinfo {volume} {15}},\ \bibinfo
  {pages} {267} (\bibinfo {year} {1977})}\BibitemShut {NoStop}%
\bibitem [{\citenamefont
  {Bicudo}(2022)}]{Bicudo2022TetraquarksPentaquarksLattice}%
  \BibitemOpen
  \bibfield  {author} {\bibinfo {author} {\bibfnamefont {P.}~\bibnamefont
  {Bicudo}},\ }\href {https://doi.org/10.48550/arXiv.2212.07793} {\emph
  {\bibinfo {title} {Tetraquarks and pentaquarks in lattice {QCD} with light
  and heavy quarks}}},\ \bibinfo {type} {Tech. Rep.}\ (\bibinfo {year} {2022})\
  \Eprint {https://arxiv.org/abs/2212.07793} {arxiv:2212.07793} \BibitemShut
  {NoStop}%
\bibitem [{\citenamefont {Atas}\ \emph {et~al.}(2023)\citenamefont {Atas},
  \citenamefont {Haase}, \citenamefont {Zhang}, \citenamefont {Wei},
  \citenamefont {Pfaendler}, \citenamefont {Lewis},\ and\ \citenamefont
  {Muschik}}]{Atas2023SimulatingOneDimensional}%
  \BibitemOpen
  \bibfield  {author} {\bibinfo {author} {\bibfnamefont {Y.~Y.}\ \bibnamefont
  {Atas}}, \bibinfo {author} {\bibfnamefont {J.~F.}\ \bibnamefont {Haase}},
  \bibinfo {author} {\bibfnamefont {J.}~\bibnamefont {Zhang}}, \bibinfo
  {author} {\bibfnamefont {V.}~\bibnamefont {Wei}}, \bibinfo {author}
  {\bibfnamefont {S.~M.-L.}\ \bibnamefont {Pfaendler}}, \bibinfo {author}
  {\bibfnamefont {R.}~\bibnamefont {Lewis}},\ and\ \bibinfo {author}
  {\bibfnamefont {C.~A.}\ \bibnamefont {Muschik}},\ }\href
  {https://doi.org/10.48550/arXiv.2207.03473} {\emph {\bibinfo {title}
  {Simulating one-dimensional quantum chromodynamics on a quantum computer:
  {Real}-time evolutions of tetra- and pentaquarks}}},\ \bibinfo {type} {Tech.
  Rep.}\ (\bibinfo {year} {2023})\ \Eprint {https://arxiv.org/abs/2207.03473}
  {arxiv:2207.03473} \BibitemShut {NoStop}%
\bibitem [{\citenamefont {Itou}\ \emph {et~al.}(2023)\citenamefont {Itou},
  \citenamefont {Matsumoto},\ and\ \citenamefont
  {Tanizaki}}]{Itou2023CalculatingCompositeParticle}%
  \BibitemOpen
  \bibfield  {author} {\bibinfo {author} {\bibfnamefont {E.}~\bibnamefont
  {Itou}}, \bibinfo {author} {\bibfnamefont {A.}~\bibnamefont {Matsumoto}},\
  and\ \bibinfo {author} {\bibfnamefont {Y.}~\bibnamefont {Tanizaki}},\ }\href
  {https://doi.org/10.48550/arXiv.2307.16655} {\emph {\bibinfo {title}
  {Calculating composite-particle spectra in {Hamiltonian} formalism and
  demonstration in 2-flavor {QED}$_{1+1\text{d}}$}}},\ \bibinfo {type} {Tech.
  Rep.}\ (\bibinfo {year} {2023})\ \Eprint {https://arxiv.org/abs/2307.16655}
  {arxiv:2307.16655} \BibitemShut {NoStop}%
\bibitem [{not({\natexlab{a}})}]{note_particle_number}%
  \BibitemOpen
  \bibinfo {note} {Because of staggered fermions, the zero flavour number
  charge sector corresponds to half filling of each quark \dof{}, whence the
  subtraction of $1/2$ from particle number densities. This also implies that
  at most $3L/2$ quarks of each specie can be hosted on a chain of lenght
  $L$.}\BibitemShut {Stop}%
\bibitem [{\citenamefont {Dalla~Torre}\ \emph {et~al.}(2016)\citenamefont
  {Dalla~Torre}, \citenamefont {Benjamin}, \citenamefont {He}, \citenamefont
  {Dentelski},\ and\ \citenamefont
  {Demler}}]{DallaTorre2016FriedelOscillationsAs}%
  \BibitemOpen
  \bibfield  {author} {\bibinfo {author} {\bibfnamefont {E.~G.}\ \bibnamefont
  {Dalla~Torre}}, \bibinfo {author} {\bibfnamefont {D.}~\bibnamefont
  {Benjamin}}, \bibinfo {author} {\bibfnamefont {Y.}~\bibnamefont {He}},
  \bibinfo {author} {\bibfnamefont {D.}~\bibnamefont {Dentelski}},\ and\
  \bibinfo {author} {\bibfnamefont {E.}~\bibnamefont {Demler}},\ }\bibfield
  {title} {\bibinfo {title} {Friedel oscillations as a probe of fermionic
  quasiparticles},\ }\href {https://doi.org/10.1103/PhysRevB.93.205117}
  {\bibfield  {journal} {\bibinfo  {journal} {Physical Review B}\ }\textbf
  {\bibinfo {volume} {93}},\ \bibinfo {pages} {205117} (\bibinfo {year}
  {2016})}\BibitemShut {NoStop}%
\bibitem [{not({\natexlab{b}})}]{note_spin_1d}%
  \BibitemOpen
  \bibinfo {note} {This is by analogy with the (3+1)D case, even though in a
  relativistic (1+1)D system there is no spin.}\BibitemShut {Stop}%
\bibitem [{\citenamefont {Vodola}\ \emph {et~al.}(2015)\citenamefont {Vodola},
  \citenamefont {Lepori}, \citenamefont {Ercolessi},\ and\ \citenamefont
  {Pupillo}}]{Vodola2015LongRangeIsing}%
  \BibitemOpen
  \bibfield  {author} {\bibinfo {author} {\bibfnamefont {D.}~\bibnamefont
  {Vodola}}, \bibinfo {author} {\bibfnamefont {L.}~\bibnamefont {Lepori}},
  \bibinfo {author} {\bibfnamefont {E.}~\bibnamefont {Ercolessi}},\ and\
  \bibinfo {author} {\bibfnamefont {G.}~\bibnamefont {Pupillo}},\ }\bibfield
  {title} {\bibinfo {title} {Long-range {Ising} and {Kitaev} models: phases,
  correlations and edge modes},\ }\href
  {https://doi.org/10.1088/1367-2630/18/1/015001} {\bibfield  {journal}
  {\bibinfo  {journal} {New Journal of Physics}\ }\textbf {\bibinfo {volume}
  {18}},\ \bibinfo {pages} {015001} (\bibinfo {year} {2015})}\BibitemShut
  {NoStop}%
\bibitem [{\citenamefont {Hastings}\ and\ \citenamefont
  {Koma}(2006)}]{Hastings2006SpectralGapExponential}%
  \BibitemOpen
  \bibfield  {author} {\bibinfo {author} {\bibfnamefont {M.~B.}\ \bibnamefont
  {Hastings}}\ and\ \bibinfo {author} {\bibfnamefont {T.}~\bibnamefont
  {Koma}},\ }\bibfield  {title} {\bibinfo {title} {Spectral {Gap} and
  {Exponential} {Decay} of {Correlations}},\ }\href
  {https://doi.org/10.1007/s00220-006-0030-4} {\bibfield  {journal} {\bibinfo
  {journal} {Communications in Mathematical Physics}\ }\textbf {\bibinfo
  {volume} {265}},\ \bibinfo {pages} {781} (\bibinfo {year}
  {2006})}\BibitemShut {NoStop}%
\bibitem [{\citenamefont {Sachdev}\ and\ \citenamefont
  {Keimer}(2011)}]{Sachdev2011QuantumCriticality}%
  \BibitemOpen
  \bibfield  {author} {\bibinfo {author} {\bibfnamefont {S.}~\bibnamefont
  {Sachdev}}\ and\ \bibinfo {author} {\bibfnamefont {B.}~\bibnamefont
  {Keimer}},\ }\bibfield  {title} {\bibinfo {title} {Quantum criticality},\
  }\href {https://doi.org/10.1063/1.3554314} {\bibfield  {journal} {\bibinfo
  {journal} {Physics Today}\ }\textbf {\bibinfo {volume} {64}},\ \bibinfo
  {pages} {29} (\bibinfo {year} {2011})}\BibitemShut {NoStop}%
\bibitem [{not({\natexlab{c}})}]{note_connected_correlators}%
  \BibitemOpen
  \bibinfo {note} {All considered operators violate flavour number
  conservation, hence the disconnected component is not present.}\BibitemShut
  {Stop}%
\bibitem [{\citenamefont {Ikeda}\ \emph {et~al.}(2023)\citenamefont {Ikeda},
  \citenamefont {Kharzeev}, \citenamefont {Meyer},\ and\ \citenamefont
  {Shi}}]{Ikeda2023DetectingCriticalPoint}%
  \BibitemOpen
  \bibfield  {author} {\bibinfo {author} {\bibfnamefont {K.}~\bibnamefont
  {Ikeda}}, \bibinfo {author} {\bibfnamefont {D.~E.}\ \bibnamefont {Kharzeev}},
  \bibinfo {author} {\bibfnamefont {R.}~\bibnamefont {Meyer}},\ and\ \bibinfo
  {author} {\bibfnamefont {S.}~\bibnamefont {Shi}},\ }\href
  {https://doi.org/10.48550/arXiv.2305.00996} {\emph {\bibinfo {title}
  {Detecting the critical point through entanglement in {Schwinger} model}}},\
  \bibinfo {type} {Tech. Rep.}\ (\bibinfo {year} {2023})\ \Eprint
  {https://arxiv.org/abs/2305.00996} {arxiv:2305.00996} \BibitemShut {NoStop}%
\bibitem [{\citenamefont {Rigobello}\ \emph {et~al.}(2021)\citenamefont
  {Rigobello}, \citenamefont {Notarnicola}, \citenamefont {Magnifico},\ and\
  \citenamefont {Montangero}}]{Rigobello2021EntanglementGeneration11d}%
  \BibitemOpen
  \bibfield  {author} {\bibinfo {author} {\bibfnamefont {M.}~\bibnamefont
  {Rigobello}}, \bibinfo {author} {\bibfnamefont {S.}~\bibnamefont
  {Notarnicola}}, \bibinfo {author} {\bibfnamefont {G.}~\bibnamefont
  {Magnifico}},\ and\ \bibinfo {author} {\bibfnamefont {S.}~\bibnamefont
  {Montangero}},\ }\bibfield  {title} {\bibinfo {title} {Entanglement
  generation in $(1+1)\mathrm{D}$ {QED} scattering processes},\ }\href
  {https://doi.org/10.1103/PhysRevD.104.114501} {\bibfield  {journal} {\bibinfo
   {journal} {Physical Review D}\ }\textbf {\bibinfo {volume} {104}},\ \bibinfo
  {pages} {114501} (\bibinfo {year} {2021})}\BibitemShut {NoStop}%
\bibitem [{\citenamefont {Belyansky}\ \emph {et~al.}(2023)\citenamefont
  {Belyansky}, \citenamefont {Whitsitt}, \citenamefont {Mueller}, \citenamefont
  {Fahimniya}, \citenamefont {Bennewitz}, \citenamefont {Davoudi},\ and\
  \citenamefont {Gorshkov}}]{Belyansky2023HighEnergyCollision}%
  \BibitemOpen
  \bibfield  {author} {\bibinfo {author} {\bibfnamefont {R.}~\bibnamefont
  {Belyansky}}, \bibinfo {author} {\bibfnamefont {S.}~\bibnamefont {Whitsitt}},
  \bibinfo {author} {\bibfnamefont {N.}~\bibnamefont {Mueller}}, \bibinfo
  {author} {\bibfnamefont {A.}~\bibnamefont {Fahimniya}}, \bibinfo {author}
  {\bibfnamefont {E.~R.}\ \bibnamefont {Bennewitz}}, \bibinfo {author}
  {\bibfnamefont {Z.}~\bibnamefont {Davoudi}},\ and\ \bibinfo {author}
  {\bibfnamefont {A.~V.}\ \bibnamefont {Gorshkov}},\ }\href
  {https://doi.org/10.48550/arXiv.2307.02522} {\emph {\bibinfo {title}
  {High-{Energy} {Collision} of {Quarks} and {Hadrons} in the {Schwinger}
  {Model}: {From} {Tensor} {Networks} to {Circuit} {QED}}}},\ \bibinfo {type}
  {Tech. Rep.}\ (\bibinfo {year} {2023})\ \Eprint
  {https://arxiv.org/abs/2307.02522} {arXiv:2307.02522} \BibitemShut {NoStop}%
\bibitem [{\citenamefont {Florio}\ \emph {et~al.}(2023)\citenamefont {Florio},
  \citenamefont {Frenklakh}, \citenamefont {Ikeda}, \citenamefont {Kharzeev},
  \citenamefont {Korepin}, \citenamefont {Shi},\ and\ \citenamefont
  {Yu}}]{Florio2023RealTimeNon}%
  \BibitemOpen
  \bibfield  {author} {\bibinfo {author} {\bibfnamefont {A.}~\bibnamefont
  {Florio}}, \bibinfo {author} {\bibfnamefont {D.}~\bibnamefont {Frenklakh}},
  \bibinfo {author} {\bibfnamefont {K.}~\bibnamefont {Ikeda}}, \bibinfo
  {author} {\bibfnamefont {D.}~\bibnamefont {Kharzeev}}, \bibinfo {author}
  {\bibfnamefont {V.}~\bibnamefont {Korepin}}, \bibinfo {author} {\bibfnamefont
  {S.}~\bibnamefont {Shi}},\ and\ \bibinfo {author} {\bibfnamefont
  {K.}~\bibnamefont {Yu}},\ }\bibfield  {title} {\bibinfo {title} {Real-time
  non-perturbative dynamics of jet production: quantum entanglement and vacuum
  modification},\ }\href {https://doi.org/10.1103/PhysRevLett.131.021902}
  {\bibfield  {journal} {\bibinfo  {journal} {Physical Review Letters}\
  }\textbf {\bibinfo {volume} {131}},\ \bibinfo {pages} {021902} (\bibinfo
  {year} {2023})},\ \Eprint {https://arxiv.org/abs/2301.11991}
  {arXiv:2301.11991} \BibitemShut {NoStop}%
\bibitem [{\citenamefont {Felser}\ \emph {et~al.}(2020)\citenamefont {Felser},
  \citenamefont {Silvi}, \citenamefont {Collura},\ and\ \citenamefont
  {Montangero}}]{Felser2020TwoDimensionalQuantum}%
  \BibitemOpen
  \bibfield  {author} {\bibinfo {author} {\bibfnamefont {T.}~\bibnamefont
  {Felser}}, \bibinfo {author} {\bibfnamefont {P.}~\bibnamefont {Silvi}},
  \bibinfo {author} {\bibfnamefont {M.}~\bibnamefont {Collura}},\ and\ \bibinfo
  {author} {\bibfnamefont {S.}~\bibnamefont {Montangero}},\ }\bibfield  {title}
  {\bibinfo {title} {Two-{D}imensional {Q}uantum-{L}ink {L}attice {Q}uantum
  {E}lectrodynamics at {F}inite {D}ensity},\ }\href
  {https://doi.org/10.1103/PhysRevX.10.041040} {\bibfield  {journal} {\bibinfo
  {journal} {Physical Review X}\ }\textbf {\bibinfo {volume} {10}},\ \bibinfo
  {pages} {041040} (\bibinfo {year} {2020})}\BibitemShut {NoStop}%
\bibitem [{\citenamefont
  {Zohar}(2021{\natexlab{b}})}]{Zohar2021QuantumSimulationLattice}%
  \BibitemOpen
  \bibfield  {author} {\bibinfo {author} {\bibfnamefont {E.}~\bibnamefont
  {Zohar}},\ }\bibfield  {title} {\bibinfo {title} {Quantum simulation of
  lattice gauge theories in more than one space dimension—requirements,
  challenges and methods},\ }\href {https://doi.org/10.1098/rsta.2021.0069}
  {\bibfield  {journal} {\bibinfo  {journal} {Philosophical Transactions of the
  Royal Society A: Mathematical, Physical and Engineering Sciences}\ }\textbf
  {\bibinfo {volume} {380}},\ \bibinfo {pages} {20210069} (\bibinfo {year}
  {2021}{\natexlab{b}})}\BibitemShut {NoStop}%
\bibitem [{\citenamefont {Magnifico}\ \emph {et~al.}(2021)\citenamefont
  {Magnifico}, \citenamefont {Felser}, \citenamefont {Silvi},\ and\
  \citenamefont {Montangero}}]{Magnifico2021LatticeQuantumElectrodynamics}%
  \BibitemOpen
  \bibfield  {author} {\bibinfo {author} {\bibfnamefont {G.}~\bibnamefont
  {Magnifico}}, \bibinfo {author} {\bibfnamefont {T.}~\bibnamefont {Felser}},
  \bibinfo {author} {\bibfnamefont {P.}~\bibnamefont {Silvi}},\ and\ \bibinfo
  {author} {\bibfnamefont {S.}~\bibnamefont {Montangero}},\ }\bibfield  {title}
  {\bibinfo {title} {Lattice quantum electrodynamics in $(3+1)$-dimensions at
  finite density with tensor networks},\ }\href
  {https://doi.org/10.1038/s41467-021-23646-3} {\bibfield  {journal} {\bibinfo
  {journal} {Nature Communications}\ }\textbf {\bibinfo {volume} {12}},\
  \bibinfo {pages} {3600} (\bibinfo {year} {2021})}\BibitemShut {NoStop}%
\bibitem [{\citenamefont {Pardo}\ \emph {et~al.}(2023)\citenamefont {Pardo},
  \citenamefont {Greenberg}, \citenamefont {Fortinsky}, \citenamefont {Katz},\
  and\ \citenamefont {Zohar}}]{Pardo2023ResourceEfficientQuantum}%
  \BibitemOpen
  \bibfield  {author} {\bibinfo {author} {\bibfnamefont {G.}~\bibnamefont
  {Pardo}}, \bibinfo {author} {\bibfnamefont {T.}~\bibnamefont {Greenberg}},
  \bibinfo {author} {\bibfnamefont {A.}~\bibnamefont {Fortinsky}}, \bibinfo
  {author} {\bibfnamefont {N.}~\bibnamefont {Katz}},\ and\ \bibinfo {author}
  {\bibfnamefont {E.}~\bibnamefont {Zohar}},\ }\bibfield  {title} {\bibinfo
  {title} {Resource-{Efficient} {Quantum} {Simulation} of {Lattice} {Gauge}
  {Theories} in {Arbitrary} {Dimensions}: {Solving} for {Gauss}' {Law} and
  {Fermion} {Elimination}},\ }\href
  {https://doi.org/10.1103/PhysRevResearch.5.023077} {\bibfield  {journal}
  {\bibinfo  {journal} {Physical Review Research}\ }\textbf {\bibinfo {volume}
  {5}},\ \bibinfo {pages} {023077} (\bibinfo {year} {2023})},\ \Eprint
  {https://arxiv.org/abs/2206.00685} {arxiv:2206.00685} \BibitemShut {NoStop}%
\bibitem [{\citenamefont {Lumia}\ \emph {et~al.}(2022)\citenamefont {Lumia},
  \citenamefont {Torta}, \citenamefont {Mbeng}, \citenamefont {Santoro},
  \citenamefont {Ercolessi}, \citenamefont {Burrello},\ and\ \citenamefont
  {Wauters}}]{Lumia2022TwoDimensionalZ2}%
  \BibitemOpen
  \bibfield  {author} {\bibinfo {author} {\bibfnamefont {L.}~\bibnamefont
  {Lumia}}, \bibinfo {author} {\bibfnamefont {P.}~\bibnamefont {Torta}},
  \bibinfo {author} {\bibfnamefont {G.~B.}\ \bibnamefont {Mbeng}}, \bibinfo
  {author} {\bibfnamefont {G.~E.}\ \bibnamefont {Santoro}}, \bibinfo {author}
  {\bibfnamefont {E.}~\bibnamefont {Ercolessi}}, \bibinfo {author}
  {\bibfnamefont {M.}~\bibnamefont {Burrello}},\ and\ \bibinfo {author}
  {\bibfnamefont {M.~M.}\ \bibnamefont {Wauters}},\ }\bibfield  {title}
  {\bibinfo {title} {Two-dimensional $\mathbb{Z}_2$ lattice gauge theory on a
  near-term quantum simulator: variational quantum optimization, confinement,
  and topological order},\ }\href {https://doi.org/10.1103/PRXQuantum.3.020320}
  {\bibfield  {journal} {\bibinfo  {journal} {PRX Quantum}\ }\textbf {\bibinfo
  {volume} {3}},\ \bibinfo {pages} {020320} (\bibinfo {year} {2022})},\ \Eprint
  {https://arxiv.org/abs/2112.11787} {arXiv:2112.11787} \BibitemShut {NoStop}%
\bibitem [{\citenamefont {Osborne}\ \emph {et~al.}(2022)\citenamefont
  {Osborne}, \citenamefont {McCulloch}, \citenamefont {Yang}, \citenamefont
  {Hauke},\ and\ \citenamefont {Halimeh}}]{Osborne2022LargeScale21d}%
  \BibitemOpen
  \bibfield  {author} {\bibinfo {author} {\bibfnamefont {J.}~\bibnamefont
  {Osborne}}, \bibinfo {author} {\bibfnamefont {I.~P.}\ \bibnamefont
  {McCulloch}}, \bibinfo {author} {\bibfnamefont {B.}~\bibnamefont {Yang}},
  \bibinfo {author} {\bibfnamefont {P.}~\bibnamefont {Hauke}},\ and\ \bibinfo
  {author} {\bibfnamefont {J.~C.}\ \bibnamefont {Halimeh}},\ }\href
  {https://doi.org/10.48550/arXiv.2211.01380} {\emph {\bibinfo {title}
  {Large-{Scale} $2+1${D} $\mathrm{U}(1)$ {Gauge} {Theory} with {Dynamical}
  {Matter} in a {Cold}-{Atom} {Quantum} {Simulator}}}},\ \bibinfo {type} {Tech.
  Rep.}\ (\bibinfo {year} {2022})\ \Eprint {https://arxiv.org/abs/2211.01380}
  {arxiv:2211.01380} \BibitemShut {NoStop}%
\bibitem [{\citenamefont {Emonts}\ and\ \citenamefont
  {Zohar}(2023)}]{Emonts2023FermionicGaussianPeps}%
  \BibitemOpen
  \bibfield  {author} {\bibinfo {author} {\bibfnamefont {P.}~\bibnamefont
  {Emonts}}\ and\ \bibinfo {author} {\bibfnamefont {E.}~\bibnamefont {Zohar}},\
  }\bibfield  {title} {\bibinfo {title} {Fermionic {Gaussian} {PEPS} in $3+1d$:
  {Rotations} and {Relativistic} {Limits}},\ }\href
  {https://doi.org/10.1103/PhysRevD.108.014514} {\bibfield  {journal} {\bibinfo
   {journal} {Physical Review D}\ }\textbf {\bibinfo {volume} {108}},\ \bibinfo
  {pages} {014514} (\bibinfo {year} {2023})},\ \Eprint
  {https://arxiv.org/abs/2304.06744} {arxiv:2304.06744} \BibitemShut {NoStop}%
\bibitem [{\citenamefont
  {Vadacchino}(2023)}]{Vadacchino2023ReviewGlueballHunting}%
  \BibitemOpen
  \bibfield  {author} {\bibinfo {author} {\bibfnamefont {D.}~\bibnamefont
  {Vadacchino}},\ }\href@noop {} {\emph {\bibinfo {title} {A review on
  {Glueball} hunting}}},\ \bibinfo {type} {Tech. Rep.}\ (\bibinfo {year}
  {2023})\ \Eprint {https://arxiv.org/abs/2305.04869} {arxiv:2305.04869}
  \BibitemShut {NoStop}%
\bibitem [{\citenamefont {Cataldi}\ \emph {et~al.}(2023)\citenamefont
  {Cataldi}, \citenamefont {Magnifico}, \citenamefont {Silvi},\ and\
  \citenamefont {Montangero}}]{Cataldi202321dSu2Yang}%
  \BibitemOpen
  \bibfield  {author} {\bibinfo {author} {\bibfnamefont {G.}~\bibnamefont
  {Cataldi}}, \bibinfo {author} {\bibfnamefont {G.}~\bibnamefont {Magnifico}},
  \bibinfo {author} {\bibfnamefont {P.}~\bibnamefont {Silvi}},\ and\ \bibinfo
  {author} {\bibfnamefont {S.}~\bibnamefont {Montangero}},\ }\href
  {https://doi.org/10.48550/arXiv.2307.09396} {\emph {\bibinfo {title}
  {(2+1){D} {SU}(2) {Yang}-{Mills} {Lattice} {Gauge} {Theory} at finite density
  via tensor networks}}},\ \bibinfo {type} {Tech. Rep.}\ (\bibinfo {year}
  {2023})\ \Eprint {https://arxiv.org/abs/2307.09396} {arxiv:2307.09396}
  \BibitemShut {NoStop}%
\bibitem [{\citenamefont {Carena}\ \emph {et~al.}(2022)\citenamefont {Carena},
  \citenamefont {Lamm}, \citenamefont {Li},\ and\ \citenamefont
  {Liu}}]{Carena2022ImprovedHamiltoniansQuantum}%
  \BibitemOpen
  \bibfield  {author} {\bibinfo {author} {\bibfnamefont {M.}~\bibnamefont
  {Carena}}, \bibinfo {author} {\bibfnamefont {H.}~\bibnamefont {Lamm}},
  \bibinfo {author} {\bibfnamefont {Y.-Y.}\ \bibnamefont {Li}},\ and\ \bibinfo
  {author} {\bibfnamefont {W.}~\bibnamefont {Liu}},\ }\bibfield  {title}
  {\bibinfo {title} {Improved {Hamiltonians} for {Quantum} {Simulations} of
  {Gauge} {Theories}},\ }\href {https://doi.org/10.1103/PhysRevLett.129.051601}
  {\bibfield  {journal} {\bibinfo  {journal} {Physical Review Letters}\
  }\textbf {\bibinfo {volume} {129}},\ \bibinfo {pages} {051601} (\bibinfo
  {year} {2022})}\BibitemShut {NoStop}%
\bibitem [{\citenamefont {Beard}\ \emph {et~al.}(1998)\citenamefont {Beard},
  \citenamefont {Brower}, \citenamefont {Chandrasekharan}, \citenamefont
  {Chen}, \citenamefont {Tsapalis},\ and\ \citenamefont
  {Wiese}}]{Beard1998DTheoryField}%
  \BibitemOpen
  \bibfield  {author} {\bibinfo {author} {\bibfnamefont {B.~B.}\ \bibnamefont
  {Beard}}, \bibinfo {author} {\bibfnamefont {R.~C.}\ \bibnamefont {Brower}},
  \bibinfo {author} {\bibfnamefont {S.}~\bibnamefont {Chandrasekharan}},
  \bibinfo {author} {\bibfnamefont {D.}~\bibnamefont {Chen}}, \bibinfo {author}
  {\bibfnamefont {A.}~\bibnamefont {Tsapalis}},\ and\ \bibinfo {author}
  {\bibfnamefont {U.-J.}\ \bibnamefont {Wiese}},\ }\bibfield  {title} {\bibinfo
  {title} {{D-Theory}: {Field} {Theory} via {Dimensional} {Reduction} of
  {Discrete} {Variables}},\ }\href
  {https://doi.org/10.1016/S0920-5632(97)00900-6} {\bibfield  {journal}
  {\bibinfo  {journal} {Nucl. Phys. B: Proc. Suppl.}\ }\bibinfo {series}
  {Proceedings of the {XVth} {International} {Symposium} on {Lattice} {Field}
  {Theory}},\ \textbf {\bibinfo {volume} {63}},\ \bibinfo {pages} {775}
  (\bibinfo {year} {1998})}\BibitemShut {NoStop}%
\bibitem [{\citenamefont {Ludl}(2010)}]{Ludl2010SystematicAnalysisFinite}%
  \BibitemOpen
  \bibfield  {author} {\bibinfo {author} {\bibfnamefont {P.~O.}\ \bibnamefont
  {Ludl}},\ }\href {https://doi.org/10.48550/arXiv.0907.5587} {\emph {\bibinfo
  {title} {Systematic analysis of finite family symmetry groups and their
  application to the lepton sector}}},\ \bibinfo {type} {Tech. Rep.}\ (\bibinfo
  {year} {2010})\ \Eprint {https://arxiv.org/abs/0907.5587} {arxiv:0907.5587}
  \BibitemShut {NoStop}%
\bibitem [{\citenamefont
  {Flyvbjerg}(1984)}]{Flyvbjerg1984GroupSpaceDecimation}%
  \BibitemOpen
  \bibfield  {author} {\bibinfo {author} {\bibfnamefont {H.}~\bibnamefont
  {Flyvbjerg}},\ }\bibfield  {title} {\bibinfo {title} {Group space decimation:
  {A} way to simulate {QCD} by the 1080 element subgroup of {SU}(3)?},\ }\href
  {https://doi.org/10.1016/0550-3213(84)90033-6} {\bibfield  {journal}
  {\bibinfo  {journal} {Nuclear Physics B}\ }\textbf {\bibinfo {volume}
  {243}},\ \bibinfo {pages} {350} (\bibinfo {year} {1984})}\BibitemShut
  {NoStop}%
\bibitem [{\citenamefont {Zwicky}\ and\ \citenamefont
  {Fischbacher}(2009)}]{Zwicky2009DiscreteMinimalFlavor}%
  \BibitemOpen
  \bibfield  {author} {\bibinfo {author} {\bibfnamefont {R.}~\bibnamefont
  {Zwicky}}\ and\ \bibinfo {author} {\bibfnamefont {T.}~\bibnamefont
  {Fischbacher}},\ }\bibfield  {title} {\bibinfo {title} {Discrete minimal
  flavor violation},\ }\href {https://doi.org/10.1103/PhysRevD.80.076009}
  {\bibfield  {journal} {\bibinfo  {journal} {Physical Review D}\ }\textbf
  {\bibinfo {volume} {80}},\ \bibinfo {pages} {076009} (\bibinfo {year}
  {2009})}\BibitemShut {NoStop}%
\bibitem [{\citenamefont {Merle}\ and\ \citenamefont
  {Zwicky}(2012)}]{Merle2012ExplicitSpontaneousBreaking}%
  \BibitemOpen
  \bibfield  {author} {\bibinfo {author} {\bibfnamefont {A.}~\bibnamefont
  {Merle}}\ and\ \bibinfo {author} {\bibfnamefont {R.}~\bibnamefont {Zwicky}},\
  }\bibfield  {title} {\bibinfo {title} {Explicit and spontaneous breaking of
  {SU}(3) into its finite subgroups},\ }\href
  {https://doi.org/10.1007/JHEP02(2012)128} {\bibfield  {journal} {\bibinfo
  {journal} {Journal of High Energy Physics}\ }\textbf {\bibinfo {volume}
  {2012}},\ \bibinfo {pages} {128} (\bibinfo {year} {2012})}\BibitemShut
  {NoStop}%
\bibitem [{\citenamefont {Bloch}\ \emph {et~al.}(2022)\citenamefont {Bloch},
  \citenamefont {Lohmayer}, \citenamefont {Schweiss},\ and\ \citenamefont
  {Unmuth-Yockey}}]{Bloch2021EffectiveMathbbz3Model}%
  \BibitemOpen
  \bibfield  {author} {\bibinfo {author} {\bibfnamefont {J.}~\bibnamefont
  {Bloch}}, \bibinfo {author} {\bibfnamefont {R.}~\bibnamefont {Lohmayer}},
  \bibinfo {author} {\bibfnamefont {S.}~\bibnamefont {Schweiss}},\ and\
  \bibinfo {author} {\bibfnamefont {J.}~\bibnamefont {Unmuth-Yockey}},\
  }\bibfield  {title} {\bibinfo {title} {Effective $\mathbb{Z}_{3}$ model for
  finite-density {QCD} with tensor networks},\ }in\ \href
  {https://doi.org/10.22323/1.396.0062} {\emph {\bibinfo {booktitle}
  {Proceedings of The 38th International Symposium on Lattice Field Theory
  {\textemdash} PoS(LATTICE2021)}}},\ Vol.\ \bibinfo {volume} {396}\ (\bibinfo
  {year} {2022})\ p.\ \bibinfo {pages} {062},\ \Eprint
  {https://arxiv.org/abs/2110.09499} {arxiv:2110.09499} \BibitemShut {NoStop}%
\bibitem [{\citenamefont {Scopa}\ \emph {et~al.}(2022)\citenamefont {Scopa},
  \citenamefont {Calabrese},\ and\ \citenamefont
  {Bastianello}}]{Scopa2022EntanglementDynamicsConfining}%
  \BibitemOpen
  \bibfield  {author} {\bibinfo {author} {\bibfnamefont {S.}~\bibnamefont
  {Scopa}}, \bibinfo {author} {\bibfnamefont {P.}~\bibnamefont {Calabrese}},\
  and\ \bibinfo {author} {\bibfnamefont {A.}~\bibnamefont {Bastianello}},\
  }\bibfield  {title} {\bibinfo {title} {Entanglement dynamics in confining
  spin chains},\ }\href {https://doi.org/10.1103/PhysRevB.105.125413}
  {\bibfield  {journal} {\bibinfo  {journal} {Physical Review B}\ }\textbf
  {\bibinfo {volume} {105}},\ \bibinfo {pages} {125413} (\bibinfo {year}
  {2022})},\ \Eprint {https://arxiv.org/abs/2111.11483} {arXiv:2111.11483}
  \BibitemShut {NoStop}%
\bibitem [{\citenamefont {Vovrosh}\ \emph {et~al.}(2022)\citenamefont
  {Vovrosh}, \citenamefont {Zhao}, \citenamefont {Knolle},\ and\ \citenamefont
  {Bastianello}}]{Vovrosh2022ConfinementInducedImpurity}%
  \BibitemOpen
  \bibfield  {author} {\bibinfo {author} {\bibfnamefont {J.}~\bibnamefont
  {Vovrosh}}, \bibinfo {author} {\bibfnamefont {H.}~\bibnamefont {Zhao}},
  \bibinfo {author} {\bibfnamefont {J.}~\bibnamefont {Knolle}},\ and\ \bibinfo
  {author} {\bibfnamefont {A.}~\bibnamefont {Bastianello}},\ }\bibfield
  {title} {\bibinfo {title} {Confinement induced impurity states in spin
  chains},\ }\href {https://doi.org/10.1103/PhysRevB.105.L100301} {\bibfield
  {journal} {\bibinfo  {journal} {Physical Review B}\ }\textbf {\bibinfo
  {volume} {105}},\ \bibinfo {pages} {L100301} (\bibinfo {year} {2022})},\
  \Eprint {https://arxiv.org/abs/2108.03976} {arXiv:2108.03976} \BibitemShut
  {NoStop}%
\bibitem [{\citenamefont {Ba{\~n}uls}\ \emph {et~al.}(2022)\citenamefont
  {Ba{\~n}uls}, \citenamefont {Heller}, \citenamefont {Jansen}, \citenamefont
  {Knaute},\ and\ \citenamefont
  {Svensson}}]{Banuls2022QuantumInformationPerspective}%
  \BibitemOpen
  \bibfield  {author} {\bibinfo {author} {\bibfnamefont {M.~C.}\ \bibnamefont
  {Ba{\~n}uls}}, \bibinfo {author} {\bibfnamefont {M.~P.}\ \bibnamefont
  {Heller}}, \bibinfo {author} {\bibfnamefont {K.}~\bibnamefont {Jansen}},
  \bibinfo {author} {\bibfnamefont {J.}~\bibnamefont {Knaute}},\ and\ \bibinfo
  {author} {\bibfnamefont {V.}~\bibnamefont {Svensson}},\ }\href
  {https://doi.org/10.48550/arXiv.2206.10528} {\emph {\bibinfo {title} {A
  quantum information perspective on meson melting}}},\ \bibinfo {type} {Tech.
  Rep.}\ (\bibinfo {year} {2022})\ \Eprint {https://arxiv.org/abs/2206.10528}
  {arxiv:2206.10528} \BibitemShut {NoStop}%
\bibitem [{\citenamefont {Knaute}(2023)}]{Knaute2023MesonContentEntanglement}%
  \BibitemOpen
  \bibfield  {author} {\bibinfo {author} {\bibfnamefont {J.}~\bibnamefont
  {Knaute}},\ }\bibfield  {title} {\bibinfo {title} {Meson content of
  entanglement spectra after integrable and nonintegrable quantum quenches},\
  }\href {https://doi.org/10.1103/PhysRevB.107.L100303} {\bibfield  {journal}
  {\bibinfo  {journal} {Physical Review B}\ }\textbf {\bibinfo {volume}
  {107}},\ \bibinfo {pages} {L100303} (\bibinfo {year} {2023})},\ \Eprint
  {https://arxiv.org/abs/2210.15682} {arXiv:2210.15682} \BibitemShut {NoStop}%
\bibitem [{\citenamefont
  {Rigobello}(2023{\natexlab{a}})}]{Rigobello2023Rgbmrc/simsio}%
  \BibitemOpen
  \bibfield  {author} {\bibinfo {author} {\bibfnamefont {M.}~\bibnamefont
  {Rigobello}},\ }\href {https://doi.org/10.5281/zenodo.7607199} {\bibinfo
  {title} {rgbmrc/simsio}} (\bibinfo {year} {2023}{\natexlab{a}})\BibitemShut
  {NoStop}%
\bibitem [{\citenamefont {Andreetto}\ \emph {et~al.}(2019)\citenamefont
  {Andreetto} \emph {et~al.}}]{Andreetto2019MergingOpenstackBased}%
  \BibitemOpen
  \bibfield  {author} {\bibinfo {author} {\bibfnamefont {P.}~\bibnamefont
  {Andreetto}} \emph {et~al.},\ }\bibfield  {title} {\bibinfo {title} {Merging
  {OpenStack}-based private clouds: the case of {CloudVeneto.it}},\ }\href
  {https://doi.org/10.1051/epjconf/201921407010} {\bibfield  {journal}
  {\bibinfo  {journal} {EPJ Web of Conferences}\ }\textbf {\bibinfo {volume}
  {214}},\ \bibinfo {pages} {07010} (\bibinfo {year} {2019})}\BibitemShut
  {NoStop}%
\bibitem [{\citenamefont {Jordan}\ and\ \citenamefont
  {Wigner}(1928)}]{Jordan1928UeberDasPaulische}%
  \BibitemOpen
  \bibfield  {author} {\bibinfo {author} {\bibfnamefont {P.}~\bibnamefont
  {Jordan}}\ and\ \bibinfo {author} {\bibfnamefont {E.}~\bibnamefont
  {Wigner}},\ }\bibfield  {title} {\bibinfo {title} {{{\"U}}ber das {Paulische}
  {{\"A}quivalenzverbot}},\ }\href {https://doi.org/10.1007/BF01331938}
  {\bibfield  {journal} {\bibinfo  {journal} {Zeitschrift Fur Physik}\ }\textbf
  {\bibinfo {volume} {47}},\ \bibinfo {pages} {631} (\bibinfo {year}
  {1928})}\BibitemShut {NoStop}%
\bibitem [{\citenamefont {Burgio}\ \emph {et~al.}(2000)\citenamefont {Burgio},
  \citenamefont {De~Pietri}, \citenamefont {Morales-T{\'e}cotl}, \citenamefont
  {Urrutia},\ and\ \citenamefont {Vergara}}]{Burgio2000BasisPhysicalHilbert}%
  \BibitemOpen
  \bibfield  {author} {\bibinfo {author} {\bibfnamefont {G.}~\bibnamefont
  {Burgio}}, \bibinfo {author} {\bibfnamefont {R.}~\bibnamefont {De~Pietri}},
  \bibinfo {author} {\bibfnamefont {H.~A.}\ \bibnamefont {Morales-T{\'e}cotl}},
  \bibinfo {author} {\bibfnamefont {L.~F.}\ \bibnamefont {Urrutia}},\ and\
  \bibinfo {author} {\bibfnamefont {J.~D.}\ \bibnamefont {Vergara}},\
  }\bibfield  {title} {\bibinfo {title} {The basis of the physical {Hilbert}
  space of lattice gauge theories},\ }\href
  {https://doi.org/10.1016/S0550-3213(99)00533-7} {\bibfield  {journal}
  {\bibinfo  {journal} {Nuclear Physics B}\ }\textbf {\bibinfo {volume}
  {566}},\ \bibinfo {pages} {547} (\bibinfo {year} {2000})}\BibitemShut
  {NoStop}%
\bibitem [{\citenamefont
  {Rigobello}(2023{\natexlab{b}})}]{Rigobello2023QhlgtModels}%
  \BibitemOpen
  \bibfield  {author} {\bibinfo {author} {\bibfnamefont {M.}~\bibnamefont
  {Rigobello}},\ }\href {https://baltig.infn.it/qpd/qhlgt-models} {\bibinfo
  {title} {{QHLGT}-models}} (\bibinfo {year} {2023}{\natexlab{b}})\BibitemShut
  {NoStop}%
\bibitem [{\citenamefont {Kordov}\ \emph {et~al.}(2023)\citenamefont {Kordov},
  \citenamefont {Horsley}, \citenamefont {Kamleh}, \citenamefont {Nakamura},
  \citenamefont {Perlt}, \citenamefont {Rakow}, \citenamefont {Schierholz},
  \citenamefont {St{\"u}ben}, \citenamefont {Young},\ and\ \citenamefont
  {Zanotti}}]{Kordov2023WeakDecayConstants}%
  \BibitemOpen
  \bibfield  {author} {\bibinfo {author} {\bibfnamefont {Z.~R.}\ \bibnamefont
  {Kordov}}, \bibinfo {author} {\bibfnamefont {R.}~\bibnamefont {Horsley}},
  \bibinfo {author} {\bibfnamefont {W.}~\bibnamefont {Kamleh}}, \bibinfo
  {author} {\bibfnamefont {Y.}~\bibnamefont {Nakamura}}, \bibinfo {author}
  {\bibfnamefont {H.}~\bibnamefont {Perlt}}, \bibinfo {author} {\bibfnamefont
  {P.~E.~L.}\ \bibnamefont {Rakow}}, \bibinfo {author} {\bibfnamefont
  {G.}~\bibnamefont {Schierholz}}, \bibinfo {author} {\bibfnamefont
  {H.}~\bibnamefont {St{\"u}ben}}, \bibinfo {author} {\bibfnamefont {R.~D.}\
  \bibnamefont {Young}},\ and\ \bibinfo {author} {\bibfnamefont {J.~M.}\
  \bibnamefont {Zanotti}},\ }\href {https://doi.org/10.48550/arXiv.2304.06095}
  {\emph {\bibinfo {title} {Weak decay constants of the pseudoscalar mesons
  from lattice {QCD}+{QED}}}},\ \bibinfo {type} {Tech. Rep.}\ (\bibinfo {year}
  {2023})\ \Eprint {https://arxiv.org/abs/2304.06095} {arxiv:2304.06095}
  \BibitemShut {NoStop}%
\bibitem [{\citenamefont {Mermin}\ and\ \citenamefont
  {Wagner}(1966)}]{Mermin1966AbsenceFerromagnetismAntiferromagnetism}%
  \BibitemOpen
  \bibfield  {author} {\bibinfo {author} {\bibfnamefont {N.~D.}\ \bibnamefont
  {Mermin}}\ and\ \bibinfo {author} {\bibfnamefont {H.}~\bibnamefont
  {Wagner}},\ }\bibfield  {title} {\bibinfo {title} {Absence of
  {Ferromagnetism} or {Antiferromagnetism} in {One}- or {Two}-{Dimensional}
  {Isotropic} {Heisenberg} {Models}},\ }\href
  {https://doi.org/10.1103/PhysRevLett.17.1133} {\bibfield  {journal} {\bibinfo
   {journal} {Physical Review Letters}\ }\textbf {\bibinfo {volume} {17}},\
  \bibinfo {pages} {1133} (\bibinfo {year} {1966})}\BibitemShut {NoStop}%
\bibitem [{\citenamefont {Hohenberg}(1967)}]{Hohenberg1967ExistenceLongRange}%
  \BibitemOpen
  \bibfield  {author} {\bibinfo {author} {\bibfnamefont {P.~C.}\ \bibnamefont
  {Hohenberg}},\ }\bibfield  {title} {\bibinfo {title} {Existence of
  {Long}-{Range} {Order} in {One} and {Two} {Dimensions}},\ }\href
  {https://doi.org/10.1103/PhysRev.158.383} {\bibfield  {journal} {\bibinfo
  {journal} {Physical Review}\ }\textbf {\bibinfo {volume} {158}},\ \bibinfo
  {pages} {383} (\bibinfo {year} {1967})}\BibitemShut {NoStop}%
\bibitem [{\citenamefont {Sachdev}(2011)}]{Sachdev2011QuantumPhaseTransitions}%
  \BibitemOpen
  \bibfield  {author} {\bibinfo {author} {\bibfnamefont {S.}~\bibnamefont
  {Sachdev}},\ }\href {https://doi.org/10.1017/CBO9780511973765} {\emph
  {\bibinfo {title} {Quantum {Phase} {Transitions}}}},\ \bibinfo {edition}
  {2nd}\ ed.\ (\bibinfo  {publisher} {Cambridge University Press},\ \bibinfo
  {address} {Cambridge},\ \bibinfo {year} {2011})\BibitemShut {NoStop}%
\end{thebibliography}
\end{document}